\documentclass[12pt]{article}
\usepackage{hyperref}
\usepackage[top=2cm,bottom=3cm,left=2cm,right=2cm]{geometry}
\usepackage{graphicx}
\usepackage{amsmath}
\usepackage{amssymb} 
\usepackage{hyperref}
\usepackage{xcolor}

\numberwithin{equation}{section}
\numberwithin{figure}{section}

\def\UHP{\mathrm{UHP}}
\def\BCFT{\mathrm{BCFT}}
\def\DBI{\mathrm{DBI}}

\graphicspath{ {./images/} }

\begin{document}

\begin{titlepage}

	\hfill \today
	\begin{center}
		\vskip 2cm
		
		{\Large \bf Symplectic structure in open string field theory III: \\ Electric field}

		\vskip 0.5cm
		
		\vskip 1.0cm
		{\large {Vin\'{\i}cius Bernardes$^{1}$, Theodore Erler$^{1}$, and Atakan Hilmi F{\i}rat$^{2}$ }}
		
		\vskip 0.5cm
		
		{\em  \hskip -.1truecm
			$^{1}$
			CEICO, FZU - Institute of Physics of the Czech Academy of Sciences \\
			No Slovance 2, 182 21, Prague 8, Czech Republic
			\\
			\vskip 0.5cm
			$^{2}$
			Center for Quantum Mathematics and Physics (QMAP),
			Department of Physics \& Astronomy, \\
			University of California, Davis, CA 95616, USA
			\\
			\vskip 0.5cm
			\tt \href{mailto:viniciusbernsilva@gmail.com}{viniciusbernsilva@gmail.com},
		\href{mailto:tchovi@gmail.com}{tchovi@gmail.com},
		 \href{mailto:ahfirat@ucdavis.edu}{ahfirat@ucdavis.edu} \vskip 5pt }
		
		\vskip 2.0cm
		{\bf Abstract}
		
	\end{center}
	\vskip 0.25cm
	\noindent
	\begin{narrower}
		\baselineskip15pt
		
		\noindent  We use a new formula for the symplectic structure on the phase space of open string field theory to evaluate the energy of a D-brane carrying a constant electric flux. This is shown to be consistent with the energy computed using the Dirac-Born-Infeld action through a generalization of the Ellwood invariant to nonpolynomial open string field theories.
	
	\end{narrower}
\end{titlepage}

\tableofcontents
\baselineskip15pt

\section{Introduction}

This is the final installment of the three papers~\cite{Bernardes:2025zzu,Bernardes:2025II} on applications of a new formula for symplectic structure~\cite{Bernardes:2025uzg} on the phase space of open bosonic string field theory (SFT).\footnote{For reviews refer to~\cite{Sen:2024nfd,Erler:2019vhl,Erler:2019loq,Erbin:2021smf, deLacroix:2017lif,Maccaferri:2023vns}.} In previous works we have investigated perturbative rolling tachyon solutions in Siegel gauge~\cite{Sen:2002nu,Cho:2023khj} and an analytic lump solution moving at constant velocity~\cite{Erler:2014eqa,Erler:2019fye}. In this work we apply the symplectic structure to a D-brane carrying a constant electric flux in bosonic string theory.

This system admits a description in terms of a perturbative open SFT solution constructed based on the marginal operator $\Psi_1 = c X^0 \partial X^1$ in Siegel gauge. Such a time-dependent solution has not been studied before explicitly so our first result is its construction. The solution is in some ways analogous to the solutions describing the Ramond-Ramond deformations in Type II closed super SFT~\cite{Cho:2018nfn,Cho:2023mhw,Kim:2024dnw,Cho:2025coy}, but is technically much simpler. Since we have already worked with Witten’s cubic open SFT~\cite{Witten:1985cc} and our analysis is perturbative, we will define the open SFT using $SL(2,\mathbb{R})$ vertices in the spirit of this contemporary discourse. These vertices turn out to be simpler to use for our computations, and illustrate how the symplectic structure operates in a nonpolynomial SFT. This may be helpful for future applications to closed SFT.

With the perturbative solution at hand, it is straightforward to evaluate the energy of a D-brane carrying a constant electric flux as an expansion in its marginality parameter $\varepsilon$,
\begin{align} \label{eq:1.1}
	E = 
	{1\over 2}\Omega_1 \varepsilon^2 
	+ {1 \over 4} \Omega_3 \varepsilon^4 
	+ \mathcal{O}\left(\varepsilon^6\right)\, ,
\end{align}
following the approaches of~\cite{Bernardes:2025zzu,Cho:2023khj}. There are important technical differences in the computations however. First, we employ the operator formalism of conformal field theory (CFT) to evaluate the correlation functions relevant for the symplectic structure. We remind that the analogous calculations have been performed using oscillator truncation method of Taylor~\cite{Taylor:2002bq} in~\cite{Bernardes:2025zzu} and analytic techniques~\cite{Erler:2019vhl} in~\cite{Bernardes:2025II}. Second, the solution is not exactly marginal, as indicated by the appearance of $L_0$-nilpotent states in the perturbative equations of motion. This forces us to consider string fields that involve multiple insertions of the bare operator $X^0$ at higher orders in $\varepsilon$. Relatedly, the correlators contain multiple insertions of $X^0$.  Dealing with them appropriately poses an additional challenge for the evaluations. Ultimately we handle them following the prescription provided in~\cite{Belopolsky:1995vi}.

The energy evaluated using the SFT symplectic structure~\eqref{eq:1.1} should be consistent with the energy determined from the boundary state considerations; and in extension, the Dirac-Born-Infeld (DBI) action~\cite{Fradkin:1985qd,Abouelsaood:1986gd,Leigh:1989jq}. In order to compare these energies we must understand how the SFT marginality parameter $\varepsilon$ is related to the DBI electric field $\varepsilon_\DBI$. This relation takes the form
\begin{align}
	\varepsilon_{\DBI} = c_1 \varepsilon + c_3 \varepsilon^3 + \mathcal{O}\left(\varepsilon^5\right) \, , 
\end{align}
where the coefficients $c_i$s can be determined using the Ellwood invariant~\cite{Ellwood:2008jh} adapted to nonpolynomial open SFTs formulated with cyclic $A_\infty$ algebras~\cite{Gaberdiel:1997ia}. After obtaining $c_i$s, we observe that the results from SFT and DBI frameworks are in agreement with each other to high degree of accuracy. This computation, along with the technical details for the perturbative solution and evaluation of the SFT symplectic form, can be found in the \texttt{Mathematica} notebook in the Supplementary material accompanying this paper.

\subsubsection*{Conventions}

We use mostly plus metric and set $\alpha'=1$. Fourier transform conventions are
\begin{equation}
	f(x) = \int \frac{d^Dk}{(2\pi)^D} e^{ik\cdot x}f(k) \, ,
	\quad \quad f(k) = \int d^D x \, e^{-ik\cdot x}f(x) \, ,
\end{equation}
where $f(x)$ is a function in position space and $f(k)$ is its Fourier transform in momentum space. The ghost correlator is normalized as 
\begin{equation} \label{eq:1.4}
	\left\langle c(z_1) c(z_2) c(z_3)\right\rangle_\UHP^\mathrm{gh} = z_{12}z_{13}z_{23} \, ,
	\quad \quad z_{ij}=z_i-z_j,
\end{equation}
and we use the left handed convention for the open string $A_\infty$ products~\cite{Erler:2019vhl}. Commutators are graded with respect to Grassmann parity.

\section{DBI action and energy} \label{sec:2}
We begin our discussion by computing the energy of a D1-brane carrying a constant electric flux using the Dirac-Born-Infeld (DBI) action~\cite{Fradkin:1985qd,Abouelsaood:1986gd,Leigh:1989jq,Polchinski:1998rq}
\begin{align} \label{eq:4.1}
	S_{\text{DBI}} = - T Z_\perp \int d^{2} x 
	\sqrt{-\det\big( \eta_{\mu \nu} + 2 \pi (F_{\DBI})_{\mu \nu}\big) } 
	\, .
\end{align}
We turn off the dilaton and Kalb-Ramond fields and take the background metric to be flat in the worldvolume directions $\mu =0,1$. The transverse directions are irrelevant for our purposes and their only effect is to multiply the action by an overall factor $Z_\perp$. Therefore $T_1 = T Z_\perp$ is the tension of the D1-brane, and in these conventions
\begin{align} \label{eq:2.2}
	T = {1 \over 2 \pi^2 g^2} \, ,
\end{align}
where $g$ is the coupling constant for the open SFT action~\cite{Sen:1999xm}. Finally, $(F_\DBI)_{\mu \nu}$ is the field strength. We often include the DBI label to the objects relevant for the boundary state/DBI description to distinguish them from their counterparts in the SFT description.

The DBI action reduces to the Maxwell action
\begin{align} \label{eq:2.3}
	S_{\text{DBI}} = - T_1 V_1 \cdot \mathrm{vol} \left(X^0\right) - {Z_\perp \over 2 g^2}  \int d^2 x \, 
	\left(F_{\DBI}\right)_{\mu \nu} \left(F_{\DBI}\right)^{\mu \nu} + \cdots\, ,
\end{align}
upon expanding~\eqref{eq:4.1} in field strengths.  We take $V_1 = \int d x^1$ to be the volume of the $\mu = 1$ direction and $\mathrm{vol} \left(X^0\right)$ to be the (infinite) volume of the temporal coordinate; hence $T_1 V_1$ is the mass of the D-brane as expected. We point out the normalization of the quadratic action above is different from the normalization descending from the SFT action by an overall factor of $2$ in our conventions; refer to the open SFT action~\eqref{eq:2.1} and the evaluation of its kinetic term for the massless fields in section 4.6 of~\cite{Sen:2024nfd}. This shows\footnote{This argument does not determine the overall sign for the field redefinition. We are going to fix this sign using the homotopy Ellwood invariant in section~\ref{sec:5.1}. An alternative approach can be found in~\cite{Coletti:2003ai}.}
\begin{align} \label{eq:10.3}
	\left(A_{\DBI}\right)_{\mu} = {1 \over \sqrt{2} } A_\mu + \cdots \, ,
\end{align}
for the field redefinition between the gauge fields. The dots stand for the terms higher order in~$A_\mu$.

The field strength can be expressed in terms of the electric field $\varepsilon^\DBI$,
\begin{align} \label{eq:2.5a}
	\left(F_{\DBI}\right)_{\mu \nu} = \begin{bmatrix}
		0 & \varepsilon_{\DBI} \\  -\varepsilon_{\DBI}&  0
	\end{bmatrix}_{\mu \nu} \, ,
\end{align}
which leads to
\begin{align} \label{eq:4.6}
	S_{\DBI} =  \int L_{\DBI} = - {T_1} \int d^2 x
	\sqrt{1 - \left(2 \pi \varepsilon_{\DBI}\right)^2} \, .
\end{align}
We can deduce the DBI symplectic form using this form of the action easily and then use it to compute energies following the approaches of~\cite{Bernardes:2025zzu,Cho:2023khj}. This is most efficiently done using the traditional covariant phase space methods~\cite{Harlow:2019yfa}, rather than using the new formula of~\cite{Bernardes:2025uzg}, since the kinetic operator of the DBI action is somewhat convoluted and the Lagrangian $L_{\DBI}$ is already available as shown above. In particular, varying $L_{\DBI}$ is straightforward:
\begin{align} \label{eq:10.4}
	\delta L_{\DBI} &= T_1 \, {4 \pi^2 \varepsilon_{\DBI} \delta \varepsilon_{\DBI}  \over \sqrt{1 -  \left( 2 \pi \varepsilon_\DBI\right)^2}} \, dt \wedge d x
	=  T_1  \, {4 \pi^2  \varepsilon^{\text{DBI}} \over \sqrt{1 - 
		 \left( 2 \pi \varepsilon_\DBI\right)^2}} \, \delta d A_{\DBI} \, .
\end{align} 
In the second equality we used
\begin{align} \label{eq:4.7}
	\varepsilon_{\DBI}\, dt \wedge dx   = F_{\DBI} = d A_{\DBI} \, .
\end{align}
Arranging,
\begin{align} \label{eq:4.9}
	\delta L_{\DBI}  =
	T_1 \left[
	-d \left(
	{4 \pi^2 \varepsilon_{\DBI} \over \sqrt{1 - \left( 2 \pi \varepsilon_\DBI\right)^2}} 
	\right) \delta A_{\DBI}
	+ d \left(
	{4 \pi^2 \varepsilon_{\DBI} \over \sqrt{1 - \left( 2 \pi \varepsilon_\DBI\right)^2}} 
	\, \delta A_{\DBI}
	\right)
	\right]\, .
\end{align}
The first term encodes the DBI equation of motion while the second term yields the symplectic potential density. Varying the expression inside the parenthesis for the latter, we can read the symplectic current
\begin{align}
	\delta \Theta_{\DBI}
	=  T_1  \, { 4 \pi^2 \over \left(1- \left( 2 \pi \varepsilon_\DBI\right)^2\right)^{3/2}} 
	\, \delta \varepsilon_{\DBI} \delta A_{\DBI} \, ,
\end{align}
and obtain the symplectic form by integrating it over a time slice:
\begin{align}
	\Omega_{\DBI} &= {T_1} \int\limits_{t=0} { 4 \pi^2 \, \delta \varepsilon_{\DBI} \, \delta A_{\DBI}  \over  \left(1- \left( 2 \pi \varepsilon_\DBI\right)^2\right)^{3/2}} 
	= {2 Z_\perp  \over g^2} \int\limits_{t=0} \delta \varepsilon_{\DBI} \, \delta A_{\DBI}  + \cdots
	\, .
\end{align}
The integrands are one-forms above. As shown in the second equality, we obtain the symplectic form of the Maxwell theory for small electric fields $\varepsilon_{\DBI}$ after taking~\eqref{eq:10.3} into account, refer to~\cite{Bernardes:2025uzg}.

A constant electric field solves the DBI equations of motion---inspect the first term in~\eqref{eq:4.9}. It can be described using the gauge field profile
\begin{align}
	A_{\DBI}(t) = \varepsilon_{\DBI}(t-t_{\DBI,0}) dx \, ,
\end{align}
where $t_{\text{DBI},0}$ is the time when the gauge field $A_{\DBI}$ vanishes. This demonstrates that the phase space for constant electric field solutions to the DBI theory can be parameterized by $(t_{\text{DBI},0}, \varepsilon_\DBI)$. The variation of the gauge field profile above with $t_{\text{DBI},0}=0$ is
\begin{align}
	\delta A_\DBI = \big( t \, \delta \varepsilon_\DBI  - \varepsilon_\DBI \, \delta t_{\text{DBI},0} \big) dx \, ,
\end{align}
from which we evaluate
\begin{align}
	\Omega_\DBI &= - T_1 V_1 \, { 4 \pi^2 \varepsilon_\DBI \over \left(1- \left( 2 \pi \varepsilon_\DBI\right)^2\right)^{3/2}} \, \delta \varepsilon_\DBI \, \delta  t_{\text{DBI},0} 
	\nonumber \\
	&=\delta t_{\text{DBI},0} 
	\, \delta \left(
	{ T_1 V_1  \over \sqrt{1- \left( 2 \pi \varepsilon_\DBI\right)^2 }}  
	\right)   \, ,
\end{align}
on this phase space. The energy following from this expression is
\begin{align} \label{eq:10.11}
	E_\DBI &= { T_1 V_1   \over \sqrt{1- \left( 2 \pi \varepsilon_\DBI\right)^2  }}  - T_1 V_1 
	\nonumber\\
	&={V \over g^2} \bigg[ {(\varepsilon_\DBI)^2} + {3 \pi^2 (\varepsilon_\DBI)^4} + \mathcal{O}\big((\varepsilon_\DBI)^6\big)  \bigg] \, ,
\end{align}
after adjusting the zero point energy so that the D1-brane without electric flux has vanishing energy. Here we introduced $V = V_1 Z_\perp$. Observe that we obtain the well-known result for the energy stored in small constant electric fields after taking~\eqref{eq:10.3} into account.

In the view of $T_1 V_1$ being the mass of the D1-brane, the similarity of the manipulations above to the standard relativistic energy-momentum manipulations is apparent. Indeed, this is not a coincidence but can be understood as a consequence of the T-duality along the (compact) $x^1$ direction~\cite{Polchinski:1998rq}: A D1-brane carrying a constant electric flux is dual to a D0-brane moving with constant velocity. In the T-dual frame $-2 \pi \varepsilon_\DBI$ is interpreted as the velocity, while $2 \pi \varepsilon_\DBI t_{\text{DBI},0}$ is interpreted as the initial position of the moving D0-brane. Thus the computation of the DBI energy above is equivalent to the computation of the spatial momentum of a D0-brane moving in $x^1$ direction in the T-dual frame. The latter has been investigated using analytic open SFT solutions in~\cite{Bernardes:2025II}.

\section{D-brane with constant electric flux}

In this section we construct a perturbative open SFT solution for a D-brane carrying a constant electric flux in Siegel gauge. The action of open SFT is
\begin{align} \label{eq:2.1}
	S[\Psi] = - {1 \over g^2} \bigg[
	{1 \over 2} \left\langle \Psi, v_1 \Psi \right\rangle
	+ {1 \over 3}\left\langle  \Psi, v_2(\Psi, \Psi) \right\rangle
	+ {1 \over 4} \left\langle \Psi, v_3(\Psi, \Psi, \Psi) \right\rangle+ \mathcal{O}\big(\Psi^5\big)
	\bigg] \, ,
\end{align}
where $v_1 = Q$ is the BRST operator, $\langle , \rangle$ is the BPZ inner product, and $g$ is the open string coupling constant. The string field $\Psi$ is ghost number one, Grassmann odd element of the Hilbert space of boundary conformal field theory (BCFT) associated with the D-brane background. We take this BCFT to be
\begin{align}
	\BCFT = \BCFT_{\text{D1}} \otimes \BCFT_\perp \otimes \BCFT_{bc} \, ,
\end{align}
where $\BCFT_{\text{D1}}$ consists of spacelike and timelike free bosons $X^\mu$ ($\mu = 0,1$) subject to Neumann boundary conditions. It encodes the D1-brane worldvolume along which the electric field is going to be turned on. The timelike $X^0$ is noncompact, while we compactify $X^1$ on a circle of volume $V_1$ to regularize the spatial volume divergence. $\BCFT_\perp$ is an arbitrary $c=24$ BCFT describing the transverse directions to this ``D1-brane'' and $\BCFT_{bc}$ is the standard $c=-26$ reparametrization ghost BCFT. The disk partition function of $\BCFT_\perp$ is assumed to be nonvanishing:
\begin{align} \label{eq:3.3}
	\langle 1 \rangle_{\text{disk}}^{\perp} = Z_\perp \neq 0 \, .
\end{align}
This factor is same as the transverse factor appearing in the DBI action~\eqref{eq:4.1}.

The open string products $v_n$s ($n \geq 2$) in~\eqref{eq:2.1} encode the elementary interactions and they satisfy the $A_\infty$ relations to ensure that the theory is gauge invariant~\cite{Gaberdiel:1997ia}. For instance, all higher products beyond $v_2$ vanish in Witten's theory~\cite{Witten:1985cc} and $A_\infty$ relations reduce to the defining relations of a differential graded associative algebra. We keep higher products in this work since we would like to work with a nonpolynomial SFT. The open string products for this generic situation can be constructed using~\emph{string vertices} $V_{n+1}$, which are given as integrals over specific regions $\mathcal{V}_{n+1}$ of the moduli space of disks with $n+1$ cyclically-ordered boundary punctures endowed with local coordinates that are sufficiently far away from degeneration. The defining relation is
\begin{align} \label{eq:3.4}
	V_{n+1}\left(\Psi_1, \cdots, \Psi_{n+1}\right) &= 
	\left\langle \Psi_1, v_n(\Psi_2, \cdots, \Psi_{n+1}) \right\rangle 
	\nonumber \\
	&= \int\limits_{\mathcal{V}_{n+1}} 
	\left\langle \mathcal{B}
	\Psi_1(w_1 = 0) \cdots \Psi_{n+1} (w_{n+1} = 0) 
	\right\rangle_\UHP \, .
\end{align}
Here $w_i$ are the local coordinates around the punctures, which are related to the coordinate $z$ on the ``uniformizing'' upper-half plane (UHP) via the maps $w_i = w_i(z)$. The string field $\Psi_i$ is inserted at the origin of the upper unit half-disk on the $w_i$-frame. In order to declutter the expressions we do not explicitly show the transformations by the local coordinates maps in the correlators. Instead, we implicitly understand that the arguments of the insertions also indicate the conformal frame used for them. The symbol $\mathcal{B}$ stands for the $b$-ghost insertions defining the measure on moduli spaces. Its construction, including the choices for the signs, is described in~\cite{Sen:2024nfd, Sen:2024npu}. We closely follow their construction after adopting to the left handed conventions. We highlight that the maps $w_i = w_i(z)$ are not arbitrary: they are chosen in a manner that $\mathcal{V}_{n+1}$ satisfy the geometric master equation~\cite{Zwiebach:1997fe}. This in turn implies that the products $v_n$ define a cyclic $A_\infty$ algebra. 

The equation of motion after varying~\eqref{eq:2.1} is
\begin{align} \label{eq:3.5}
	v_1 \Psi + v_2(\Psi, \Psi) + v_3(\Psi, \Psi, \Psi) + \mathcal{O}\big(\Psi^4\big) = 0 \, .
\end{align}
Upon making the perturbative ansatz 
\begin{align} \label{eq:2.5}
	\Psi = \varepsilon \Psi_1 + \varepsilon^2 \Psi_2 + \varepsilon^3 \Psi_3 + \mathcal{O}(\varepsilon^4) \, ,
\end{align}
for its solutions, we have
\begin{subequations} \label{eq:2.7}
\begin{align}
	&\mathcal{O}\left(\varepsilon^1\right):  
	\quad \quad
	0 = v_1 \Psi_1 \, , \\ \label{eq:3.7b}
	&\mathcal{O}\left(\varepsilon^2\right): 
	\quad \quad
	0 = v_1 \Psi_2  + v_2 (\Psi_1, \Psi_1) \, , \\
	&\mathcal{O}\left(\varepsilon^3\right):  
	\quad \quad
	0 = v_1 \Psi_3 + v_2(\Psi_1, \Psi_2) +  v_2(\Psi_2, \Psi_1) + v_3 (\Psi_1, \Psi_1, \Psi_1) \, . \\
	\nonumber
	&\quad\vdots \hspace{5cm} \vdots \hspace{5cm} \vdots
\end{align}
\end{subequations}
These equations can be solved order-by-order in $\varepsilon$. Observe that the first equation demands that the string field $\Psi_1$ is BRST-closed.

We want to describe a D-brane carrying a constant electric flux as a perturbative open SFT solution~\eqref{eq:2.5}. We proceed by fixing the Siegel gauge $b_0 \Psi = 0$ and turning on a constant electric field in the $\mu = 1$ direction. That is, we consider the weight-$0$ primary
\begin{align} \label{eq:2.8}
	\Psi_1 = c X^0 \partial X^1(0, 0) | 0 \rangle  =- {i \over \sqrt{2} } c_1 x^0 \alpha_{-1}^1 | 0 \rangle \, ,
	\quad \quad
	Q \Psi_1 = 0 \, ,
\end{align}
and construct the solution based on~\eqref{eq:2.7} while assuming that the state $c X^0 \partial X^1$ (and $c \partial X^1$) only appears at $\mathcal{O}(\varepsilon)$. Here $|0 \rangle$ is the $SL(2,\mathbb{R})$-invariant vacuum. We can motivate the form of $\Psi_1$ by observing that it provides the following expectation value for the SFT gauge field (i.e., the coefficient of $c_1 \alpha_{-1}^\mu|0 \rangle $) in the $\mu =1$ direction:
\begin{align} \label{eq:3.9}
	A_1(t) = - {i \over \sqrt{2}} \, \varepsilon  t + \mathcal{O} (\varepsilon^2) \, ,
\end{align}
which gives rise to a constant electric field at the leading order in $\varepsilon$. For this reason we often call $\varepsilon$ SFT electric field. Note that the marginal operator~\eqref{eq:2.8} provides the only~\emph{single-valued} perturbative Siegel gauge solution describing a constant electric field since $X^1$ BCFT is assumed to be compact. There is another solution for the electric field generated from the marginal operator $c X^1 \partial X^0$ if $X^1$ is noncompact instead. This has a linear profile in space and would be multi-valued when $X^1$ is compactified. Perhaps such a solution can be given a meaning in the patch-by-patch description of Frenkel and Kim~\cite{Frenkel:2025wko}, but this will be discussed elsewhere.

At second order the solution can be expressed as
\begin{align} \label{eq:3.10}
	\Psi_2 = - {b_0 \over L_0} (1- \mathbb{P}) v_2(\Psi_1, \Psi_1) +  \psi_2 \, .
\end{align}
Here $\psi_2 = \mathbb{P} \Psi_2$ and $\mathbb{P}$ is the projector onto~\emph{$L_0$-nilpotent states}---states $\psi$ satisfying $L_0^N \psi = 0$ for a finite $N \geq 0$. These are essentially weight-$0$ states containing polynomial functions of spacetime coordinates for which the action of the propagator $1/L_0$ is a priori ambigious. Plugging~\eqref{eq:3.10} to~\eqref{eq:3.7b}, we see that $\Psi_2$ solves the equation of motion as long as
\begin{align} \label{eq:2.11}
	Q \psi_2 + \mathbb{P} v_2(\Psi_1, \Psi_1) = 0 \, ,
\end{align}
after using
\begin{align}
	\left[Q, {b_0 \over L_0}(1-\mathbb{P}) \right] = 1-\mathbb{P} \, .
\end{align}
Similarly, $L_0$-nilpotent states $\psi_n = \mathbb{P} \Psi_n$ must be considered separately in the solution at higher orders and satisfy equations akin to~\eqref{eq:2.11}. We collectively refer them as~\emph{obstruction terms} since they represent obstructions to the exact marginality of $\Psi_1$~\eqref{eq:2.8}.

The $L_0$-nilpotent states in this work always take the form
\begin{align} 
	\psi = - {i \over \sqrt{2} } c_0 c_1 p\left(x^0\right) \alpha_{-1}^1 |0 \rangle \, ,
\end{align}
where $p(x^0)$ is some polynomial of the temporal position zero mode $x^0$. The projector $\mathbb{P}$ acting on them can be explicitly written as
\begin{align} \label{eq:2.13}
	\mathbb{P} = 
	 {1 \over V}
	 \int\limits_{-\infty}^\infty {d E \over 2 \pi} 
	c_0 c_1\alpha_{-1}^1| E \rangle \langle -E | c_{-1} \alpha_{1}^1 \, , 
\end{align}
where
\begin{align} \label{eq:3.14}
	| E \rangle = e^{i E X^0(0,0)} |0 \rangle \, , \quad \quad
	\langle E | c_{-1} c_0 c_1 |E' \rangle = 2 \pi V \delta(E + E') \, ,
\end{align}
and $V = V_1 Z_\perp $ as in~\eqref{eq:10.11}. Using this presentation for $\mathbb{P}$ in~\eqref{eq:2.11}, we find
\begin{align}
	Q \psi_2 
	= -  {i \sqrt{2}  \over V} \int\limits_{-\infty}^\infty  {d E \over 2 \pi} 
	c_0 c_1\alpha_{-1}^1 | E \rangle \cdot 
	V_3 \left(c e^{-i E X^0} \partial X^1, c X^0 \partial X^1, c X^0 \partial X^1\right)  \, .
\end{align}
But note that the right hand side vanishes because the correlator of three $\partial X^1$s is zero. Therefore the second order obstruction term can be chosen to vanish and
\begin{align} \label{eq:2.16}
	\psi_2 = 0 \quad  \implies \quad \Psi_2 = - {b_0 \over L_0} (1- \mathbb{P}) v_2(\Psi_1, \Psi_1) \, .
\end{align}
This argument is independent of the choice of string vertices and can be generalized to establish that the obstruction term can be chosen to be trivial at every even order i.e., $\psi_n = 0$ for even $n$.

At third order the solution is
\begin{align} \label{eq:3.17}
	\Psi_3 = - {b_0 \over L_0} (1- \mathbb{P}) 
	\big[ v_2(\Psi_1, \Psi_2) + v_2(\Psi_2, \Psi_1) + v_3(\Psi_1, \Psi_1,\Psi_1) \big] + \psi_3 \, ,
\end{align}
where the obstruction term $\psi_3 = \mathbb{P} \Psi_3$ obeys
\begin{align} \label{eq:3.18}
	Q \psi_3 + \mathbb{P} \big[ v_2(\Psi_1, \Psi_2) + v_2(\Psi_2, \Psi_1) + v_3(\Psi_1, \Psi_1,\Psi_1) \big] = 0 \, .
\end{align}
There are three distinct terms upon distributing the projector $\mathbb{P}$~\eqref{eq:2.13} inside the square bracket above. First two of them involve the string propagator by the virtue of containing $\Psi_2$~\eqref{eq:2.16} and they can be represented as a moduli integral over the quartic Feynman region obtained from sewing two cubic vertices. On the other hand, the last one involves the quartic vertex. Altogether, they can be understood as the following integral over the entire moduli space:
\begin{align} \label{eq:3.19}
	Q \psi_3 = & -{i \sqrt{2} \over V}  \int\limits_{-\infty}^\infty  {d E \over 2 \pi} c_0 c_1 \alpha_{-1}^1 |E \rangle \\
	&\int_0^1 { -du \over \rho_3(u) }
	\left\langle c e^{-i E X^0} \partial X^1 (w_1=0) c X^0 \partial X^1 (w_2=0) X^0 \partial X^1(w_3=0)  c X^0 \partial X^1(w_4=0)   
	\right\rangle_\UHP \, ,
	\nonumber
\end{align}
with an important caveat regarding the treatment of the integration around $u = 0$ and $u = 1$ that we are going to address shortly. Here $w_i$ ($i=1,2,3,4$) are the local coordinates around the punctures, which are canonically placed at $z= \infty,1,u,0$ respectively on the uniformizing UHP. Depending on cross ratio $u$, these local coordinates are either obtained from sewing two cubic vertices or by the choice of local coordinates for the quartic vertex.

Note that the third vertex operator in~\eqref{eq:3.19} is~\emph{integrated}. This follows from acting the $b$ ghosts associated with the cross ratio deformations to the insertions: applying $\mathcal{B}$ annihilates the insertions except for the component of $\mathcal{B}$ proportional to the mode $b_{-1}$. Such a mode exclusively appears at the third puncture, as one can easily establish by adapting the argument in Appendix A of~\cite{Yang:2005rx} to open strings. In effect, the $c$ ghost insertion at the third puncture gets replaced with $-du/\rho_3$, where $\rho_3 = \rho_3(u)$ is the~\emph{mapping radius} associated with the local coordinate map around this integrated puncture. Recall that the mapping radii is defined by
\begin{align} \label{eq:3.21}
	\rho_i \equiv {d z(w_i) \over d w_i} \bigg|_{w_i = 0} \, ,
\end{align}
for open strings (there is an extra division by $z^2$ when we map to $z = \infty$). We highlight that the orientation form in the moduli space of disks with four boundary punctures is given by $-du$ due to the left handed conventions, so the replacement factor above contains an additional sign in contrast to the sign conventions of~\cite{Sen:2024nfd}, see equation (3.82) there.

Before evaluating~\eqref{eq:3.19}, we need to explain how the bare operator $X^0$ should be treated in the correlators. The most straightforward way to deal with it is by replacing each $X^0$ with a plane wave operator $e^{i E X^0}$ as in~\cite{Belopolsky:1995vi} (also see~\cite{Kim:2026kex}),
\begin{align} \label{eq:3.20}
	\left\langle \cdots X^0 (w_i = 0) \cdots \right\rangle_{\UHP}
	=
	\left( - i {\partial \over \partial E}\right) 
	\left\langle \cdots e^{i E X^0} (w_i = 0) \cdots \right\rangle_{\UHP} \bigg|_{E = 0} \, ,
\end{align}
and evaluating the correlator accordingly before differentiating with respect to $E$. We emphasize that the insertion is no longer a weight-0 primary after the replacement, but instead carries a dependence on the local coordinates. This dependence often persists in the final result and it is important to keep track of this data carefully in this prescription. Also note that the final result is expected to contain derivatives of the delta function.

We are now ready to evaluate the correlator~\eqref{eq:3.19}. We find
\begin{align}
	&Q \psi_3 =  {i \sqrt{2} \over V}  \int\limits_{-\infty}^\infty  {d E_1 \over 2 \pi} c_0 c_1 \alpha_{-1}^1 |E_1 \rangle 
	\prod_{j=2}^4 \bigg(- i {\partial \over \partial E_j}\bigg) 
	\left[\int_0^1 du \,
	F(E_i; u) \,
	2 \pi V \delta\left(- E_1 + \sum_{j=2}^4 E_j \right)
	\right]_{E_j = 0} \, ,
\end{align}
after relabeling $E \to E_1$ and introducing the function
\begin{align} \label{eq:2.22}
	F(E_1, E_2, E_3, E_4; u) \equiv 
	\left( \prod_{i=1}^4 \rho_i(u)^{-E_i^2} \right) 
	\bigg( {1 \over 4 u^2} + {1 \over 4 (1-u)^2} + {1\over 4}\bigg) 
	 (1-u)^{-2 E_2 E_3} u^{-2 E_3 E_4} \, .
\end{align}
Differentiating with respect to energies $E_2,E_3,E_4$ and setting them to zero then gives
\begin{align}
	&Q \psi_3 = {i \sqrt{2}}  \int\limits_{-\infty}^\infty  {d E_1} \, c_0 c_1 \alpha_{-1}^1 |E_1 \rangle \\ 
	&\hspace{0.5in} \times {i }\int_0^1 du \,
	\rho_1(u)^{-E_1^2}
	\bigg({1 \over 4 u^2} + {1 \over 4(1-u)^2} +{1 \over 4} \bigg)
	\bigg(  \log \left(u^2(1-u)^2\right)  \, \delta'(E_1) -  \delta'''(E_1) \bigg)
	\nonumber \, ,
\end{align}
where $'$ refers to the differentiation with respect to the argument of the delta function. Integrating over energy $E_1$ yields
\begin{align} \label{eq:3.24}
	Q \psi_3 = &\Bigg[ -{i \over \sqrt{2} } c_0 c_1 x^0 \alpha_{-1}^1 |0 \rangle \Bigg]
	 \Bigg[- {1 \over 2}\int_0^1 du
	 \bigg({1 \over u^2} + {1 \over (1-u)^2} +1\bigg)
	 \bigg( \log u^2(1-u)^2 + 3 \log \rho_1(u)^2 \bigg) \Bigg] \nonumber \\
	 & +\Bigg[-{i \over \sqrt{2} } c_0 c_1 \left(x^0\right)^3  \alpha_{-1}^1 |0 \rangle \Bigg]
	\Bigg[ - {1 \over 2}\int_0^1  du
	 \bigg({1 \over u^2} + {1 \over (1-u)^2} +1\bigg) \Bigg]\, .
\end{align}
Accordingly, let us introduce the moduli integrals
\begin{subequations} \label{eq:3.25}
\begin{align}
	&I_1 \equiv -{1 \over 2}\int_0^1 du
	\bigg({1 \over u^2} + {1 \over (1-u)^2} +1\bigg) 
	\bigg(\log u^2(1-u)^2 + 3 \log \rho_1(u)^2 \bigg) \, ,
	\\
	&I_3 \equiv  -{1 \over 2} \int_0^1 du
	\bigg({1 \over u^2} + {1 \over (1-u)^2} +1\bigg) \, .
\end{align}
\end{subequations}
As already hinted below~\eqref{eq:3.19}, these integrals appear to be problematic and diverge towards $u=0$ and $u=1$ when they are taken at their face value. Nevertheless, they can be assigned consistent finite values based on their origin in~\eqref{eq:3.18} and the Feynman decomposition of the moduli space. Before explaining the procedure it is beneficial to commit to specific string vertices. We choose to adopt $SL(2,\mathbb{R})$ vertices because of their perturbative simplicity~\cite{Erbin:2023rsq}. More precisely, we adopt $SL(2,\mathbb{R})_{\lambda}$ vertices with~\emph{stubs}, where $\lambda \geq 3$ denotes the stub parameter.\footnote{For recent treaties on stubs see~\cite{Erbin:2023hcs,Erler:2023emp,Firat:2024ajp,Schnabl:2023dbv,Schnabl:2024fdx,Stettinger:2024hkp,Maccaferri:2024puc,Chiaffrino:2021uyd}.} Refer to appendix~\ref{sec:A} for more details on our specific choice of (cubic and quartic) vertices and the associated Feynman decomposition at the quartic order.

The evaluation of the integral $I_1$ requires the local coordinate data for the entire moduli space (in particular the mapping radius $\rho_1(u)$ as shown in figure~\ref{fig:rho}), while only the decomposition of the moduli space matters for the integral $I_3$. So we begin by decomposing their integration regions into two parts: those coming from the Feynman region and those coming from the vertex region. The divergences are always contained within the Feynman region, so let us focus on there. In the Feynman region it is natural to use the sewing parameter $0 < q < 1$ as the modulus and express the cross ratio in terms of $q$, $u = u(q)$, see~\eqref{eq:A.9}-\eqref{eq:A.9a}. Expanding the integrands in $q$ after such a change of variable, the divergent contributions to the moduli integrals take the form
\begin{align} \label{eq:3.27}
	\int_0^1 dq \, {(\log q)^n \over q^m}  \, , 
\end{align}
for some integers $n \geq 0$ and $m \geq 1$. The presence of these terms in the $q$-expansion indicates that the propagator $1/L_0$ coming from the solution $\Psi_2$~\eqref{eq:2.16} is incorrectly represented in~\eqref{eq:3.18} using Schwinger parametrization~\cite{Sen:2019jpm}. Note that the states containing the zero mode $x^0$ may also run along the string propagator due to the form of $\Psi_1$~\eqref{eq:2.16}, which give rise to potential $\log q$ powers appearing in the numerator above. 

According to~\cite{Sen:2019jpm} the divergent integrals~\eqref{eq:3.27} should not be interpreted directly, but through their analytically-continued form
\begin{align} \label{eq:3.27a}
	\int_0^1 dq \, {(\log q)^n \over q^m}  \to - {n! \over (m-1)^{n+1} } \, ,
\end{align}
or equivalently, through their representations using $1/L_0$ before the Schwinger parametrization is made. This works as long as $m > 1$, but the integral truly diverges when $m=1$. Fortunately, the presence of the $L_0$-nilpotent projector $\mathbb{P}$ in the solution $\Psi_2$~\eqref{eq:2.16} discards these weight-$0$ terms. It can be checked that the only divergences are of the kind $m=2$ and $n=0,1$ for the integrals $I_1, I_3$. These tachyonic divergences are replaced with $-1$ according to~\eqref{eq:3.27a}.

With this ``SFT prescription'', it is straightforward to evaluate the integrals $I_1$ and $I_3$ using $SL(2,\mathbb{R})_\lambda$ vertices and their Feynman diagrams. This is presented in the \texttt{Mathematica} notebook in the Supplementary material accompanying this paper. We report\footnote{There is in fact a closed-form expression for $I_1$ but its form is unenlightening.}
\begin{align}
	I_1(\lambda) \approx 6.03564 \, , \quad \quad I_3(\lambda) = 0 \, .
\end{align}
The integral $I_3$ exactly vanishes while $I_1$ is independent of the stub parameter $\lambda$---both of these facts are expected but nontrivial, inspect~\eqref{eq:3.25}. After noticing
\begin{align} \label{eq:3.29}
	Q \cdot \left( c \big(X^0\big)^3 \partial X^1(0,0) \right) |0\rangle 
	&= 3! \, \bigg(-{i \over \sqrt{2}} c_0 c_1 x^0  \alpha_{-1}^1 |0 \rangle \bigg) \, ,
\end{align}
we can choose the first nontrivial obstruction term to be
\begin{align} \label{eq:3.30}
	\psi_3 = K c \big(X^0\big)^3 \partial X^1 
	\, , \quad \quad 
	K \equiv {I_1 \over 3!} \approx 1.00594 \, ,
\end{align}
via~\eqref{eq:3.24}. We do not need any contribution proportional to $(X^0)^5$ at this order since $I_3 = 0$.

\section{SFT symplectic structure and energy} \label{sec:4}

In this section we find the energy of a D-brane carrying a constant electric flux using the symplectic structure of~\cite{Bernardes:2025uzg}. The present discussion follows the computation of the energy of rolling tachyon solutions in~\cite{Bernardes:2025zzu} closely. Except for the presence of higher string products and a nontrivial obstructions term~\eqref{eq:3.30}, our considerations are conceptually analogous and we often adapt the corresponding results without deriving them explicitly. 

The symplectic structure of nonpolynomial open SFT is
\begin{align} \label{eq:4.1a}
	\Omega = &-{1 \over g^2} \bigg[ {1 \over 2} \left\langle \delta \Psi, [Q, \sigma] \delta \Psi \right\rangle 
	+ \left\langle \Psi, v_2(\sigma \delta \Psi, \delta \Psi) - v_2 (\delta \Psi, \sigma \delta \Psi) \right\rangle \\
	&+  {1 \over 2}\left\langle \Psi, 
	 2v_3 (\sigma \delta \Psi, \delta \Psi, \Psi) 
	- 2v_3 (\delta \Psi, \sigma \delta \Psi, \Psi) 
	- v_3 (\sigma \delta \Psi, \Psi, \delta \Psi)
	+ v_3 (\delta \Psi, \Psi, \sigma \delta \Psi) \right\rangle
	+\mathcal{O}\big(\Psi^3\big)
	\bigg]
	\nonumber \, .
\end{align}
We remind that $\Psi$ is Grassmann odd, so $\delta \Psi$ is Grassmann even. As usual, $\sigma$ above is the~\emph{sigmoid} operator with the boundary conditions
\begin{align} \label{eq:4.2}
	\lim_{t \to - \infty} \sigma = 0 \, , \quad \quad
	\lim_{t \to \infty} \sigma = 1 \, .
\end{align}
This version of $\Omega$ can be derived from the symplectic structure of~\cite{Bernardes:2025uzg} by expanding the kinetic operator $Q_\Psi$ around the background $\Psi$ and arranging the interaction terms suitably while being mindful about the $L_\infty$ vs. $A_\infty$ difference for the string products and the sign differences due to the suspension~\cite{Erler:2024lje}. Recall that $v_n$s are $A_\infty$ products, whereas the $L_\infty$ products appearing in $\Omega$ are given by the sums over different cyclic orderings of $v_n$ with appropriate Koszul signs. 

We highlight that arranging the formula as above requires using cyclicity and the manipulations are subtle because of the presence of possibly unaccounted boundary contributions.\footnote{For recent investigations of the boundary contributions in SFT, and possibly related matters, see~\cite{Frenkel:2025wko,Kim:2026kex,Firat:2024kxq,Stettinger:2024uus,Maccaferri:2025orz,Maccaferri:2025onc,Erler:2022agw,Mazel:2025fxj,Mamade:2025jbs,Mamade:2025htb,Choi:2026ljv,BernardesWIP,GeorgWIP}.} The proper argument involves~\emph{tau regulator}~\cite{Bernardes:2025uzg}. Nevertheless, tau regulator drops out at the end since the interaction terms after the arrangements become supported only at finite times: the $L_\infty$ products are graded-symmetric in the~\emph{suspended} sign convention and they vanish when two~\emph{Grassmann even} $\delta \Psi$s are inputted---like in $\Omega$ when $t \to \infty$, inspect~\eqref{eq:4.1a}-\eqref{eq:4.2}. This argument demands that the temporal volume divergences are regularized such that an exact cancellation occurs at the level of the $L_\infty$ products however. This is what we shall assume.

In this work, the sigmoid is assumed to be
\begin{align} \label{eq:4.3}
	\sigma = \int\limits_{-\infty}^\infty {d E\over 2 \pi} \sigma(E) e^{i E x^0} \, , 
\end{align}
where $x^0$ is the time coordinate of the open string center of mass:
\begin{align} \label{eq:4.4}
	x^0 = {1 \over \pi} \int_0^\pi d\theta X^0 \left(e^{i \theta}, e^{-i \theta}\right) \, .
\end{align}
The limits~\eqref{eq:4.2} imply that the Fourier modes $\sigma(E)$ endow a simple pole at $E=0$, and in particular
\begin{align} \label{eq:4.5}
	\lim_{E \to 0} \dot{\sigma}(E) = 1 \, ,
\end{align}
where the dot indicates the time derivative in momentum space, which is equivalent to multiplication by $i E $. The sigmoid~\eqref{eq:4.3} acts homogeneously for the rest of BCFT. 

Let us expand the symplectic structure $\Omega$ in the marginality parameter $\varepsilon$ now. We need the expansion of $\delta \Psi$ in $\varepsilon$ for this. After restoring the origin of the time coordinate $t_0$ in the solution~\eqref{eq:2.5} by $X^0 \to X^0 - t_0$, this expansion takes the form
\begin{align} \label{eq:4.6a}
	\delta \Psi = \delta \varepsilon \left( \Psi_1 + 2 \varepsilon \Psi_2 + 3 \varepsilon^2 \Psi_3 + \mathcal{O}\left(\varepsilon^4\right) \right) 
	-\delta t_0 \left(\varepsilon \dot{\Psi}_1 + \varepsilon^2 \dot{\Psi}_2 + \varepsilon^3 \dot{\Psi}_3 + \mathcal{O}\left(\varepsilon^4\right)\right) \, ,
\end{align}
when $t_0 = 0$. The dot on the string fields indicates the time derivative, which is provided by the action of the zero mode of the translational current:
\begin{align}
	i p_0 = \oint {d w \over 2 \pi i} \partial X^0 (w) \, .
\end{align}
In particular $\dot{\Psi}_1 = c \partial X^1 $. The symplectic structure $\Omega$, and in extension the energy $E$, expands in the marginality parameter $\varepsilon$ as\footnote{The $E$ appearing in the Fourier decompositions as in~\eqref{eq:4.3} should not be confused with the $E$~\eqref{eq:4.8} denoting the energy of a D-brane carrying a constant electric flux. Their difference is going to be obvious from the context.}
\begin{align} \label{eq:4.8}
	\Omega = \delta t_0 \delta \varepsilon \left[
	\varepsilon \Omega_1 + \varepsilon^2 \Omega_2 + \varepsilon^3 \Omega_3 + \mathcal{O}\left(\varepsilon^4\right)
	\right]
	\quad \implies \quad
	E = {\varepsilon^2 \over 2} \Omega_1 + {\varepsilon^3 \over 3} \Omega_2 + {\varepsilon^4 \over 4} \Omega_3 +\mathcal{O}\left(\varepsilon^5\right)\, ,
\end{align}
after adjusting the zero point energy to be $E(\varepsilon = 0) = 0$ like in section~\ref{sec:2}. The coefficients $\Omega_i$ follow from inserting the expansions~\eqref{eq:2.5} and~\eqref{eq:4.6a} in~\eqref{eq:4.1a}. They are given by
\begin{subequations}
\begin{align}
	\label{eq:4.10a}
	&\Omega_1 = {1 \over g^2} \left\langle \dot{\Psi}_1, [Q,\sigma] \Psi_1 \right\rangle \, , \\
	\label{eq:4.10b}
	&\Omega_2 = {1 \over g^2} \bigg[
	2 \left\langle \dot{\Psi}_1, [Q, \sigma] \Psi_2 \right\rangle
	+ \left\langle \dot{\Psi}_2, [Q, \sigma] \Psi_1 \right\rangle
	\\ \nonumber
	&\hspace{0.75in}
	+ \left\langle \Psi_1, 
	v_2(\dot{\Psi}_1, \sigma \Psi_1) + v_2(\sigma \Psi_1, \dot{\Psi}_1)
	-v_2 (\sigma \dot{\Psi}_1, \Psi_1) - v_2 (\Psi_1, \sigma \dot{\Psi}_1) 
	\right\rangle 
	\bigg] \, ,
\\ 
	\label{eq:4.10c}
	& \Omega_3 = {1 \over g^2} \bigg[
	3 \left\langle \dot{\Psi}_1 , [Q, \sigma] \Psi_3 \right\rangle 
	+ 2 \left\langle \dot{\Psi}_2 , [Q,\sigma] \Psi_2 \right\rangle
	+ \left\langle \dot{\Psi}_3, [Q, \sigma] \Psi_1 \right\rangle
	\\ \nonumber
	&\hspace{0.75in}
	+ \left\langle \Psi_2, v_2(\dot{\Psi}_1, \sigma \Psi_1) + v_2(\sigma \Psi_1, \dot{\Psi}_1)
	- v_2(\sigma \dot{\Psi}_1, \Psi_1) - v_2(\Psi_1, \sigma  \dot{\Psi}_1) \right\rangle
	\\ \nonumber
	&\hspace{0.75in}
	+ \left\langle \Psi_1, v_2(\dot{\Psi}_2, \sigma \Psi_1) +  v_2(\sigma \Psi_1, \dot{\Psi}_2)
	-v_2(\sigma  \dot{\Psi}_2, \Psi_1) -  v_2(\Psi_1, \sigma  \dot{\Psi}_2) \right\rangle 
	\\ \nonumber
	&\hspace{0.75in}
	+ 2 \left\langle \Psi_1, v_2(\dot{\Psi}_1, \sigma \Psi_2) + v_2(\sigma \Psi_2, \dot{\Psi}_1)
	-v_2(\sigma \dot{\Psi}_1,  \Psi_2) - v_2(\Psi_2, \sigma  \dot{\Psi}_1) \right\rangle
	\\ \nonumber
	&\hspace{0.75in}
	-{1 \over 2} \bigg\langle \Psi_1, 
	2v_3(\sigma \dot{\Psi}_1, \Psi_1, \Psi_1)
	- 2 v_3(\dot{\Psi}_1, \sigma \Psi_1, \Psi_1)
	+v_3(\sigma \dot{\Psi}_1, \Psi_1, \Psi_1)
	-v_3(\dot{\Psi}_1, \Psi_1, \sigma \Psi_1)
	\\ \nonumber
	&\hspace{1.2in}
	-2v_3(\sigma \Psi_1, \dot{\Psi}_1, \Psi_1)
	+ 2 v_3(\Psi_1, \sigma \dot{\Psi}_1, \Psi_1)
	- v_3(\sigma \Psi_1, \Psi_1, \dot{\Psi}_1)
	+ v_3(\Psi_1, \Psi_1, \sigma \dot{\Psi}_1)
	\bigg\rangle
	\bigg] \, ,
\end{align}
\end{subequations}
after adapting them from the analogous expressions derived in equation (4.17) of~\cite{Bernardes:2025zzu} and including the additional contributions involving the three-product $v_3$ to $\Omega_3$. Note that $\Psi_3$ in~\eqref{eq:4.10c} contains the nontrivial obstruction term, see~\eqref{eq:3.17}.

\subsection{Energy to leading order $\varepsilon^2$}

We now evaluate the energy~\eqref{eq:4.8} to the leading order in $\varepsilon$ expansion in order to set the stage for the latter computations. Focus on $\Omega_1$~\eqref{eq:4.10a}. We first replace $X^0$ in $\Psi_1$ with the plane wave vertex operator:
\begin{align}
	\Psi_1 &= \left( -i {\partial \over \partial E'} \right)
	\left( c \partial X^1 e^{i E' X^0} \right)_{E'=0} \, .
\end{align}
After acting with the operator $[Q , \sigma]$,
\begin{align}
	[Q,\sigma] \Psi_1 = \left(-i {\partial \over \partial E'} \right)\Bigg[ \int\limits_{-\infty}^\infty {d E \over 2 \pi} \sigma(E) \big( - (E+E')^2 + (E')^2 \big) \partial c\,  c e^{i (E+E') X^0} \Bigg]_{E'=0} \, .
\end{align}
It is easier to evaluate the BPZ product for $\Omega_1$ before taking $E'$ derivative:
\begin{align}
	\Omega_1 &= {1 \over g^2} \left( -i {\partial \over \partial E'} \right) \Bigg[\int\limits_{-\infty}^\infty {d E \over 2 \pi} \sigma(E) \Big( - (E+E')^2 + (E')^2 \Big)
	\left\langle c \partial X^1, \partial c \, c \partial X^1 e^{i (E+E')X^0} \right\rangle \Bigg]_{E'=0} 
	\\ \nonumber
	&= {1 \over g^2} \left( -i {\partial \over \partial E'} \right) 
	\Bigg[\int\limits_{-\infty}^\infty {d E \over 2 \pi} \sigma(E) \Big( - (E+E')^2 + (E')^2 \Big)
	\left(-{1 \over 2} \right) 2 \pi V \delta (E+ E')\Bigg]_{E'=0} 
	\, .
\end{align}
This shows
\begin{align}
	\Omega_1 &= -{V \over 2g^2} \int\limits_{-\infty}^\infty dE \, \sigma(E) \left(
	2 i E \delta(E) + i E^2 \delta' (E)
	\right) \, .
\end{align}
We remind that $\sigma(E)$ has a pole at $E=0$ so the integrand above is ambigious when it is taken at its face value, see the discussion in~\cite{Bernardes:2025zzu}. This pole should be removed by canceling it against the rest of the integrand before anything else in order to implement the symplectic structure in the momentum space correctly. So we multiply/divide the integrand by $i E$ and use the regular function $\dot{\sigma}(E) = i E \sigma(E)$ instead of using $\sigma(E)$ directly.  Then the apparent pole at $E=0$ explicitly cancels and we find
\begin{align} \label{eq:4.13}
	\Omega_1 &= -{V \over 2g^2} \int\limits_{-\infty}^\infty {d E } \,
	\dot{\sigma}(E) \big(2  \delta(E) + E \delta'(E)  \big)
	\\ \nonumber
	&= - {V \over 2g^2} \big( 2 \dot{\sigma}(0) - 0 \dot{\sigma}'(0) - \dot{\sigma}(0) \big)
	\nonumber \\
	&= - {V \over 2 g^2} \nonumber \, ,
\end{align}
where we used $\dot{\sigma}(0) = 1$~\eqref{eq:4.5} in the final line. We note that $\dot{\sigma}'(0) = {d \dot{\sigma} (0)/ dE}$ term comes with a vanishing coefficient above. This is a reflection of the independence of the symplectic structure from sigmoid: only the boundary conditions~\eqref{eq:4.2} (and consequentially~\eqref{eq:4.5}) matter in $\Omega$, which means nontrivial contributions are always proportional to $\dot{\sigma}(0)$, and not to its derivatives. We also point out in passing that there cannot be any ``nonzero energy terms'' (i.e., terms proportional to $\sigma(E)$ for $E \neq 0$) at the intermediate steps like in~\cite{Bernardes:2025zzu} because the profile of the electric field solution is polynomial in temporal coordinate---at least to any finite order in $\varepsilon$.

The final result~\eqref{eq:4.13} demonstrates that the energy~\eqref{eq:4.8} at the leading order evaluates to
\begin{align} \label{eq:4.15}
	E = - {V \over 4 g^2} \varepsilon^2 + \mathcal{O}\left(\varepsilon^3\right) \, .
\end{align}
This is consistent with the electromagnetic energy computed in~\eqref{eq:10.11} earlier. In order to see this, we just relate the parameters $\varepsilon_{\DBI}$ and $\varepsilon$ via
\begin{align} \label{eq:4.15a}
	\varepsilon_{\DBI} &= \left( {1 \over \sqrt{2} }  \right) \left(- { i \over \sqrt{2} } \right)\varepsilon + \mathcal{O}\left(\varepsilon^2\right)
	\\ \nonumber
	&= - {i \over 2} \varepsilon + \mathcal{O}\left(\varepsilon^2\right) \, ,
\end{align}
where the first factor comes from the field redefinition~\eqref{eq:10.3} while the second factor comes from the fact that the SFT gauge field $A_\mu$ in~\eqref{eq:10.3} is the coefficient of the mode $c_1 \alpha_{-1}^\mu$, rather than $c \partial X^\mu$, as already highlighted above~\eqref{eq:3.9}.

Proceeding with the subleading order, we find $\Omega_2 = 0$. This immediately follows from the structure of the correlators in~\eqref{eq:4.10b}. Each contains three $\partial X^1$ insertions alone, hence they all vanish. Thus the energy~\eqref{eq:4.15} is in fact correct to $\mathcal{O}(\varepsilon^4)$. This argument can be extended to show $\Omega_n = 0$ for all even $n$. Indeed, this is what is expected from the symmetry grounds: changing the direction of the electric field does not affect the energy, therefore no odd powers in $\varepsilon$ should appear. We note that the mechanism behind the cancellation at even orders is somewhat distinct from the mechanism behind the analogous cancellation for the rolling tachyons solutions~\cite{Bernardes:2025zzu}.

\subsection{Energy to order $\varepsilon^4$}

Let us proceed with the next nontrivial order and focus on $\Omega_3$~\eqref{eq:4.10c}. We begin by noting that we always ignore terms proportional to derivatives of $\dot{\sigma}(E)$ in our calculations because these do not matter for the final result as a consequence of the conservation of $\Omega$; see the discussion below~\eqref{eq:4.13}. In other words, the nontrivial contributions to the symplectic structure are always proportional $\dot{\sigma}(0) = 1$ and any ``trivial'' terms proportional to something else appearing at the intermediate steps are discarded without explicitly demonstrating their explicit cancellations at whole.

Given this remark, let us focus on $\Psi_1\textendash \Psi_3$ cross terms in $\Omega_3$ first. This is
\begin{align} \label{eq:4.17}
	\Omega_3^{\Psi_1\textendash \Psi_3} = {1 \over g^2} \bigg[
	3 \left\langle \dot{\Psi}_1 , [Q, \sigma] \Psi_3\right\rangle 
	+ \left\langle \dot{\Psi}_3, [Q, \sigma] \Psi_1 \right\rangle 
	\bigg] \, .
\end{align}
Recall that $\Psi_3$~\eqref{eq:3.17} consists of a term annihilated by the $L_0$-nilpotent projector $\mathbb{P}$ and a nontrivial obstruction term $\psi_3$. We can argue the former does not contribute, as it involves considering overlaps of nonsingular states with the states
\begin{align} \label{eq:4.18}
	(1- \mathbb{P}){b_0 \over L_0} [Q, \sigma] \dot{\Psi}_1 \, ,
	\quad \quad
	(1- \mathbb{P}){b_0 \over L_0} [Q, \sigma] \Psi_1 \, , 
\end{align}
which happen to vanish. The appearance of these states can be deduced from the structure of~\eqref{eq:3.17} and using cyclicity in~\eqref{eq:4.17}. We highlight that using cyclicity does not introduce any boundary terms in this manipulation since $[Q,\sigma]$ is supported at finite times. 

Let us illustrate why the ``non-obstruction'' portion of $\Psi_3 $ cannot contribute to $\Omega_3^{\Psi_1\textendash \Psi_3}$ in more detail. We first notice
\begin{align} \label{eq:4.19}
	{b_0 \over L_0} [Q, \sigma] \cdot c \partial X^1 e^{i E' X^0}
	=  \int\limits_{-\infty}^\infty {d E \over 2\pi} \sigma(E)
	\Bigg( {-(E+E')^2 + (E')^2 \over - (E+E')^2} \Bigg) c \partial X^1 e^{i (E+E')X^0} \, ,
\end{align}
which implies
\begin{subequations} \label{eq:4.20}
\begin{align}
	&(1- \mathbb{P}){b_0 \over L_0} [Q, \sigma] \dot{\Psi}_1 
	= \int\limits_{-\infty}^\infty {d E \over 2\pi} \dot{\sigma}(E) \bigg({ 1- \delta_{E} \over i E}\bigg)c \partial X^1 e^{i EX^0} 
	\, , \\
	&(1- \mathbb{P}){b_0 \over L_0} [Q, \sigma] \Psi_1 
	= \int\limits_{-\infty}^\infty {d E \over 2\pi} \dot{\sigma}(E) \bigg({ 1- \delta_{E} \over i E}\bigg)
	c X^0 \partial X^1 e^{i EX^0} \, .
\end{align}
\end{subequations}
Here $\delta_{E}$ is the~\emph{Kronecker} delta for the states with $E = 0$, which appears due to the presence of the projector $\mathbb{P}$ on the left hand side. We divided/multiplied the integrands by $i E$ to make the behavior at $E=0$ manifest. The apparent singularity at $E=0$ inside the parenthesis' above should be removed before anything else for the proper implementation of the symplectic structure in the momentum space. In order to do this explicitly, we regulate the Kronecker delta by $\delta_{E}\to e^{\Lambda p_0^2}$ and understand it as the $\Lambda \to \infty$ limit. This yields
\begin{align} \label{eq:4.21}
	{ 1- \delta_{E} \over i E}
	&= \lim_{\Lambda \to \infty} { 1- e^{-\Lambda E^2} \over i E}
	= \lim_{\Lambda \to \infty} \bigg( - i \Lambda E + {i \over 2} \Lambda^2 E^3 + \cdots \bigg) \, ,
\end{align}
in~\eqref{eq:4.20}. The function inside the parenthesis, along with its derivatives, vanishes at $E = 0$ for all $\Lambda$, therefore it is identically zero upon removing the regulator. This means that the contributions from $E=0$ are absent in the Fourier integral when we consider the overlaps of~\eqref{eq:4.20} with the regular states. In contrast, the limit~\eqref{eq:4.21} clearly evaluates to $1/i E$ when $E \neq 0$ so the contributions are still possible from the nonzero energy.\footnote{Another way to state is that $(1-\delta_{E})/iE$ is~\emph{the principal value} for $1/iE$ and the integrals~\eqref{eq:4.20} should be understood in the sense of Cauchy principal value for which the singularity at $E=0$ is skipped over.} As remarked below~\eqref{eq:4.13}, however, these contributions are absent for the perturbative electric field solution. Thus the non-obstruction portion of $\Psi_3$ do not contribute to $\Omega_3$.

The argument above demonstrates that we just need to be concerned with
\begin{align} \label{eq:4.22}
	\Omega_3^{\Psi_1\textendash \Psi_3} = {1 \over g^2} \bigg[
	3 \left\langle \dot{\Psi}_1 , [Q, \sigma] \psi_3 \right\rangle 
	+ \left\langle \dot{\psi}_3, [Q, \sigma] \Psi_1 \right\rangle 
	\bigg]  \, ,
\end{align}
where only the obstruction term in $\Psi_3$ appears in $\Omega_3^{\Psi_1\textendash \Psi_3}$. These terms may potentially contribute. However, it turns out that they are trivial:~\eqref{eq:4.22} is proportional to $\dot{\sigma}''(0)$. In order to argue for this, consider
\begin{align}
		 \left \langle \dot{\Psi}_1 , [Q, \sigma] \psi_3 \right\rangle  
		&=  - \left\langle \psi_3,  [Q, \sigma] \dot{\Psi}_1 \right\rangle 
		\\ \nonumber
		&= - \int\limits_{-\infty}^\infty {d E \over 2\pi} \dot{\sigma}(E)
		(i E) 
		\left\langle K c \left(X^0\right)^3 \partial X^1, \partial c \, c \partial X^1 e^{i E X^0} \right\rangle \, ,
\end{align}
where we adapted~\eqref{eq:4.19}, removed the apparent singularity at $E=0$, and used the obstruction term~\eqref{eq:3.30}. No boundary term is induced in the first line above given that the commutator $[Q,\sigma]$ is supported at finite times. Replacing $(X^0)^3$ with a plane wave vertex operator, this overlap evaluates to
\begin{align}
	 \left\langle \dot{\Psi}_1 , [Q, \sigma] \psi_3 \right\rangle  &= 
	 - K \bigg(-i {\partial \over \partial E'}\bigg)^3 
	 \Bigg[ \int\limits_{-\infty}^\infty {d E \over 2\pi} \dot{\sigma}(E) (i E) \bigg(-{1 \over 2} \bigg) 2 \pi V \delta(E' + E) 
	 \Bigg]_{E' = 0}
	 \\ \nonumber
	 &= - {K \over 2} V \int\limits_{-\infty}^\infty {d E} \, \dot{\sigma}(E) E \delta'''(E) 
	 \\ \nonumber
	 &= {3 K \over 2} V \dot{\sigma}''(0) \, . \nonumber
\end{align}
Analogously, we can express
\begin{align}
	\left\langle \dot{\psi}_3, [Q, \sigma] \Psi_1 \right\rangle 
	&= \bigg(- i {\partial \over \partial E'} \bigg)
	\bigg(- i {\partial \over \partial E''} \bigg)^2
	\Bigg[ \int\limits_{-\infty}^\infty {d E \over 2\pi} 
	\dot{\sigma}(E) 
	\left( { - (E+E')^2 + (E')^2 \over i E} \right)
	\\ \nonumber
	& \hspace{2.5in}
	\left\langle 3 K c e^{i E'' X^0} \partial X^1, \partial c \, c \partial X^1 e^{i(E+E')X^0} \right\rangle
	\Bigg]_{\substack{E'=0 \\ E''=0}} \, ,
\end{align}
which evaluates to
\begin{align}
	\left\langle \dot{\psi}_3, [Q, \sigma] \Psi_1 \right\rangle 
	&= -{3K} \bigg(- i {\partial \over \partial E'} \bigg) \Bigg[
	\int\limits_{-\infty}^\infty {d E \over 2\pi} \dot{\sigma}(E) 
	\bigg( { - (E+E')^2 + (E')^2 \over i E} \bigg)
	\bigg(-{1 \over 2} \bigg)2 \pi V \delta''(E + E')\Bigg]_{E'=0}
	\nonumber \\
	&= {3K \over 2} V \int\limits_{-\infty}^\infty {d E} \, \dot{\sigma}(E) 
	\big( 2 \delta''(E) + E \delta'''(E) \big) 
	\nonumber \\
	&= - {3 K \over 2} V \dot{\sigma}''(0)
	\, .
\end{align}
Both of the terms in~\eqref{eq:4.22} are then proportional to $\dot{\sigma}''(0)$ as claimed. Hence
\begin{align}
	\Omega_3^{\Psi_1 \textendash \Psi_3}\,  \text{   is trivial} \, ,
\end{align}
in the sense that it does not contain anything proportional to $\dot{\sigma}(0) = 1$. Somewhat surprisingly, the value of the coefficient of the obstruction term $K$~\eqref{eq:3.30} turns out~\emph{not} to be essential for evaluating $\Omega_3$. The precise value of $K$ is going to be needed for the consistency checks and field redefinitions in the upcoming sections however.

Let us consider $\Psi_2 \textendash \Psi_2$ contribution $\Omega_3^{\Psi_2 \textendash \Psi_2}$ to $\Omega_3$ next. This can be arranged into
\begin{align} \label{eq:4.28}
	\Omega_3^{\Psi_2 \textendash \Psi_2} = {2 \over g^2}
	\bigg\langle v_2(\dot{\Psi}_1, \Psi_1) + v_2(\Psi_1, \dot{\Psi}_1),
	\bigg[\sigma, {b_0 \over L_0} (1 - \mathbb{P}) \bigg] v_2 (\Psi_1, \Psi_1) \bigg\rangle \, ,
\end{align}
following the arguments in section 4.3 of~\cite{Bernardes:2025zzu}. Similarly, we can also arrange the ``cubic'' contributions $\Omega_3^{\text{cubic}}$ to $\Omega_3$ as
\begin{align} \label{eq:4.29}
	\Omega_3^{\text{cubic}} =
	{1 \over g^2} \bigg[
	&- \bigg\langle v_2(\Psi_1, \Psi_1) , {b_0 \over L_0} (1-\mathbb{P}) \sigma_+ (\dot{\Psi}_1, \Psi_1) \bigg\rangle
	\nonumber \\
	&
	+ 2 \bigg\langle \sigma_-(\dot{\Psi}_1, \Psi_1), {b_0 \over L_0}(1-\mathbb{P}) v_2(\Psi_1, \Psi_1) \bigg\rangle
	\nonumber \\
	&+ {1 \over 2} \bigg\langle  v_2(\dot{\Psi}_1, \Psi_1) + v_2(\Psi_1, \dot{\Psi}_1), {b_0 \over L_0} (1-\mathbb{P}) \sigma_+ (\Psi_1, \Psi_1) \bigg\rangle \bigg]  \, ,
\end{align}
by introducing the products
\begin{subequations}
\begin{align}
	&\sigma_+ (A, B) \equiv 2 \sigma \big( v_2(A,B) - (-1)^{A B} v_2(B,A) \big)
	-  \big( v_2(\sigma A,B) - (-1)^{A B} v_2(B,\sigma A)  \big)
	\\ \nonumber
	&\hspace{3.12in}- \big( v_2(A,\sigma B) - (-1)^{A B} v_2(\sigma B,A) \big) \, ,
	\\
	&\sigma_-(A,B) \equiv \big( v_2(\sigma A,B) - (-1)^{A B} v_2(B,\sigma A)  \big)
	- \big( v_2(A,\sigma B) - (-1)^{A B} v_2(\sigma B,A) \big) \, .
\end{align}
\end{subequations}
Here $(-1)^A$ stands for the Grassmann parity of $A$. Note that these products are supported at finite times, which is essential to obtain a finite and unambiguous results as we shall witness in the next subsection. Finally, we can present the ``quartic'' terms $\Omega^{\text{quartic}}_3$ in $\Omega_3$ as
\begin{align}
	\Omega^{\text{quartic}}_3 = - {1 \over g^2} \big\langle
	\Psi_1&, v_3(\sigma \dot{\Psi}_1, \Psi_1, \Psi_1)
	 + v_3(\Psi_1, \sigma \dot{\Psi}_1, \Psi_1)
	+ v_3(\Psi_1, \Psi_1, \sigma \dot{\Psi}_1)
	\\ \nonumber
	&\hspace{-0.1in}- v_3(\dot{\Psi}_1, \Psi_1, \sigma  \Psi_1)
	 - v_3(\sigma \Psi_1,\dot{\Psi}_1, \Psi_1)
	 - v_3(\Psi_1, \sigma\Psi_1, \dot{\Psi}_1)  \big\rangle \, .
\end{align}
We need to introduce tau regulator~\cite{Bernardes:2025uzg} in~\eqref{eq:4.10c} and use cyclicity for this arrangement---as we did for the cubic terms above implicitly. Nevertheless, tau regulator can be removed at the end as the final result become localized at finite times: the terms at the top and bottom cancel as $t \to \infty$ in the view of the discussion below~\eqref{eq:4.2}.

\subsection{Evaluations and results} \label{sec:4.3}

\begin{figure}[t]
	\centering
	\includegraphics[scale=.59]{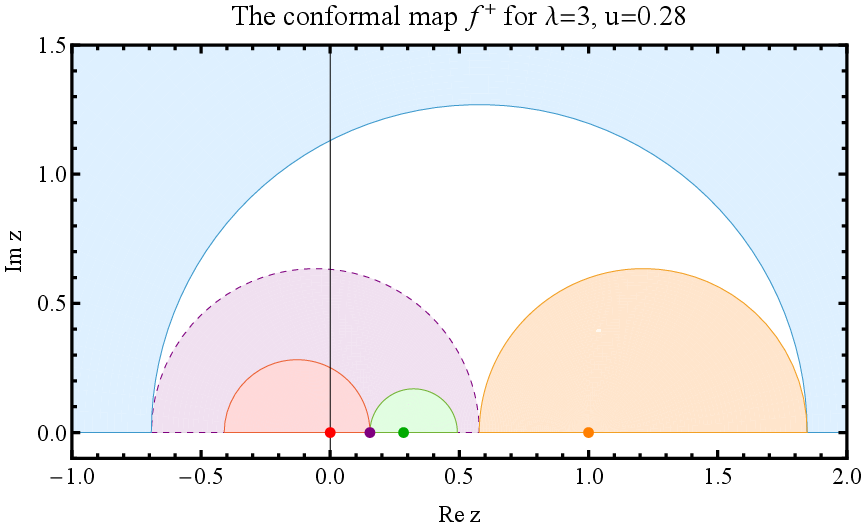}
	\includegraphics[scale=.59]{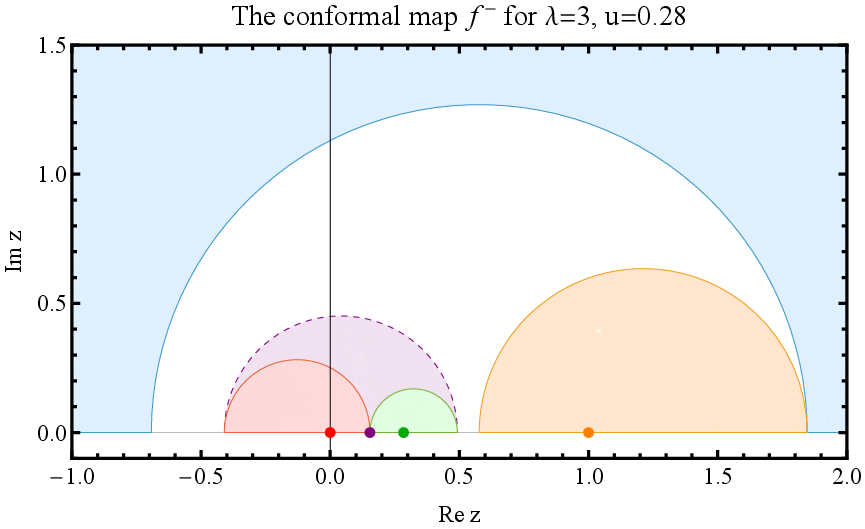}
	\caption{\label{fig:f}The conformal maps $f^+$ (\emph{left}) and $f^-$ (\emph{right}) when $\lambda = 3$ and $q = 0.75$, shown together with the local coordinates for the Feynman diagram. They are characterized by conformally mapping the upper unit half-disk to the \textcolor{purple}{purple} region on the UHP. The origin is mapped to the point $z = u_0$~\eqref{eq:4.37anew}.}
\end{figure}

We finally evaluate the nontrivial contributions to $\Omega_3$ in this subsection. Let us begin with the $\Psi_2$-$\Psi_2$ cross term $\Omega_3^{\Psi_2 \textendash \Psi_2}$~\eqref{eq:4.28}. We can rewrite it as
\begin{align} \label{eq:4.32}
	&\Omega_3^{\Psi_2 \textendash \Psi_2} = {2 \over g^2}
	\prod_{j=2}^4 \bigg(-i {\partial \over \partial E_j}\bigg)
	\bigg\langle
	v_2 \left(c \partial X^1 e^{i E_2 X^0}, c \partial X^1  e^{i E_2 X^0}\right),
	\\ \nonumber
	&\hspace{3in}
	\bigg[\sigma, {b_0 \over L_0} (1 - \mathbb{P}) \bigg] 
	v_2\left(c \partial X^1  e^{i E_3 X^0}, c \partial X^1  e^{i E_4 X^0}\right)
	\bigg\rangle \bigg|_{E_j = 0} \, ,
\end{align}
after introducing plane wave vertex operators according to the prescription~\eqref{eq:3.20}. More explicitly,
\begin{align} \label{eq:4.33new}
	&\Omega_3^{\Psi_2 \textendash \Psi_2} = {2 \over g^2}
	\int\limits_{-\infty}^\infty {d E \over 2 \pi} \, \sigma(E)
	\prod_{j=2}^4 \bigg(-i {\partial \over \partial E_j}\bigg) 
	\int\limits_{0}^1 {d q \over q}
	\bigg\langle
	v_2 \left(c \partial X^1 e^{i E_2 X^0}, c \partial X^1  e^{i E_2 X^0}\right),
	\\ \nonumber
	&\hspace{2.5in}
	b_0 \bigg[e^{i E x^0}, q^{L_0}(1 - \mathbb{P}) \bigg] 
	v_2\left(c \partial X^1  e^{i E_3 X^0}, c \partial X^1  e^{i E_4 X^0}\right)
	\bigg\rangle 
	\bigg|_{E_j = 0}  \, .
\end{align}
Note that $e^{i E x^0}$ above is~\emph{not} a vertex operator at a puncture, but rather a collection of $X^0$s---inspect~\eqref{eq:4.4}. Recalling the construction of Feynman diagrams by sewing fixture as explained in appendix~\ref{sec:A}, we can convert this overlap into the following moduli integral:
\begin{align} \label{eq:3.34new}
	&\Omega_3^{\Psi_2 \textendash \Psi_2} = {2 \over g^2}
\int\limits_{-\infty}^\infty {d E \over 2 \pi} \, \sigma(E)
\prod_{j=2}^4 \bigg(-i {\partial \over \partial E_j}\bigg) 
\int\limits_{0}^{u_- (\lambda)} {d u \over \rho_3 } \,
\bigg\langle
(f^+ - f^-) \circ e^{i E x^0}
\\ \nonumber
&
c \partial X^1 e^{i E_2 X^0} (w_1 = 0) \, 
c \partial X^1 e^{i E_2 X^0} (w_2 = 0) \,
\partial X^1  e^{i E_3 X^0} (w_3 = 0) \,
c \partial X^1  e^{i E_4 X^0} (w_4 = 0)
\bigg\rangle_\UHP 
\bigg|_{E_j = 0}  \, ,
\end{align}
where the boundary $u = u_\pm(\lambda)$ for the Feynman region is given in~\eqref{eq:A.16}. The maps $f^\pm$ above encode the conformal transformations of $e^{i E x^0}$ insertions when we express them on the uniformizing UHP whose punctures are placed at $z=\infty,1,u,0$.  Examples for these maps are shown in figure~\ref{fig:f}. They depend on the moduli and we obtain a difference between two distinct maps $f^\pm$ corresponding to two ends of the propagator as $e^{i E x^0}$ acts before/after $q^{L_0}(1 - \mathbb{P}) $ in~\eqref{eq:4.33new}.

\begin{figure}[t]
	\centering
	\includegraphics[scale=.59]{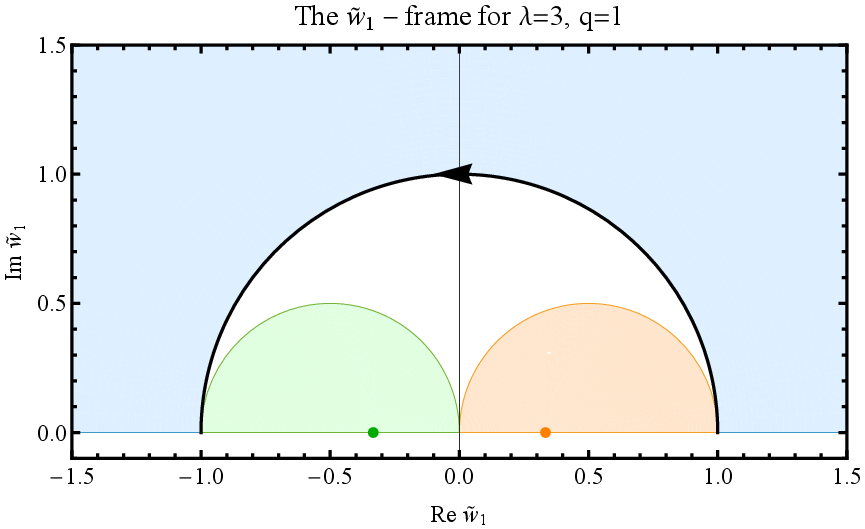}
	\includegraphics[scale=.59]{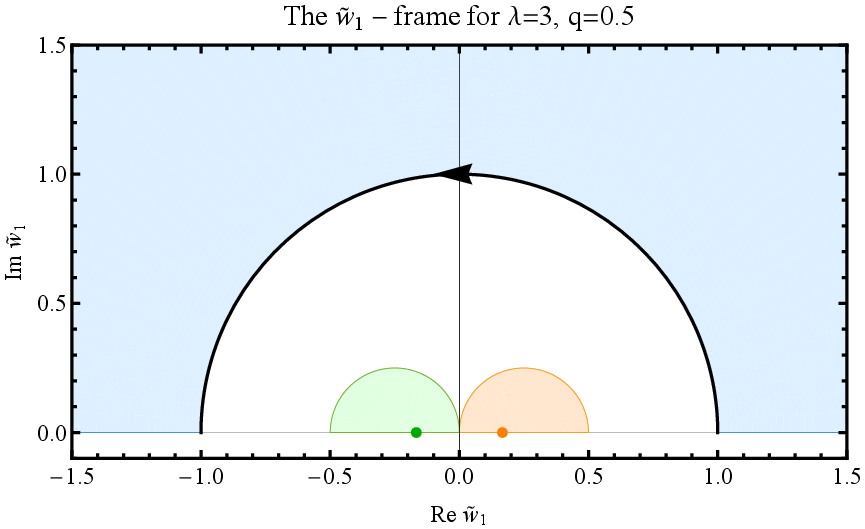}
	\caption{\label{fig:w1tilde}The local coordinates on the $\widetilde{w}_1$-frame when $\lambda = 3$. The black curve (i.e., the upper unit half-circle) is the contour for the zero mode integral~\eqref{eq:4.4} in the correlator~\eqref{eq:4.34}. It is orientated~\emph{counter-clockwise} due to the inversion and subsequent reparametrization.}
\end{figure}

It is useful to give more details on the construction of the maps $f^\pm$ shown in figure~\ref{fig:f} before we carry on with our evaluation. For this, we first determine the conformal frame where the position zero mode $x^0$ takes the simple form~\eqref{eq:4.3}. Consider the following (possibly singular) overlap:
\begin{align} \label{eq:4.33}
	\left\langle A, e^{i E x^0} q^{L_0} v_2(B, C) \right\rangle  &= \left\langle q^{L_0} e^{i E x^0} A, v_2(B, C) \right\rangle  
	\\
	&= \left\langle (q^{L_0}  e^{i E x^0} A)\bigg(-{1 \over \widetilde{w}_1} = 0 \bigg) v_2(B,C) (\widetilde{w}_1 = 0) \right\rangle_{\UHP} \nonumber
	\, .
\end{align}
We used cyclicity in the first line and the definition of the BPZ product in the second line.  This overlap is same as the correlator of $(e^{i E x^0} A),B,C$ inserted on UHP using appropriate local coordinate maps by the definition~\eqref{eq:3.4}, i.e.,
\begin{align} \label{eq:4.34}
	\left\langle A, e^{i E x^0} q^{L_0}  v_2(B, C) \right\rangle  &=
	\left\langle (e^{i E x^0} A) \left({w_1 \over q} = 0\right) B (w_2 = 0) C(w_3=0)\right\rangle_{\UHP} 
	\, ,.
\end{align}
Here we have attached a stub of length $- \log q$ to the first local coordinates by evaluating the operator $q^{L_0}$ on the state. This immediately shows 
\begin{align} \label{eq:4.37new}
	\widetilde{w}_1 = - {q \over w_1} \, ,
\end{align}
from which we can also read the local coordinates $\widetilde{w}_1(w_2) $ and $\widetilde{w}_1(w_2) $ that must be used for $B$ and $C$ when they are presented as vertex operator insertions on the $\widetilde{w}_1$-frame. In particular, we note $B$ and $C$ must be placed at $\widetilde{w}_1 = \pm q/\lambda$. The resulting geometry on the $\widetilde{w}_1$-frame is illustrated in figure~\ref{fig:w1tilde}.

It is possible to restore $e^{i E x^0}$ as a collection of bulk $X^0$ insertions in~\eqref{eq:4.34} with $A,B,C$ insertions. This is rather difficult to implement generally since it requires carefully keeping track of how the position zero mode $x^0$ is expressed on an arbitrary conformal frame. Fortunately, a drastic simplification occurs when the $\widetilde{w}_1$-frame above is adapted: the zero mode is given by~\eqref{eq:4.4} without any modification and the integration variable $\theta \in [0,\pi]$ is the standard angular parametrization. These features follow directly from the definition~\eqref{eq:4.4} and the relation~\eqref{eq:4.37new}. The resulting integration contours are shown in black in figure~\ref{fig:w1tilde}.

\begin{figure}[t]
	\centering
	\includegraphics[scale=.59]{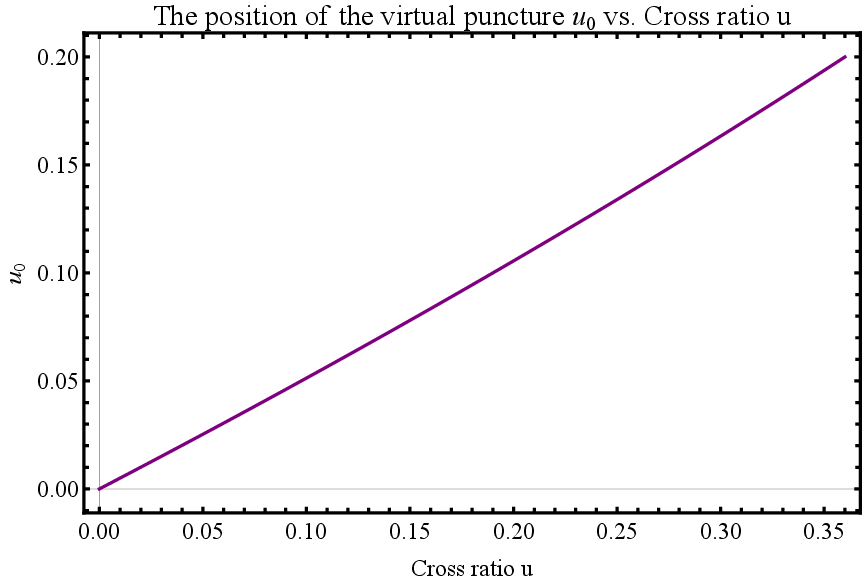}
	\includegraphics[scale=.59]{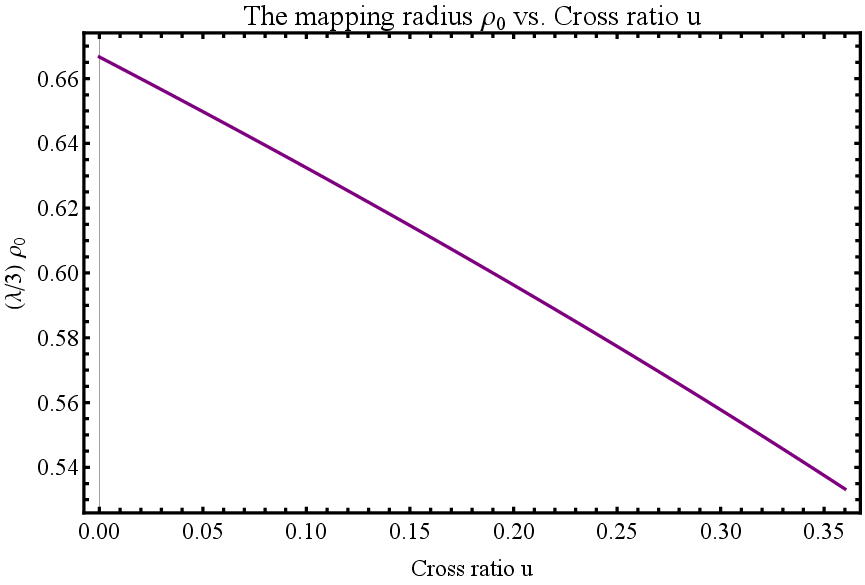}
	\caption{\label{fig:virt}The position $u_0$ and mapping radius $\rho_0$ given in~\eqref{eq:4.38new}. They are plotted over the maximum extent of the lower Feynman region, see~\eqref{eq:A.18}. Observe $0< u_0 < u$ and $0 < \rho_0 < 1$.}
\end{figure}

For our purposes, the argument above implies that the maps $f^\pm$ from $(- q/w_3)$-and $(-1/w_3)$-frames to the uniformizing UHP must be used for the conformal transformations of $e^{i E x^0}$ insertions upon expressing $\Omega_3^{\Psi_2 \textendash \Psi_2} $ as in~\eqref{eq:3.34new}. Indeed, this is what is shown in figure~\ref{fig:f}. The explicit expression for the maps $f^\pm$ can be found straightforwardly using the sewing analysis presented in appendix~\ref{sec:A}. Here we only report the following:
\begin{subequations} \label{eq:4.38new}
	\begin{align}
		&f^\pm(0) = {2 q \over q+ \lambda^2}=1 - \sqrt{1-u} \equiv u_0  \label{eq:4.37anew}\, , 
		\\
		&\partial f^+(0) = -{2 \over \lambda} {q - \lambda^2 \over q+ \lambda^2} = {2 \over \lambda} \sqrt{1-u} \equiv \rho_0\, ,
		\\
		&\partial f^- (0) = q \rho_0 \, .
	\end{align}
\end{subequations}
Note that both $f^\pm$ maps the origin to the same point on the uniformizing UHP. We call this point~\emph{virtual puncture}. The position of the virtual puncture $u_0$ and the associated mapping radius $\rho_0$ as a function of cross ration are shown in figure~\ref{fig:virt}. Take note of the relation between $\partial f^+(0)$ and $\partial f^-(0)$.

We are now ready to evaluate $\Omega_3^{\Psi_2 \textendash \Psi_2}$. In order to do this we need to understand how to handle insertions of $e^{i E x^0}$ in correlators. We discuss this in appendix~\ref{sec:D} extensively. In nutshell, we derive a formula~\eqref{eq:C.24a} exchanging the complicated insertion $e^{i E x^0}$ with a new plane wave vertex operator carrying an energy $E$ at the virtual puncture placed at $z=f^\pm(0) = u_0$. Adapting it to the situation at hand and evaluating the resulting correlator containing only plane wave vertex operators, we find
\begin{align} \label{eq:4.41a}
	&\Omega_3^{\Psi_2 \textendash \Psi_2} = {2 V \over g^2}
	\int\limits_{-\infty}^{\infty} {dE} \, \dot{\sigma}(E) 
	\prod_{j=2}^4 \bigg(-i {\partial \over \partial E_j}\bigg) 
	\Bigg(
	\int\limits_0^{u_-(\lambda)} du \, \delta\left( E + 2 E_2 + E_3 + E_4 \right) F(E_2, E_2, E_3, E_4)
	\nonumber \\ 
	&
	\times {1 \over iE }\bigg[ 
	q^{2 E (E_3 + E_4)}
	- q^{-E^2}
	\bigg]
	\rho_0^{-E^2}\lambda^{-2 E (E_3 + E_4)}   (1-u_0)^{-2 E E_2} (u-u_0)^{-2 E E_3} u_0^{-2 E E_4}
	\Bigg)_{E_j = 0}  \, ,
\end{align}
where we used the function $F(E_i;u)$~\eqref{eq:2.22} to collect the expression. Observe that the factors coming from the formula~\eqref{eq:C.24a} differ for the two components of the commutator in~\eqref{eq:4.32}. Thanks to this structure, the singularity at $E=0$ is removable and we have multiplied/divided the integrand by $iE$ to replace $\sigma(E)$ with $\dot{\sigma}(E)$ and make this feature manifest like before. The explicit cancellation of the singularity at $E=0$ is not available after taking $E_j$ derivatives unlike before, however, given that the function inside the square bracket is not polynomial in $E$. Instead, we understand the $E=0$ value of the integrand (and its derivatives with respect to $E$) by filling in the removable singularity at $E=0$ with the value of its $E \to 0$ limit as in~\cite{Bernardes:2025zzu}. 

We note in passing that the $E$-dependent factor in~\eqref{eq:4.41a} grows as an inverted Gaussian at high energies---we have $0 < \rho_0 < 1$, see figure~\ref{fig:virt}. This is a reflection of~\emph{transgressive locality} of SFT~\cite{Bernardes:2025zzu}. Despite this peculiar notion of nonlocality, the Fourier integrals such as~\eqref{eq:4.41a} remain well-defined in the limit $E \to \pm \infty$ thanks to the presence of the (derivatives of the) delta functions at $E=0$ resulting from the perturbative electric field solution.

We can express the rest of the contributions to $\Omega_3$ in a way amenable to evaluations by adapting a similar procedure. For the cubic contributions $\Omega_3^{\text{cubic}}$~\eqref{eq:4.29}, we find
\begin{subequations} \label{eq:4.40}
\begin{align}
	\label{eq:4.41b}
	&\Omega_3^{\sigma_+, I} = -{V \over g^2}
	\int\limits_{-\infty}^{\infty} {dE} \, \dot{\sigma}(E) 
	\prod_{j=1}^3 \bigg(-i {\partial \over \partial E_j}\bigg) 
	\Bigg(
	\int\limits_0^{u_-(\lambda)} du  \, \delta\left( E + E_1 + E_2 + 2 E_3 \right)
	\\ \nonumber
	&\hspace{0.25in}\times
	{1 \over iE }\bigg[ 
	2 \, q^{- E^2} \rho_0^{- E^2} \lambda^{-4 E E_3}
	(1-u_0)^{-2 E E_2} (u-u_0)^{-2 E E_3} u_0^{-2 E E_3}
	F(E_1, E_2, E_3, E_3)
	\nonumber \\
	&\hspace{2.25in}
	-F(E_1, E_2, E_3 + E, E_3)
	-F(E_1, E_2, E_3, E_3 + E)
	\bigg]
	\Bigg)_{E_j = 0} \, ,
	\nonumber
\end{align}
and
\begin{align}
	\Omega_3^{\sigma_-} = {2  V\over g^2}
	\int\limits_{-\infty}^{\infty} {dE} \,  \dot{\sigma}(E) 
	&\prod_{j=2}^4 \bigg(-i {\partial \over \partial E_j}\bigg)
	\Bigg(
	\int\limits_0^{u_-(\lambda)} du \, \delta\left( E +E_3 + E_4 \right)
	\\ \nonumber
	&
	\times {1 \over iE }\bigg[ 
	F(-E_2 + E, E_2, E_3, E_4)
	- F(-E_2, E_2+ E, E_3, E_4)
	\bigg]
	\Bigg)_{E_j = 0}  \, ,
\end{align}
and
\begin{align}
	\label{eq:4.41c}
	&\Omega_3^{\sigma_+, II} = {V \over g^2}
	\int\limits_{-\infty}^{\infty} {dE} \, \dot{\sigma}(E) 
	\prod_{j=2}^4 \bigg(-i {\partial \over \partial E_j}\bigg) 
	\Bigg(
	\int\limits_0^{u_-(\lambda)} du  \, \delta\left( E + 2 E_2 + E_3 + E_4 \right)
	\\ \nonumber
	&\hspace{0.25in}\times
	{1 \over iE }\bigg[ 
	2 \, q^{- E^2} \rho_0^{- E^2} \lambda^{-2 E (E_3 + E_4)}
	(1-u_0)^{-2 E E_2} (u-u_0)^{-2 E E_3} u_0^{-2 E E_4}
	F(E_2 , E_2, E_3, E_4)
	\\ \nonumber
	& \hspace{2.25in}
	-F(E_2, E_2, E_3 + E, E_4)
	-F(E_2, E_2, E_3, E_4 + E)
	\bigg]
	\Bigg)_{E_j = 0} \, ,
	\nonumber
\end{align}
\end{subequations}
while the quartic contributions $\Omega_3^{\text{quartic}}$ can be collected into the following moduli integral over the vertex region~\eqref{eq:A.13}:
\begin{align} \label{eq:4.41}
	\Omega_3^{\text{quartic}}  = {V \over g^2}
	\int\limits_{-\infty}^{\infty} {dE} \, &\dot{\sigma}(E) 
	\prod_{j=2}^4 \bigg(-i {\partial \over \partial E_j}\bigg) 
	\Bigg(
	\int\limits_{u_-(\lambda)}^{u_+ (\lambda)} du  \, \delta\left( E + E_2 + E_3 + E_4 \right)
	\\ \nonumber
	\times {1 \over iE }&\bigg[ 
	F(E_2 , E, E_3 , E_4)
	+F(E_2 , E_3, E , E_4)
	+F(E_2 , E_3, E_4 , E)
	\nonumber \\
	&- F(E_2, 0,  E_3, E_4 + E)
	- F(E_2, E_3 + E,  0, E_4 )
	-F(E_2 , E_3, E_4 +E , 0)
	\bigg]
	\Bigg)_{E_j = 0} \, .
	\nonumber
\end{align}
Note the orientation for this integral. The functions in the square brackets above vanish when $E \to 0$ similar to~\eqref{eq:4.41a}.

\begin{figure}[p!]
	\centering
	\includegraphics[scale=0.59]{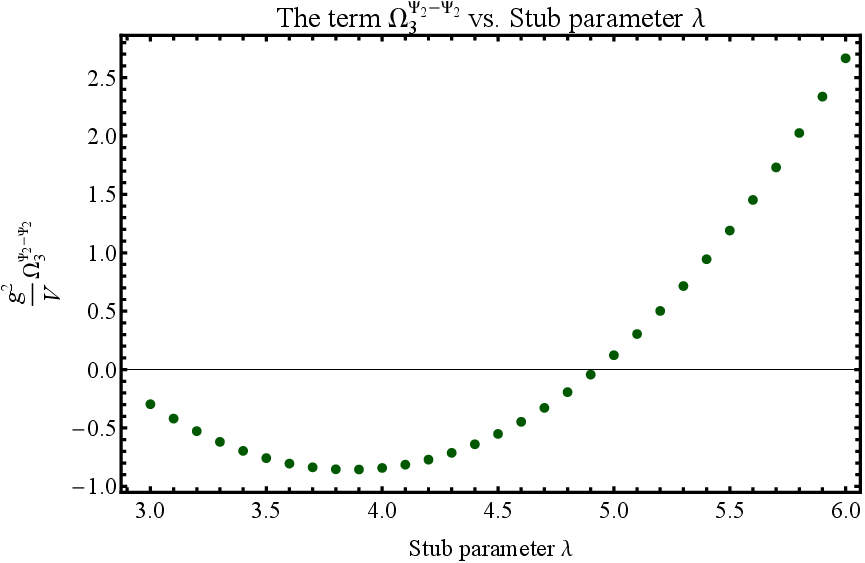}
	\includegraphics[scale=0.59]{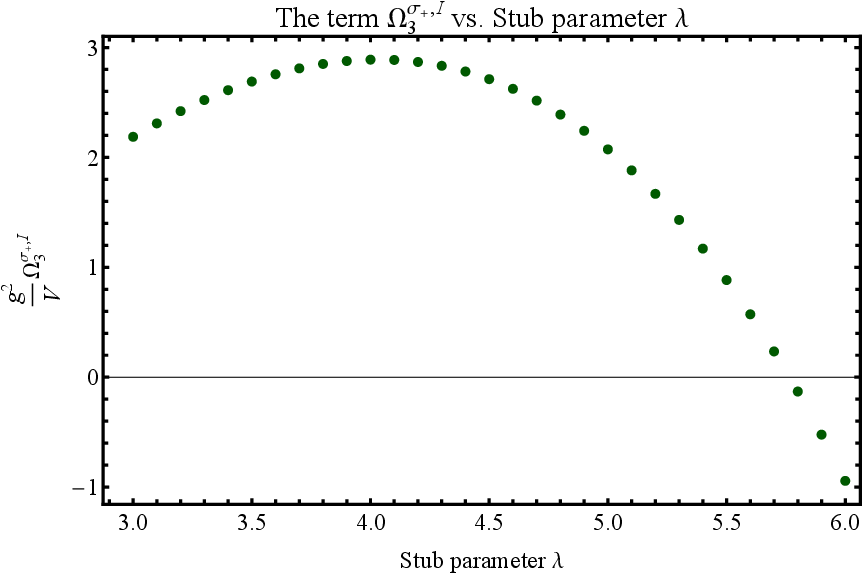}
	\includegraphics[scale=0.59]{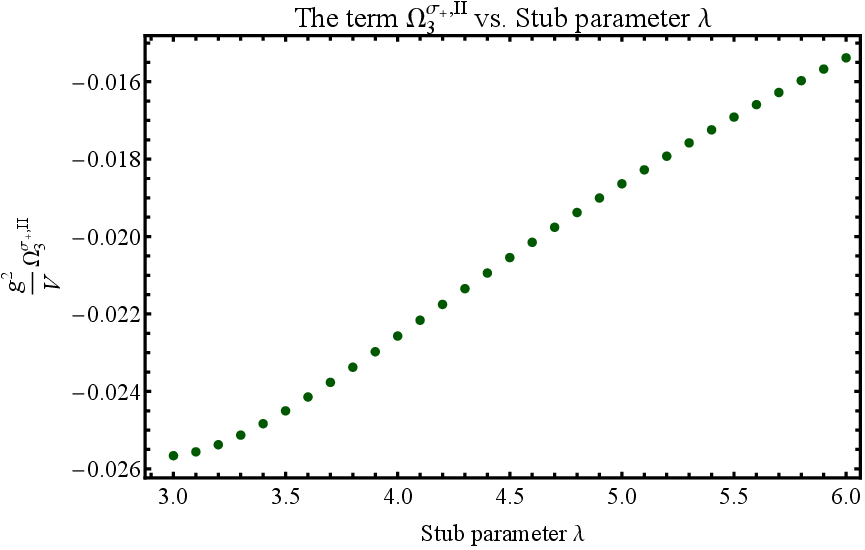}
	\includegraphics[scale=0.59]{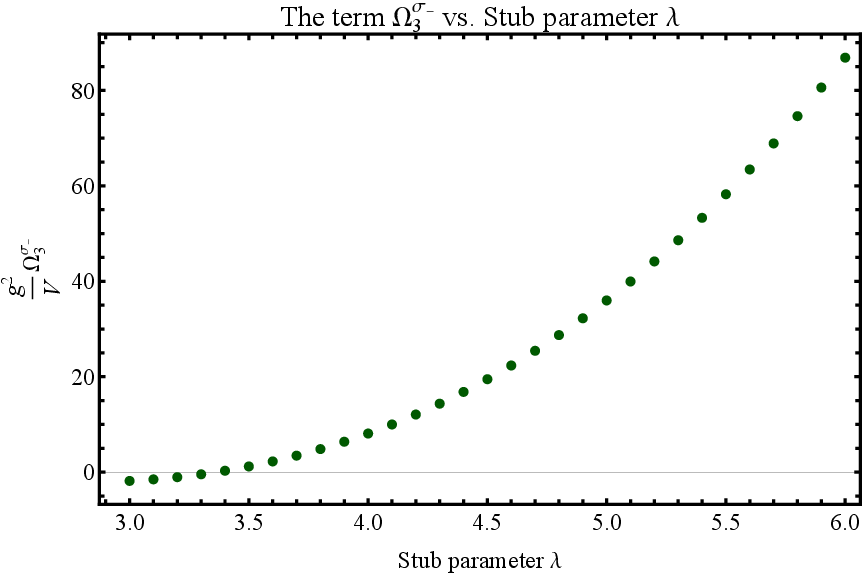}
	\includegraphics[scale=0.59]{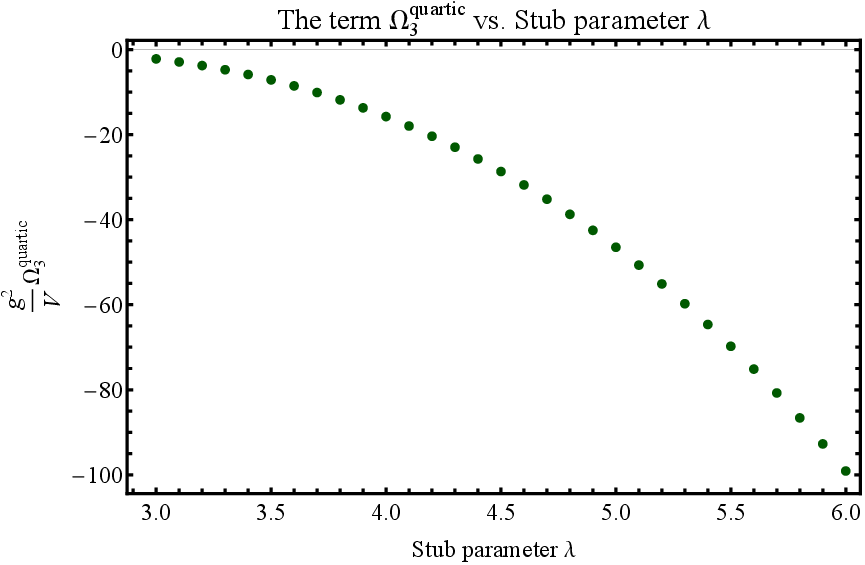}
	\caption{\label{fig:Omega3}The nontrivial contributions to the symplectic form at third order $\Omega_3$  from~\eqref{eq:4.41a}, \eqref{eq:4.40}, and~\eqref{eq:4.41} as a function of the stub parameter~$\lambda$.}
\end{figure}
\begin{figure}[t]
	\centering
	\includegraphics[scale=1]{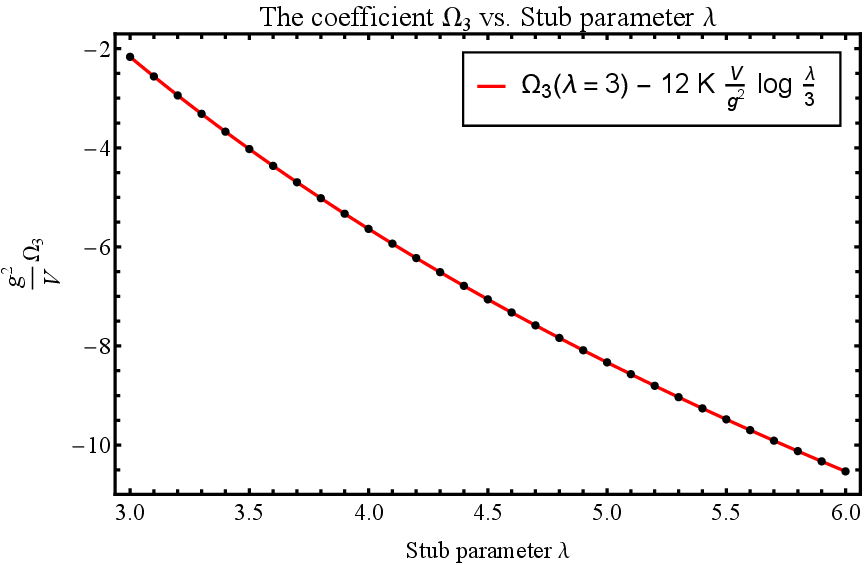}
	\caption{\label{fig:E4}The symplectic form at third order $\Omega_3$ as a function of the stub parameter $\lambda$. The red curve is the flow of $\Omega_3$ under the stub deformation with the initial condition fixed by $\lambda = 3$.}
\end{figure}

 
The next order of business is to differentiate with respect to $E_j$s and set them to zero. The resulting expressions are somewhat involved, so we omit reporting them here explicitly. Instead, we just explain our procedure. After eliminating $E_j$, we obtain Fourier integrals that contain 
\begin{align}
	\delta(E) \, , \quad 
	\delta'(E) \, , \quad
	\delta''(E) \, , \quad
	\delta'''(E) \, .
\end{align}
These integrals can be evaluated trivially since the distributions above always multiply functions that are regular at $E=0$ by our discussion below~\eqref{eq:4.41a}. The resulting expressions contain terms proportional to the derivatives of $\dot{\sigma}(E)$ in general, but we discard them by the conservation of $\Omega$ as before. After the dust settles, we are just left with moduli integrals proportional to $\dot{\sigma}(0)=1$. The numerical evaluation of these moduli integrals is straightforward. The only subtlety is that the divergences toward $u = 0$ in~\eqref{eq:4.41a} and \eqref{eq:4.40} should be treated according to the SFT prescription~\eqref{eq:3.27a}. The results are shown in figure~\ref{fig:Omega3}. Adding them up, we find $\Omega_3$ as a function of the stub parameter $\lambda$. This is shown in figure~\ref{fig:E4}. The explicit expressions for the integrals and the details for their numerical evaluations can be found in the \texttt{Mathematica} notebook in the Supplementary material accompanying this paper.

We can check the consistency of the final result $\Omega_3 = \Omega_3(\lambda)$ by investigating how it ``flows'' under the stub deformations. In order to do this, we first recall the relation between two solutions $\Psi, \Psi'$ of open SFT using different stub parameters $\lambda, \lambda'$~\cite{Bernardes:2025zzu,Erler:2023emp}:
\begin{align} \label{eq:4.46}
	\Psi = \bigg({\lambda \over \lambda'} \bigg)^{-L_0} \Psi' \, .
\end{align}
Inspecting the non-obstruction part of the perturbative solution~\eqref{eq:2.5}, the ``stub operator'' above takes out stubs of length $\log \lambda' /3$ and adds back stubs of length $\log \lambda / 3$ geometrically as long as $\lambda, \lambda' \geq 3$. On the other hand, its action on the obstruction term is somewhat more nontrivial:
\begin{align}
	\bigg({\lambda \over \lambda'} \bigg)^{-L_0}  \psi_3 &= 
	\bigg(-i {\partial \over \partial E} \bigg)^3 
	\bigg[ \bigg({\lambda \over \lambda'} \bigg)^{E^2} 
	\, K c \partial X^1 e^{i E X^0} \bigg]_{E=0}
	\\ \nonumber 
	&= K c \left(X^0\right)^3 \partial X^1 - 6 K \log \bigg({\lambda  \over \lambda'} \bigg) c X^0 \partial X^1 \, .
\end{align}
This implies that the perturbative electric field solutions of two open SFTs using different $SL(2,\mathbb{R})_\lambda$ vertices constructed as in~\eqref{eq:2.8} relate to each other only after the field redefinition
\begin{align}
	\varepsilon(\lambda) = \varepsilon(\lambda') 
	- 6 K \log \bigg({\lambda  \over \lambda'} \bigg) \varepsilon(\lambda') ^3 + \mathcal{O}\left(\varepsilon^5\right) \, .
\end{align}
Substituting it into the expansion~\eqref{eq:4.8}, we then find that the symplectic structure at third order $\Omega_3 = \Omega_3(\lambda)$ obeys by the following equation
\begin{align} \label{eq:4.50}
	\Omega_3(\lambda) = \Omega_3 (\lambda = 3) - 12 K {V \over g^2} \log {\lambda \over 3}  \, ,
\end{align} 
where we assume the ``initial condition" is fixed using the extremal stub parameter $\lambda = 3$. The resulting flow is the~\textcolor{red}{red} curve shown in figure~\ref{fig:E4}. This agrees with $\Omega_3(\lambda)$ computed using the SFT symplectic form within an accuracy of $10^{-11}$ after constructing an appropriate fit for the latter. 

For the record, we explicitly report the energy~\eqref{eq:4.8} evaluated using the SFT symplectic form with $SL(2,\mathbb{R})_3$ vertices:
\begin{align} \label{eq:4.49}
	E(\lambda = 3 ) = {V \over g^2}  \bigg[ - {1 \over 4} \varepsilon^2
	+ \alpha_4 \varepsilon^4 + \mathcal{O}\left(\varepsilon^6\right) \bigg]  
	\, , \quad \quad
	\alpha_4 \approx -0.5410655 \, .
\end{align}
The numerical value of $\alpha_4$ is not determined by the flowing argument above and we still need to independently confirm that it is consistent with the DBI result~\eqref{eq:10.11}. This necessitates determining the higher order terms in the field redefinition~\eqref{eq:4.15a}. We investigate this in the next section.

\section{Homotopy Ellwood invariant} \label{sec:5.1}

In this section we find the field redefinition between the open SFT and DBI descriptions of a D-brane carrying a constant electric flux. A crucial ingredient is provided by the homotopy-algebraic generalization of the Ellwood invariant in Witten's open SFT~\cite{Ellwood:2008jh} to nonpolynomial open SFTs based on cyclic $A_\infty$ algebras. This object has previously appeared in~\cite{Maccaferri:2021lau} in the context of super SFT, however there appears to be no explicit discussion on its connection to boundary states along the lines of~\cite{Ellwood:2008jh} in the literature. We develop this connection partially as part of our application.

We begin our discussion by introducing~\emph{the (homotopy) Ellwood invariant} 
\begin{align} \label{eq:5.1}
	\Gamma^{\mathcal{O}}[\Psi] =  \left\langle \Psi, e_0^{\mathcal{O}} \right\rangle
	+ {1 \over 2} \left\langle \Psi, e_1^{\mathcal{O}} \Psi \right\rangle
	+ {1 \over 3} \left\langle \Psi, e_2^{\mathcal{O}} \left(\Psi, \Psi \right) \right\rangle
	+ \mathcal{O} \left( \Psi^4 \right)  \, ,
\end{align}
for which~\emph{the Ellwood products} $e_n^{\mathcal{O}}$ ($n \geq 0$) are defined by
\begin{align} \label{eq:5.2}
	E_{n+1}^{\mathcal{O}}\left(\Psi_1, \cdots \Psi_{n+1} \right) 
	&=
	\left\langle \Psi_1, e_n^{\mathcal{O}}\left(\Psi_2, \cdots \Psi_{n+1} \right) \right\rangle
	\\ \nonumber
	&= \int\limits_{\mathcal{E}_{n+1}} \left\langle
	\mathcal{B} \mathcal{O}(\xi =0 ) \Psi_1(w_1=0) \cdots \Psi_{n+1}(w_{n+1}=0)
	\right\rangle_\UHP\, .
\end{align}
Observe the similarities between these objects and~\eqref{eq:2.1}-\eqref{eq:3.4}. There are few important differences however. One difference is the presence of the 0-th product $e_0^\mathcal{O}$ in~\eqref{eq:5.1} relative to the action~\eqref{eq:2.1}.  Another, a somewhat more crucial, difference is the insertion of the~\emph{bulk} vertex operator $ \mathcal{O}$ in the correlator~\eqref{eq:5.2}. We assume~$\mathcal{O}$ to be Grassmann even, level-matched, and annihilated by the closed string BRST operator $Q_c$:
\begin{align} \label{eq:5.3}
	Q_c \cdot \mathcal{O} = 0 
	\, ,  \quad \quad
	\left(b_0 - \overline{b}_0\right) \mathcal{O} = 
	\left(L_0 - \overline{L}_0\right) \mathcal{O} = 0
	\, .
\end{align} 
Accordingly, the integration regions $\mathcal{E}_{n+1}$ in~\eqref{eq:5.2} are subsets of the moduli spaces of disks with $n+1$ cyclically-ordered boundary punctures, and a~\emph{bulk} puncture, with non-overlapping local coordinates that are sufficiently far away from degeneration. The local coordinate $0 < |\xi| < 1$ around the bulk puncture can be defined up to a phase by the level-matching. In fact, it is even admissible for $\xi$ to be singular as long as $\mathcal{O}$ is a weight-$(0,0)$ primary. We call $\mathcal{E}_{n+1}$s~\emph{(geometric) Ellwood vertices}. Similar to string vertices $\mathcal{V}_n$, the choice of Ellwood vertices $\mathcal{E}_{n+1}$ is not arbitrary: they should obey a geometric master equation to ensure the gauge invariance of $\Gamma^\mathcal{O}[\Psi]$, as we shall discuss shortly. For more details on this geometric master equation, see section 3.3 of~\cite{Zwiebach:1997fe}. Lastly, the $b$-ghost insertions in~\eqref{eq:5.2} proceed as in~\eqref{eq:3.4}, with the exception that we do not include the normalization factor $\eta_c = - 1/(2\pi i)$ due to the bulk insertion in order to agree with the standard conventions of the Ellwood invariant literature~\cite{Ellwood:2008jh,Kiermaier:2008qu,Kudrna:2012re}.

Before developing the theory of~\eqref{eq:5.1} further let us comment on how $\Gamma^{\mathcal{O}}[\Psi]$ reduces to the standard Ellwood invariant and provides its homotopy-algebraic generalization. Similar to how the action of Witten's theory truncates at the cubic term, $\Gamma^{\mathcal{O}}[\Psi]$ can be truncated to its leading term in Witten's theory while satisfying the relevant geometric master equation after choosing a specific $\mathcal{E}_1$~\cite{Zwiebach:1990az,Zwiebach:1992bw}. This is the vertex studied by Ellwood~\cite{Ellwood:2008jh}, which originally appeared in~\cite{Shapiro:1987gq,Shapiro:1987ac} (also see~\cite{Hashimoto:2001sm,Gaiotto:2001ji}). On the other hand, such a truncation is not available in a generic open SFT and, naturally, we should expect $\Gamma^{\mathcal{O}}[\Psi]$ to contain higher order terms. Their associated products can be shown to obey certain homotopy coherent relations as a consequence of the relevant geometric master equation and, in this precise sense, $\Gamma^{\mathcal{O}}[\Psi]$ is the ``homotopy" Ellwood invariant.

The homotopy Ellwood invariant~\eqref{eq:5.1} has few important properties. Arguably, the most important among these is that it is indeed an~\emph{invariant},
\begin{align} \label{eq:5.4}
	 \Gamma^{\mathcal{O}}[\Psi_\ast + \delta_\Lambda \Psi_\ast] = \Gamma^{\mathcal{O}}[\Psi_\ast] 
	 + \mathcal{O} \left( \Lambda^2 \right) \, ,
\end{align}
as long as the bulk insertion $\mathcal{O}$ is supported at finite times and $\Psi = \Psi_\ast$ solves the equation of motion. Here
\begin{align}
	\delta_{\Lambda} \Psi = v_1 \Lambda &+ v_2 (\Psi, \Lambda) - v_2 (\Lambda, \Psi)
	\\ \nonumber
	&+ v_3 (\Psi, \Psi, \Lambda) - v_3 (\Psi, \Lambda, \Psi) + v_3(\Lambda, \Psi, \Psi) 
	+ \mathcal{O} \left(\Psi^3\right) \, ,
\end{align}
is an infinitesimal gauge transformation of open SFT with a Grassmann even and ghost number $0$ gauge parameter $\Lambda$.  In this work we consider~\emph{small} gauge transformations. That is, we assume the gauge transformations are continuously connected to the identity. There are no unaccounted spatial boundary and/or topological contributions in~\eqref{eq:5.4} as a consequence.

The invariance~\eqref{eq:5.4} follows from the structural similarity between the homotopy Ellwood invariant and the open SFT action. Analogous to how the gauge invariance of the action~\eqref{eq:2.1} is a consequence of the geometric master equation~\cite{Sen:2024nfd}, the homotopy Ellwood invariant is gauge invariant as a consequence of the relevant geometric master equation satisfied by $\mathcal{E}_{n}$s and $\mathcal{V}_{n}$s. This argument requires $\Psi=\Psi_\ast$ to satisfy the equation of motion and the bulk insertion $\mathcal{O}$ to be BRST-closed, which guarantees the BRST identity on the disk remains applicable. Crucially, establishing the gauge invariance this fashion also requires the Ellwood products $e_n^\mathcal{O}$ defined in~\eqref{eq:5.2} to be cyclic---otherwise there may be unaccounted~\emph{temporal} boundary contributions on the right hand side of~\eqref{eq:5.4}. We can ensure the cyclicity of  $e_n^\mathcal{O}$  by having the bulk insertion $\mathcal{O}$ to be level-matched and supported at finite times.

This discussion shows $\Gamma^\mathcal{O}[\Psi_\ast]$ is an~\emph{observable} of open SFT whenever $\Psi=\Psi_\ast$ solves the equation of of motion~\eqref{eq:3.5}. In fact, the Ellwood invariant provides a pairing between the open SFT solutions and the semi-relative closed string cohomology. In order to show this is well-defined, we must establish
\begin{align}
	\Gamma^{Q_c \cdot \chi}[\Psi_\ast] = 0 \, 
	\quad \text{where} \quad
	\left(b_0 - \overline{b}_0 \right) \chi = \left(L_0 - \overline{L}_0 \right) \chi = 0 \, .
\end{align}
A quick inspection shows that deforming the BRST current contour surrounding the state $\chi$ towards the boundary of the disk is permitted, which leads to two distinct contributions in $\Gamma^{Q_c \cdot \chi}[\Psi_\ast] $: the terms with the open string BRST charge $Q$ acting on $\Psi_\ast$ and the moduli integrals of total derivatives over the Ellwood vertices. These two types of terms cancel each other according to the relevant geometric master equation after using the equation of the motion~\eqref{eq:3.5} for the former and the Stokes theorem for the latter. More details for an argument along these lines can be found in~\cite{Moeller:2010mh,Moeller:2011zz}. In mathematical terms, this pairing establishes an isomorphism between~\emph{the cyclic cohomology}~\cite{loday2013cyclic} of the cyclic $A_\infty$ algebra underlying the open SFT and the semi-relative BRST cohomology of its closed string background.

We are now ready to state the homotopy-algebraic generalization of the so-called Ellwood conjecture relating the boundary states to the open SFT solutions:
\begin{align} \label{eq:5.7}
	\left\langle B_\ast | \left(c_0 - \overline{c}_0 \right) | \mathcal{O} \right\rangle_{S^2}  
	- \left\langle B_0| \left(c_0 - \overline{c}_0 \right)  | \mathcal{O} \right\rangle_{S^2}    
	= - 4 \pi i \,\Gamma^{\mathcal{O}}[\Psi_\ast] \, .
\end{align}
Here $| B_0 \rangle$ is the (ghost number $3$) boundary state associated with the background BCFT, while $| B_\ast \rangle$ is the boundary state associated with the BCFT corresponding to the open SFT solution $\Psi = \Psi_\ast$. We put subscript $S^2$ (for sphere) on the BPZ product on the left hand side to indicate that these overlaps use closed string conventions. We will not attempt to give an argument for ``the homotopy Ellwood conjecture''~\eqref{eq:5.7} in this paper. Nevertheless, we point out that Ellwood gave an argument for~\eqref{eq:5.7} in Witten's SFT in his landmark work~\cite{Ellwood:2008jh} and it should be possible to transfer this argument to nonpolynomial open SFTs by developing a field redefinition along the lines of~\cite{Hata:1993gf}. This will be presented elsewhere.

\subsection{Electric field boundary state}

Let us now study each side of the relation~\eqref{eq:5.7} for a D-brane carrying a constant electric flux in order to find the field redefinition between the parameters $\varepsilon$ and $\varepsilon_\DBI$. In this subsection we focus on the boundary state (i.e., left hand) side of~\eqref{eq:5.7}. The Ellwood invariant side is developed in the next subsection. 

The closed string insertion $\mathcal{O}$ appears in both side of~\eqref{eq:5.7} so it is a perfect initial point for our investigations. We fix it to be a Kalb-Ramond state
\begin{align} \label{eq:5.8}
	\mathcal{O} = \int\limits_{-\infty}^\infty {d E \over 2\pi} f(E)
	c \overline{c}
	\epsilon_{\mu \nu} \partial X^\mu \overline{\partial} X^{\nu} e^{i k \cdot X} \, ,
\end{align}
with the following polarization $\epsilon_{\mu \nu}$ and momentum $k_\mu$ ($\mu,\nu = 0,1,2$):
\begin{align} \label{eq:5.9}
	\epsilon_{\mu \nu}   = \begin{bmatrix}
		0 & 1 & 0 \\ -1 & 0 & -1 \\\ 0 & 1 & 0
	\end{bmatrix}_{\mu \nu}
	\, , 
	\quad \quad
	k_{\mu} = \begin{bmatrix}
		E  \\ 0 \\ E
	\end{bmatrix}_\mu \, .
\end{align}
The ``wave function'' $f(E)$ is chosen to be a Schwartz function---one may take this to be a Gaussian to be concrete. Note that BCFT$_\perp$ must contain a free~\emph{noncompact} boson $X^2$ in order to consider this state. We assume that it indeed does and separate it from the rest of  the BCFT$_\perp$ moving forward. We further demand that $X^2$ is subject to Dirichlet boundary conditions and place the D-brane at the origin $x_0^2 = 0$. These choices make sure that $\mathcal{O}$ is supported at finite times, ensuring the absence of boundary terms spoiling gauge invariance~\eqref{eq:5.4}. We highlight
\begin{align} 
	\epsilon_{\mu \nu} = - \epsilon_{\nu \mu}
	\, , \quad \quad
	\epsilon_{\mu \nu} k^\nu =\epsilon_{\mu \nu} k^\mu =  0 
	\, , \quad \quad
	k^2 = 0 \, ,
\end{align}
which makes $\mathcal{O}$ a weight-$(0,0)$ primary, thus we do not need to keep track of its local coordinates in diagrams. This simplification was the primary motivation behind including an additional $X^2$ BCFT to our considerations. The motivation for using an antisymmetric polarization is going to be apparent in the next subsection.

Now, consider the left hand side of~\eqref{eq:5.7} with the Kalb-Ramond state~\eqref{eq:5.8}. First, we focus on the overlap of $(c_0 - \overline{c}_0)| \mathcal{O} \rangle$ with $| B_\ast \rangle$, which we assume it to be the boundary state associated with a D1-brane carrying a constant electric flux. Such a boundary state (and its T-dual) is investigated in~\cite{DiVecchia:1999mal,DiVecchia:1999fje}. Adapting it to the bosonic theory and our conventions, we have\footnote{We take $x^1$ direction to be non-compact in this computation for brevity. We are going to compactify it after evaluating the relevant overlaps. The result remains unchanged if $x^1$ is compactified from the get-go.}
\begin{align} \label{eq:5.11}
	| B_\ast \rangle = \mathcal{N}\left(\varepsilon_\DBI\right) \exp 
	\bigg[- \sum_{n=1}^\infty  {1 \over n} \alpha^{\mu}_{-n} \Omega_{\mu \nu}
	\left(\varepsilon_\DBI\right) \overline{\alpha}^\nu_{-n} 
	&+ \sum_{n=1}^\infty \left( c_{-n} \overline{b}_{-n} + \overline{c}_{-n} b_{-n} \right) \bigg]
	\\ \nonumber
	&{1 \over 2}\left( c_0 + \overline{c}_0 \right)  c_1 \overline{c}_1
	\int\limits_{-\infty}^\infty {d k_2 } \, e^{i k_2 X^2(0,0)}
	|0 \rangle
	\otimes | B_\perp \rangle \, ,
\end{align}
where
\begin{align} \label{eq:5.12}
	\Omega_{\mu \nu}
	&= \left( {1 - F_\DBI \over 1 + F_\DBI} \right)_{\mu \nu}
	\\ \nonumber
	&= {1 \over 1 - \left(2 \pi \varepsilon_\DBI\right)^2}
	\begin{bmatrix}
		-\left(1 + \left(2 \pi \varepsilon_\DBI\right)^2 \right) & -4 \pi \varepsilon_\DBI & 0 \\
		4 \pi \varepsilon_\DBI & 1 +  \left(2 \pi \varepsilon_\DBI\right)^2 &0 \\
		0 & 0 & -\left(1 - \left(2 \pi \varepsilon_\DBI\right)^2\right)
	\end{bmatrix}_{\mu \nu} \, ,
\end{align}
is the matrix intermixing the left-and right-moving oscillators in the presence of the electric field. We only show the $\mu =0,1,2$ and ghost components of the boundary state explicitly while suppressing the transverse component of the boundary state $| B_\perp \rangle$. The normalization factor $\mathcal{N}(\varepsilon_\DBI)$ depends on the electric field nontrivially and it is given by
\begin{align} \label{eq:5.13}
	\mathcal{N}(\varepsilon_\DBI) =  {\kappa T \over 2} \sqrt{1 - \left(2 \pi \varepsilon_\DBI\right)^2 } \, .
\end{align}
Here $T$ is as given in~\eqref{eq:2.2} and $\kappa = 8 \pi G_N$ is the gravitational constant. Note that the boundary state takes a somewhat familiar form involving the Neumann/Dirichlet boundary conditions (with the correct normalization) when $\varepsilon_\DBI = 0$~\cite{Polchinski:1998rq,DiVecchia:1999mal}.

In the light of these remarks, we evaluate the overlap:
\begin{align} \label{eq:5.14}
	\langle B_\ast | \left(c_0 - \overline{c}_0 \right) | \mathcal{O} \rangle_{S^2}  
	&= \bigg(-{2  \over g_s T }\bigg) \times \bigg( - {i \over \sqrt{2}} \bigg)^2 
	\times
	\bigg({\kappa T \over 2} \sqrt{1 - (2 \pi \varepsilon_\DBI)^2} \bigg)
	\nonumber \\
	&\hspace{1.5in}\times\int\limits_{-\infty}^\infty {d E \over 2 \pi} \, f(E) \,
	{-8\pi \varepsilon_\DBI\over 1 - (2 \pi \varepsilon_\DBI)^2}  \, 2 \pi V \delta(E)
	\, .
\end{align}
It is helpful to explain the origin of various factors appearing above. The first factor comes from the normalization of the closed string correlators, which effectively takes the form~\cite{Sen:2024nfd,Sen:2024npu}\footnote{Refer to section 5.2 in~\cite{Sen:2024nfd} for the derivation of this normalization relative to its open string counterpart~\eqref{eq:3.14}. We take $\eta_c \to 1$ as in~\eqref{eq:5.2}. It can be checked that restoring $\eta_c$ in the arguments does not alter the results.}
\begin{align} \label{eq:5.15}
	\left\langle k | c_{-1} \overline{c}_{-1} c_0 \overline{c}_0 c_1 \overline{c}_1 |k'\right\rangle^{\text{gh},X^{0,1,2}}_{S^2} = 
	\bigg( - {2 \over g_s T} \bigg) (2\pi)^3 \delta^{(3)}(k + k') \, ,
\end{align}
for the relevant part of the CFT in our conventions. Here $g_s$ is~\emph{closed} string coupling constant and it is related to the gravitational constant $\kappa$ by
\begin{align}
	\kappa = 2 g_s \, ,
\end{align}
see equation (4.123) in~\cite{Sen:2024nfd}. The second factor in~\eqref{eq:5.14} is due to the normalization of the Kalb-Ramond state~\eqref{eq:5.8} and the third factor is due to the boundary state normalization~\eqref{eq:5.13}. Finally, the Fourier integral in the second line is a consequence of evaluating the closed string overlap in the view of the matrix~\eqref{eq:5.12}. For this, we used
\begin{align}
	\left\langle B_\perp | 0 \right\rangle^\perp_{S^2} = Z_\perp \, ,
\end{align}
coming from the disk partition function of BCFT$_\perp$~\eqref{eq:3.3}, and then combined it with the rest of the factors coming from~\eqref{eq:5.15} ($2 \pi \delta(E)$, $2 \pi \delta(0) = V_1$, $2 \pi = Z_2$) to produce $2 \pi V \delta(E)$. Here $V = V_1 Z_2 Z_\perp$ which previously appeared in~\eqref{eq:10.11} and~\eqref{eq:3.14}. Simplifying all of these terms, we find
\begin{align} \label{eq:5.16}
	\left\langle B_\ast | \left(c_0 - \overline{c}_0 \right) | \mathcal{O} \right\rangle_{S^2}  
	&= -V f(0) \, { 8 \pi \varepsilon_\DBI \over \sqrt{1 - \left(2 \pi \varepsilon_\DBI\right)^2} } 
		\\ \nonumber 
	&= 
	-V f(0) \left( 8 \pi  \varepsilon_\DBI + 16 \pi^3 \left(\varepsilon_\DBI \right)^3 +  \mathcal{O} \left( \left( \varepsilon_\DBI \right)^5 \right) \right)
	\, .
\end{align}
This result further implies 
\begin{align}
	\left\langle B_0 | \left(c_0 - \overline{c}_0 \right) | \mathcal{O} \right\rangle_{S^2}    = 0 \, ,
\end{align}
after setting $\varepsilon_\DBI = 0 $. Indeed, this overlap vanishes after presenting it as a UHP correlator and observing that it contains either one $\partial X^1$ or $\overline{\partial} X^1$ insertion without the corresponding momenta. Thus~\eqref{eq:5.16} constitutes the entirety of the left hand side of~\eqref{eq:5.7} in our application. We highlight that~\eqref{eq:5.16} agrees with equation (4.11) in~\cite{Ishibashi:2018ynb} after a Wick rotation and including the overall $-4 \pi i$ factor in~\eqref{eq:5.7}. 

\subsection{Field redefinition}

Now we turn our attention to the right hand side of~\eqref{eq:5.7}. Given a perturbative SFT solution $\Psi$~\eqref{eq:2.5}, we can expand the associated Ellwood invariant $\Gamma^\mathcal{\mathcal{O}}[\Psi]$ in $\varepsilon$ as
\begin{align}
	\Gamma^\mathcal{\mathcal{O}}[\Psi] = \varepsilon \Gamma_1^\mathcal{O} +
	\varepsilon^2 \Gamma_2^\mathcal{O} + \varepsilon^3 \Gamma_3^\mathcal{O} + \mathcal{O}\left(\varepsilon^4\right) \, ,
\end{align}
where
\begin{subequations} \label{eq:5.18}
\begin{align}
	&\Gamma_1^\mathcal{O} = \left\langle \Psi_1, e_0^\mathcal{O} \right\rangle \, ,
	\\
	&\Gamma_2^\mathcal{O} =\left \langle \Psi_2, e_0^{\mathcal{O}} \right\rangle 
	+ {1 \over 2} \left\langle \Psi_1, e_1^{\mathcal{O}} \Psi_1 \right\rangle  \, , \label{eq:5.18b}
	\\
	&\Gamma_3^\mathcal{O} = 
	\left\langle \Psi_3, e_0^{\mathcal{O}} \right\rangle 
	+ {1 \over 2} \left\langle \Psi_2, e_1^{\mathcal{O}} \Psi_1 \right\rangle
	+ {1 \over 2} \left\langle \Psi_1, e_1^{\mathcal{O}} \Psi_2 \right\rangle
	+ {1 \over 3} \left\langle \Psi_1, e_2^{\mathcal{O}} (\Psi_1, \Psi_1) \right\rangle \, .
	\label{eq:5.18c} 
\end{align}
\end{subequations}
We investigate each of these contributions for the perturbative electric field solution and compare them with the other side of~\eqref{eq:5.14}. Begin with $\Gamma_1^\mathcal{O}$:
\begin{align} \label{eq:5.20a}
	\Gamma_1^\mathcal{O} &= 
	-\int\limits_{-\infty}^\infty {dE \over 2 \pi} f(E) \epsilon_{\mu \nu}
	\left\langle c \overline{c} \partial X^\mu \overline{\partial} X^\nu e^{i k \cdot X}(\xi = 0) c X^0 \partial X^1 (w_1 = 0) \right\rangle_\UHP
	\\ \nonumber 
	&=
	-\left(-i {\partial \over \partial E'} \right)
	\left[ \int\limits_{-\infty}^\infty {dE \over 2 \pi} f(E)  \epsilon_{\mu \nu} \rho_1^{-(E')^2}
	\left\langle c \overline{c}  \partial X^\mu \overline{\partial} X^\nu e^{i k \cdot X} (z = i) c e^{i E' X^0}\partial X^1 (z = 0) \right\rangle_\UHP \right]_{E' = 0} \, .
\end{align}
In the second line we have replaced $X^0$ with the plane wave vertex operator according to~\eqref{eq:3.20} and then mapped everything to the uniformizing UHP. This has introduced the mapping radius $\rho_1$ for the local coordinate map $w_1(z)$ around the boundary puncture of the first Ellwood vertex $\mathcal{E}_1$. An overall sign is included in this expression as part of the ghost insertions $\mathcal{B}$, as highlighted in the equation (3.83) of~\cite{Sen:2024nfd} and explained in~\cite{Sen:2024npu}. After employing the doubling trick,\footnote{It is important to be mindful about the branches in the correlators of chiral plane vertex operators, otherwise the integrands may contain incorrect relative phases. We choose to orient UHP of $X^0$ correlators~\emph{oppositely} to those of $X^2$. This cancels the phases consistently in the doubling trick without needing to keep track of them explicitly.\label{fn:8}}
\begin{align} 
	\Gamma_1^\mathcal{O} &= -\left(-i {\partial \over \partial E'} \right)
	\Bigg[\int\limits_{-\infty}^\infty {dE \over 2 \pi} f(E)  
	\rho_1^{-(E')^2} (2 i) \bigg( {1 \over 2} \bigg)
	\bigg( 
	{i (2 E') \over 2(i-0)} + {i E \over 2 (i -(-i))}
	- {i (2 E') \over 2(-i-0)} - {i E \over 2(-i-i)} 
	\nonumber \\ 
	 &- {- i (-E) \over 2(i-(-i))} 
	+ {-i (-E) \over 2(-i-i)} 
	\bigg) 2^{-E^2}
	2 \pi V \delta(2 E + 2 E') \Bigg]_{E'=0}
	 \, ,
\end{align}
where the parenthesis' show the results from evaluating the ghost, $X^1$, and $X^0\textendash X^2$ correlators respectively. Observe that the second and fourth terms in the first line cancel the terms in the second line for the $X^0 \textendash X^2$ correlator. This implies
\begin{align} \label{eq:5.20}
	\Gamma_1^\mathcal{O}  &=- i V \left(-i {\partial \over \partial E'} \right) 
	\left[ \int\limits_{-\infty}^\infty dE f(E)2^{-E^2}\rho_1^{-(E')^2} E' \delta(E+E') \right]_{E' = 0}
	\\ \nonumber 
	& = - V \int\limits_{-\infty}^\infty dE f(E) 2^{-E^2} \delta(E)
	 \\ \nonumber
	&= -V f(0) \, .
\end{align}
The dependence on the first Ellwood vertex $\mathcal{E}_1$ entirely drops out, which is expected from~\eqref{eq:5.20a} since the boundary insertion is a weight-0 primary.  We observe that~\eqref{eq:5.20} is nonzero thanks to the polarization structure provided by the Kalb-Ramond state~\eqref{eq:5.8}, thus the field redefinition can be found starting at this order. We find
\begin{align} \label{eq:5.23}
	  -8 \pi \varepsilon_\DBI + \mathcal{O}\left(\left(\varepsilon_\DBI \right)^3\right) = - 4 \pi i 
	  \left( -\varepsilon + \mathcal{O}\left(\varepsilon^2\right)\right)
	 \quad \implies \quad
	 \varepsilon_\DBI  = - {i \over 2} \varepsilon + \mathcal{O}\left(\varepsilon^2\right) \, ,
\end{align}
for all open SFTs as in~\eqref{eq:4.15a} before; but with the fixed overall sign now. This relation also holds to $\mathcal{O}(\varepsilon^3)$ since $\Gamma_2^{\mathcal{O}}$~\eqref{eq:5.18b} contains three $\partial X^1$s after expressing it as a correlator, which vanishes. In fact, $\Gamma_n^\mathcal{O} = 0$ for all even $n$ by a similar reasoning.

\begin{figure}[t]
	\centering
	\includegraphics[scale=.59]{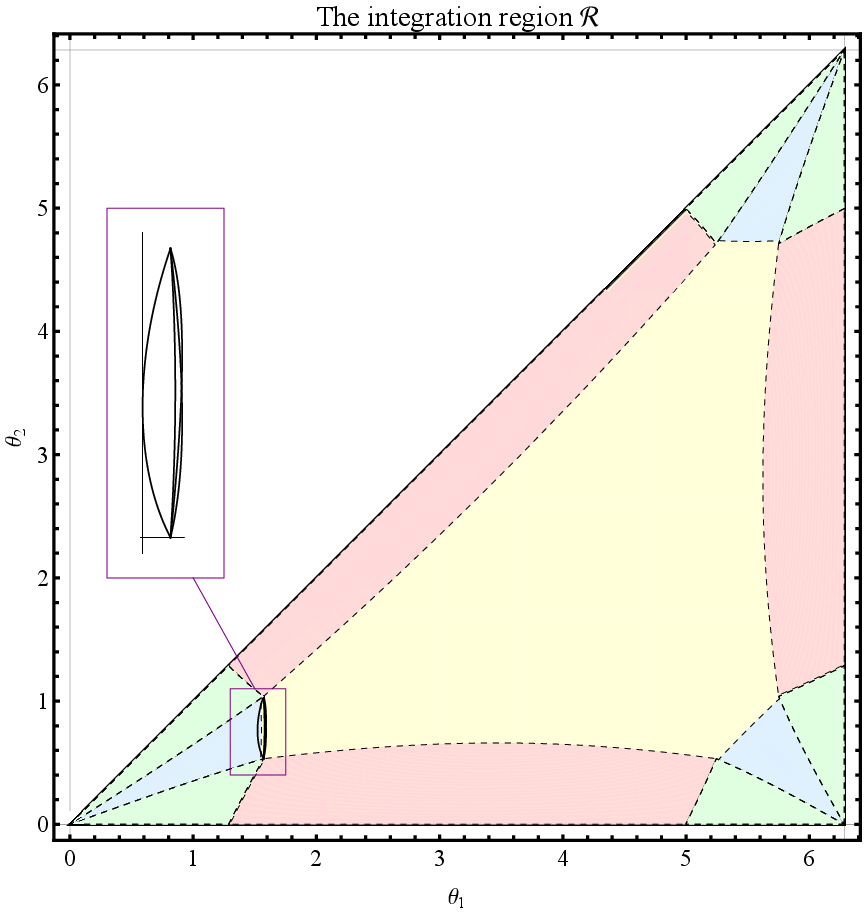}
	\hfill
	\raisebox{12mm}{
		\includegraphics[scale=.5]{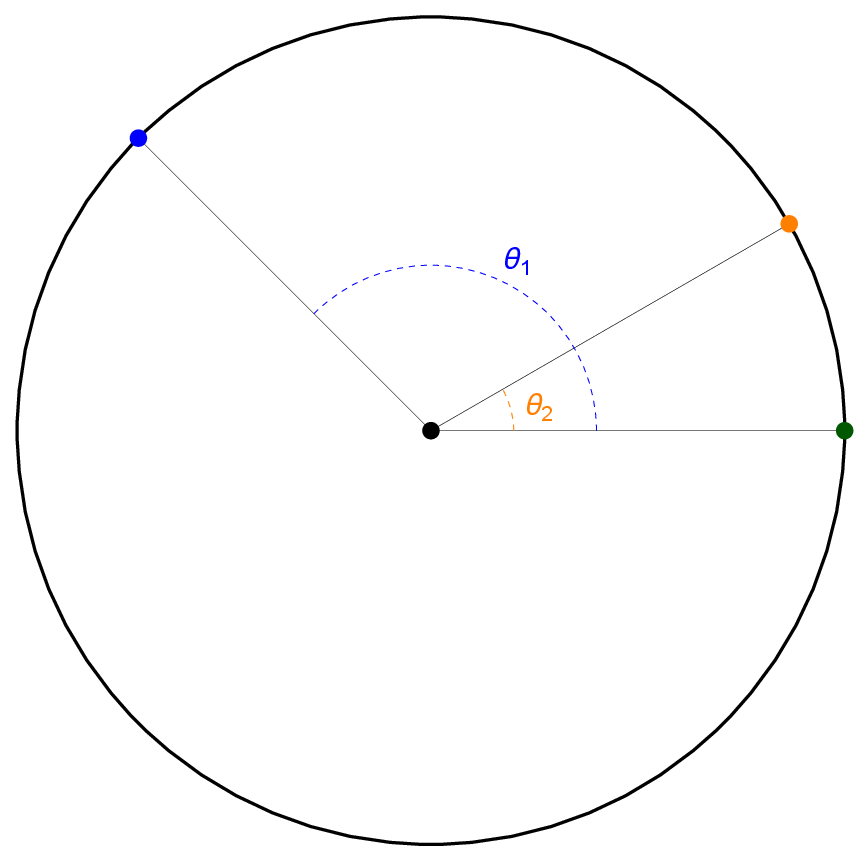}
	}
	\caption{The integration region $\mathcal{R}$ together with its Feynman decomposition (\emph{left}) and our conventions for the parametrization of the disk conformal frame (\emph{right}). The~\textcolor{blue}{blue} and~\textcolor{yellow}{yellow} regions in $\mathcal{R}$ should be understood in the sense of ``generalized sections''~\cite{deLacroix:2017lif} and they lead to weighted averages of the associated moduli integrals over them. An example for a distinct decomposition resulting from generalized sections for a component of the~\textcolor{blue}{blue} region is highlighted in the~\textcolor{purple}{purple} box.
		\label{fig:decomposition}}
\end{figure}

Therefore we consider $\Gamma_3^{\mathcal{O}}$~\eqref{eq:5.18c} next. This order gets contributions from the obstruction and non-obstruction part of $\Psi_3$~\eqref{eq:3.17}, so let us decompose
\begin{align}
	\Gamma_3^\mathcal{O} = \Gamma_3^{\mathcal{O},\text{obs}}  + \Gamma_3^{\mathcal{O},\text{non-obs}} \, ,
\end{align}
where $\Gamma_3^{\mathcal{O},\text{obs}}$ is the first Ellwood vertex $E_1^\mathcal{O}$ acting on the obstruction term $\psi_3$~\eqref{eq:3.30},
\begin{align}
	 \Gamma_3^{\mathcal{O},\text{obs}}  
	 =
	 - K \int\limits_{-\infty}^{\infty} {dE \over 2 \pi} f(E) \epsilon_{\mu \nu}
	 \left\langle c \overline{c} \partial X^\mu \overline{\partial} X^\nu e^{i k \cdot X} (\xi = 0)
	 c \left(X^0\right)^3 \partial X^1 (w_1=0) \right\rangle_\UHP  \, ,
\end{align}
while $\Gamma_3^{\mathcal{O},\text{non-obs}}$ is the non-obstruction contributions that are collected into the following moduli integral using the perturbative electric field solution:
\begin{align} \label{eq:5.26}
	\Gamma_3^{\mathcal{O},\text{non-obs}} = -{1 \over 3} 
	\int\limits_{-\infty}^\infty &{d E \over 2 \pi} f(E) \epsilon_{\mu \nu}
	\int\limits_{\mathcal{R}}  \left( { -d u_1\over \rho_1(u_i)} \right)
	\left( { -d u_2\over \rho_2(u_i)} \right) 
	\\ \nonumber
	&\hspace{-0.25in}\left\langle
	c \overline{c} \partial X^\mu \overline{\partial} X^{\nu} e^{i k \cdot X} (\xi = 0)
	X^0 \partial X^1 (w_1 = 0) 
	X^0 \partial X^1 (w_2 = 0) 
	c X^0 \partial X^1 (w_3 = 0) 
	\right\rangle_\UHP
	\, .
\end{align}
Here $\rho_j(u_i)$ are the mapping radii for the third Ellwood vertex $\mathcal{E}_3$. The relevant integration region  $\mathcal{R}$ , i.e., the moduli space of disks with one bulk and three boundary punctures, is illustrated in figure~\ref{fig:decomposition}. We parametrize $\mathcal{R}$ using the angles of the punctures on the disk conformal frame, whose conventions are also shown in the figure. 

We stress that the integral~\eqref{eq:5.26} potentially diverges towards the boundary of the integration region $\partial \mathcal{R}$ due to the collision of the insertions. These divergences should be handled according to the SFT prescription~\eqref{eq:3.27a} once again. Running this procedure here requires understanding the Feynman decomposition of $\mathcal{R}$ with a choice of Ellwood vertices $\mathcal{E}_1$ and $\mathcal{E}_2$ compatible with $SL(2,\mathbb{R})_\lambda$ vertices---the so-called $SL(2,\mathbb{R})_\lambda$ Ellwood vertices. Their construction, and our conventions for them, are explained in appendix~\ref{app:B}. For our purposes it is sufficient to fix $\lambda =3$ and that is what we do henceforth. The Feynman decomposition resulting from our choices for the Ellwood vertices when $\lambda = 3$ is shown in figure~\ref{fig:decomposition}. 

We consider $\Gamma_3^{\mathcal{O},\text{obs}}$ and  $\Gamma_3^{\mathcal{O},\text{non-obs}}  $ separately. We can evaluate the obstruction contribution similar to~\eqref{eq:5.20} and find
\begin{align} \label{eq:5.29}
	\Gamma_3^{\mathcal{O},\text{obs}} 
	 &=- i K V \left(-i {\partial \over \partial E'} \right)^3 
	 \left[
	 \int\limits_{-\infty}^{\infty} {d E } f(E) 2^{-E^2} E' \delta (E+E')
	 \right]_{E' = 0}
	 \nonumber \\
	 & = 3KV \left( - f(0) \log 4 + f''(0) \right) \, ,
\end{align}
where $\rho_1 = 1$ with our choice for the first Ellwood vertex $\mathcal{E}_1$~\eqref{eq:B.1}. Observe that this term contains $f''(0)$, which should cancel against a similar term in $\Gamma_3^{\mathcal{O}, \text{non-obs}}$ given that the boundary state side of~\eqref{eq:5.16} do not contain derivatives of $f$. We shall argue for this cancellation momentarily. 

On the other hand, the non-obstruction part of $\Gamma_3^\mathcal{O}$ is equal to
\begin{align} \label{eq:5.30}
	\Gamma_3^{\mathcal{O},\text{non-obs}} &= -{1 \over 3} 
	\prod_{j=1}^3 \bigg( - i { \partial \over \partial E_j}\bigg)
	\Bigg[ \int\limits_{-\infty}^\infty {d E \over 2 \pi} f(E) \epsilon_{\mu \nu} 
	\int\limits_{\mathcal{R}}  { d u_1 } {d u_2 }
	\left( \prod_{j=1}^3 \rho_j(u_1,u_2)^{-E_j^2} \right)
	\\ \nonumber
	&\hspace{-0.7in} \times\bigg\langle
	c \overline{c} \partial X^\mu \overline{\partial} X^{\nu} e^{i k \cdot X} (z = i)
	e^{i E_1 X^0} \partial X^1 (z_1 = u_1) 
	e^{i E_2 X^0} \partial X^1 (z_2 = u_2) 
	c e^{i E_3 X^0} \partial X^1 (z_3 = 0) 
	\bigg \rangle_\UHP \Bigg]_{E_j = 0}
	\nonumber \\
	&= -{V \over 3 } 
	\prod_{j=1}^3 \bigg( - i { \partial \over \partial E_j}\bigg)
	\left[\int\limits_{-\infty}^\infty {d E } f(E) 
	\int\limits_{\mathcal{R}}  { d u_1 } {d u_2 } \, 
	G(E,E_j;u_i) 
	\, \delta\left(2 E +2 \sum_{j=1}^3 E_j\right)
	\right]_{E_j = 0}
	\nonumber \, ,
\end{align}
where the function $G$ is defined by
\begin{align} \label{eq:5.31}
	&G(E,E_j;u_i) \equiv (2i) \times
	2\mathrm{Re}\Bigg[
	\bigg( {-1 \over 4(u_1 -u_2)^2 } + {1 \over 4(i + u_1 )^2 u_2^2}  + {1 \over 4 u_1^2 (i + u_2)^2 }  \bigg)
	\bigg( {E_1 \over 1 -i u_1} +  {E_2 \over 1 -i u_2} + E_3\bigg)
	\Bigg]
	\nonumber \\
	&\times \left( \prod_{j=1}^3 \rho_j(u_1, u_2)^{-E_j^2} \right)
	2^{-E^2}
	(1+u_1^2)^{- E E_1} (1+u_2^2)^{- E E_2}
	(u_1 - u_2)^{-2 E_1 E_2}
	u_1^{-2 E_1 E_3} u_2^{-2 E_2 E_3} \, ,
\end{align}
after evaluating the correlator on the UHP using the doubling trick. The origin of terms above is relatively clear.

It is rather involved to report the expressions explicitly after this point, so we just explain the resulting structure in words. More details and explicit expressions can be found in the \texttt{Mathematica} notebook in the Supplementary material. After differentiating~\eqref{eq:5.30} with respect to $E_j$s and setting them to zero, we are left with terms proportional to 
\begin{align} \label{eq:5.32}
	\delta(E) \, ,\quad
	E \delta' (E) \, , \quad
	 \delta''(E) \, .
\end{align}
Note that the derivatives of the function $G$ always produces an extra factor of $E$ for $\delta'(E)$ terms. Evaluating the Fourier integral then yields a pair of moduli integrals proportional to $f(0)$ and $f''(0)$. These integrals do not contain the mapping radii $\rho_j(u_i)$ of the third Ellwood vertex $\mathcal{E}_3$, which is consistent with the form of the original expression, inspect~\eqref{eq:5.26}. Thus, these moduli integrals can be evaluated without specifying the vertex $\mathcal{E}_3$ as long as the Feynman decomposition of $\mathcal{R}$ is implemented to treat their divergences consistently---see appendix~\ref{app:B} for more details. Doing this numerically, we report
\begin{align}
	\Gamma_3^{\mathcal{O}, \text{non-obs}} &\approx
	0.151568 V f(0) -3.01782 V \left(- f(0) \log 4 + f''(0) \right)
	\, .
\end{align}
In the light of the numerical value for the obstruction coefficient $K$~\eqref{eq:3.30}, the term proportional to $-f(0)\log4 + f''(0)$ indeed cancels against~\eqref{eq:5.29} in $\Gamma_3^{\mathcal{O}}$ within an accuracy of $10^{-6}$. Therefore,
\begin{align}
	\Gamma_3^{\mathcal{O}} = \gamma V f(0) \, , 
	\quad \quad
	\gamma \approx 0.151568 \, ,
\end{align}
and from the relation~\eqref{eq:5.7}
\begin{align}
	-8 \pi  \varepsilon_\DBI - 16 \pi^3 \left(\varepsilon_\DBI\right)^3 +  \mathcal{O} \left(\left(\varepsilon_\DBI\right)^5\right) =
	- 4 \pi i \left( -\varepsilon + \gamma \varepsilon^3 +  \mathcal{O}\left(\varepsilon^5\right) \right) \, ,
\end{align}
we obtain
\begin{align}
	\varepsilon_\DBI = - {i \over 2} \varepsilon 
	+ c_3 \varepsilon^3 +
	\mathcal{O}\left(\varepsilon^5\right)  \, , 
	\quad \quad
	c_3 = - {i \over 4} (\pi^2 - 2 \gamma)\approx -2.39162 i \, ,
\end{align}
for the field redefinition between the DBI and the $SL(2,\mathbb{R})_3$ open SFT electric fields. Finally, substituting it to the DBI energy~\eqref{eq:10.11}, our final result is
\begin{align}
	E_\DBI = {V \over g^2} \bigg[ - {1 \over 4} \varepsilon^2 
	+ \alpha_{\DBI,4} \varepsilon^4 +
	\mathcal{O}\left(\varepsilon^6\right) \bigg] \, , 
	\quad \quad
	\alpha_{\DBI,4} =  {\gamma \over 2} - {\pi^2 \over 16} 
	\approx -0.5410662 \, .
\end{align}
This agrees with the energy $E(\lambda = 3)$ computed using the SFT symplectic form in~\eqref{eq:4.49} with a stellar degree of accuracy:
\begin{align}
	\text{error} = \left|1 - { \alpha_4 \over \alpha_{\DBI,4}} \right| \approx 0.0001 \%  \, .
\end{align}

\subsection*{Acknowledgments}

TE and AHF would like to thank David Gross for hospitality at the KITP while carrying out part of this work. TE thanks Mat\v{e}j Kudrna and Jakub Vo\v{s}mera for discussions of the electric field boundary state. AHF thanks Manki Kim and Mukund Rangamani for conversations. The work of VB and TE was supported by the European Structural and Investment Funds and the Czech Ministry of Education, Youth and Sports (project No. FORTE—CZ.02.01.01/00/22\_008/0004632). The work of AHF is supported by the U.S. Department of Energy, Office of Science, Office of High Energy Physics of U.S. Department of Energy under grant Contract Number DE-SC0009999, and the funds from the University of California. This research was supported in part by grant NSF PHY-2309135 and the Gordon and Betty Moore Foundation Grant No. 2919.02 to the Kavli Institute for Theoretical Physics~(KITP).

\appendix
\section{$SL(2,\mathbb{R})$ string vertices} \label{sec:A}

In this appendix, we construct the local coordinates for the (cubic and quartic) $SL(2,\mathbb{R})_\lambda$ vertices and (quartic) Feynman diagrams, as well as explain the Feynman decomposition of the moduli space of disks with four cyclically-ordered boundary punctures. More details on these local coordinates (and string vertices in general) can be found in~\cite{Sen:2024nfd,Erbin:2023rsq,Sen:2019jpm}. Refer to the~\texttt{Mathematica} notebook in the Supplementary material for the explicit expressions of the quantities that have not been reported here.

The cubic $SL(2,\mathbb{R})_\lambda$ vertex for open SFT is defined in terms of the following local coordinates around the boundary punctures $z=\infty, 1, 0$,
\begin{align} \label{eq:A.1}
	w_1(z) = \lambda \, {1 \over -2 z + 1} 
	\, , \quad
	w_2(z) = \lambda \, {z-1 \over z + 1}
	\, , \quad
	w_3(z) = \lambda \, {z \over -z + 2} 
	\, ,
\end{align}
respectively for $0 < |w_i |< 1$ and $\mathrm{Im} \, w_i \geq 0$. Here $z$ is the coordinate on the uniformizing UHP. The inverse maps are easy to find,
\begin{align} \label{eq:A.2}
	z (w_1) = {w_1 - \lambda \over 2 w_1}
	\, , \quad
	z(w_2) = {- w_2 - \lambda \over w_2 - \lambda}
	\, , \quad
	z(w_3) = {2 w_3 \over w_3 + \lambda} \, ,
\end{align}
and from which we read the mapping radii~\eqref{eq:3.21} associated with punctures to be $2/\lambda$. The parameter $\lambda$ is~\emph{the stub parameter} and we assume $\lambda \geq 3$ to ensure the local coordinates~\eqref{eq:A.1} are non-overlapping. Two instances for the resulting geometry on the uniformizing UHP are shown in figure~\ref{fig:SL(2,R)}. Note that~\eqref{eq:A.1} is invariant under the global conformal transformation
\begin{align} \label{eq:A.3}
	\tau (z) = {z - 1 \over z} \, ,
\end{align}
implementing the cyclic permutation $\infty \to 1 \to 0 \to \infty$ of punctures. That is, $\tau(w_i(z)) = w_{i+1}(z)$ upon taking the subscript addition to be modulo 3.
\begin{figure}[t]
	\centering
	\includegraphics[scale=.59]{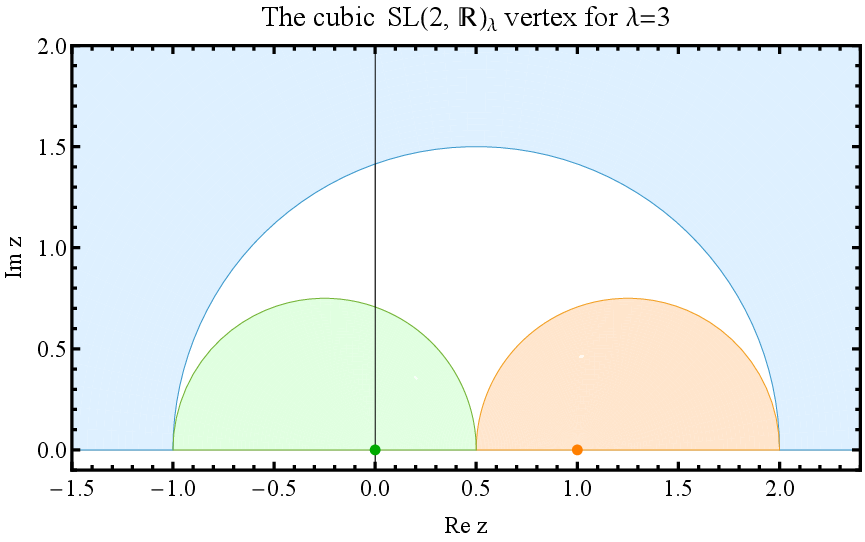}
	\includegraphics[scale=.59]{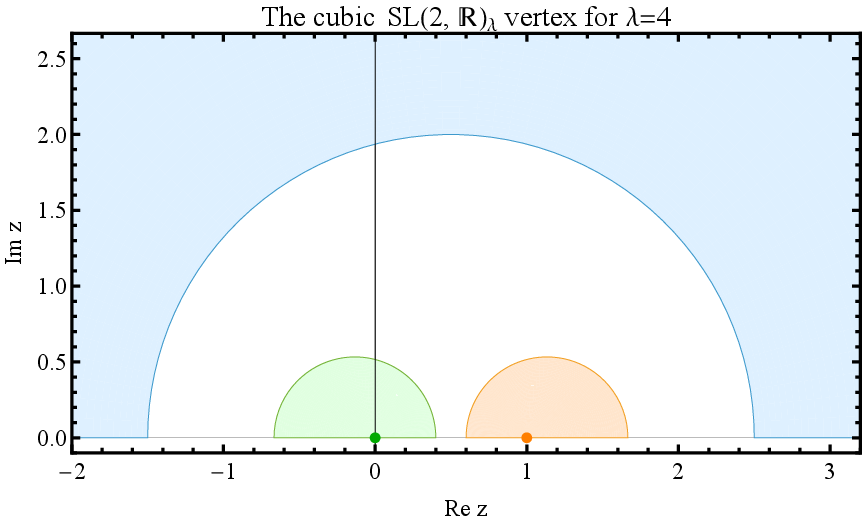}
	\caption{\label{fig:SL(2,R)}The local coordinates for the cubic $SL(2,\mathbb{R})_\lambda$ vertex when~$\lambda=3$~(\emph{left}) and $\lambda =4$~(\emph{right}). Increasing the stub parameter $\lambda$ retracts the local coordinates towards the punctures.}
\end{figure} 

We can construct the local coordinates for the quartic $SL(2,\mathbb{R})_\lambda$ Feynman diagrams using \eqref{eq:A.1}. Let us briefly summarize how this proceeds. Imagine two thrice-punctured UHPs with coordinates $z,z'$ whose punctures are placed at $z,z' = \infty,1,0$. Using the~\emph{sewing fixture}, we can identify the local coordinates $w_3,w_3'$ around the punctures $z,z'=0$ via
\begin{align} \label{eq:A.4}
	w_3(z) w_3(z') = - q \, ,
	\quad
	0 < q < 1 \, ,
\end{align}
and obtain a combined UHP with four boundary punctures. Here $q$ is called~\emph{the sewing parameter}. Observe that the left hand side of~\eqref{eq:A.4} is proportional to $\lambda^2$, which implies that any expression involving $q$ and $\lambda$ can be expressed as a single-variable function of $q/\lambda^2$. In particular, this means that we can work with the restricted sewing parameter $0 < q <9/\lambda^2$ and scale each local coordinate by $\lambda/3$ instead of working with $q$ and $\lambda$ separately. However we do not do this unless stated otherwise.

Given the maps~\eqref{eq:A.2}, we can solve for $z' = z'(z)$ in~\eqref{eq:A.4},
\begin{align} \label{eq:A.5}
	z'(z) = {2 q \, (-2 + z) \over q \, (-2 + z) + z \lambda^2} \, ,
\end{align}
and use it as the global coordinates on the combined UHP. The four punctures are placed at
\begin{align} \label{eq:A.6}
	z' = \infty, 1, {2 q \over q + \lambda^2} , {2 q \over q^2 - \lambda^2} \, ,
\end{align}
where the positions of the last two punctures are obtained from mapping the non-identified punctures at $z=\infty, 1$ to the $z'$-frame using~\eqref{eq:A.5}. It is beneficial to place the punctures~\eqref{eq:A.6} to their canonical positions of $\{\infty,1,u,0 \}$ and subsequently use the cross ratio (i.e., the position of the third puncture) $0 < u < 1$ to parameterize the moduli space. We do this by composing~\eqref{eq:A.5} with the affine map
\begin{align} \label{eq:A.7}
	\widetilde{z}(z') = {(-q+\lambda^2) \, z' + 2 q \over q + \lambda^2} \, ,
\end{align}
that fixes the punctures at $z'=\infty,1$, while mapping the fourth puncture in~\eqref{eq:A.6} to $z' = 0$. The resulting cross ratio is
\begin{align} \label{eq:A.9}
	u = {4 \, q \, \lambda^2 \over (q+\lambda^2)^2}
	\quad \implies \quad
	q = \lambda^2 \, {2 -2 \sqrt{1-u} - u \over u}  \, .
\end{align}
Given the maps~\eqref{eq:A.5} and~\eqref{eq:A.7}, the local coordinates $w_i^-(z)$ around the punctures $z=\infty,1,u,0$ can be straightforwardly found. The resulting geometry is shown in figure~\ref{fig:SL(2,R)F}. The minus superscript on $w_i^-(z)$ refers that the surfaces generated with this procedure produce the ``lower-end'' of the moduli space, i.e.,~\eqref{eq:A.4} leads to four-punctured UHPs whose cross-ratios are closer to $u = 0$. The opposite situation when the cross ratio is closer to $u =1$ can be characterized in an analogous manner and leads to the local coordinates $w_i^+(z)$ , which are also shown in figure~\ref{fig:SL(2,R)}. The relation between the cross ratio $u$ and the sewing parameter $q$ for this situation is
\begin{align} \label{eq:A.9a}
	u = \left( { q-\lambda^2 \over  q+\lambda^2} \right)^2 
	\quad \implies \quad
	q = \lambda^2 \, {1 - \sqrt{u} \over 1 + \sqrt{u}}  \, .
\end{align}
\begin{figure}[t]
	\centering
	\includegraphics[scale=.59]{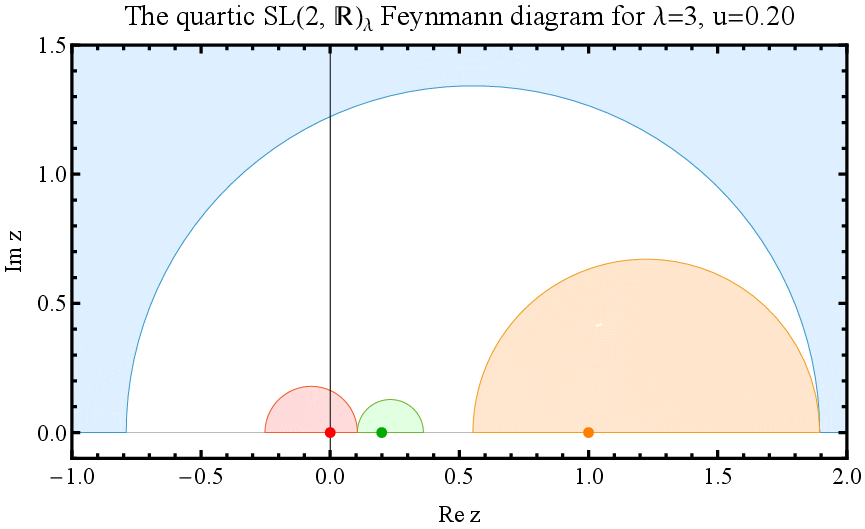}
	\includegraphics[scale=.59]{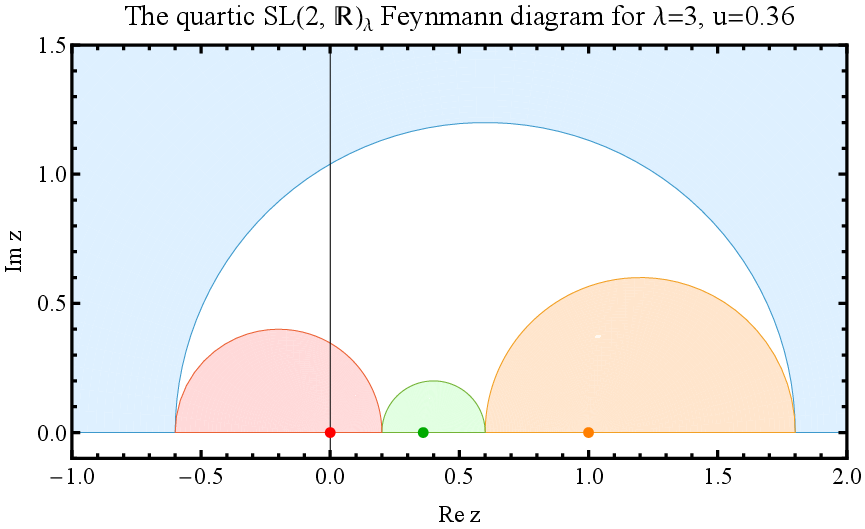}
	\includegraphics[scale=.59]{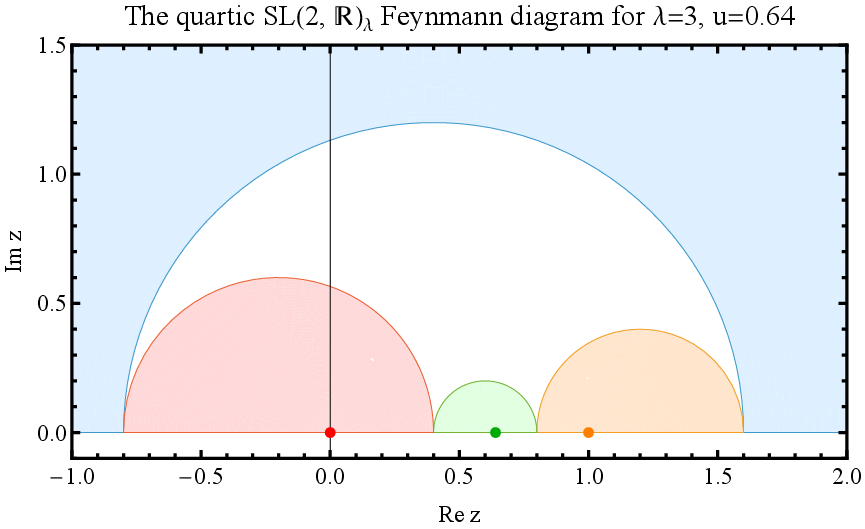}
	\includegraphics[scale=.59]{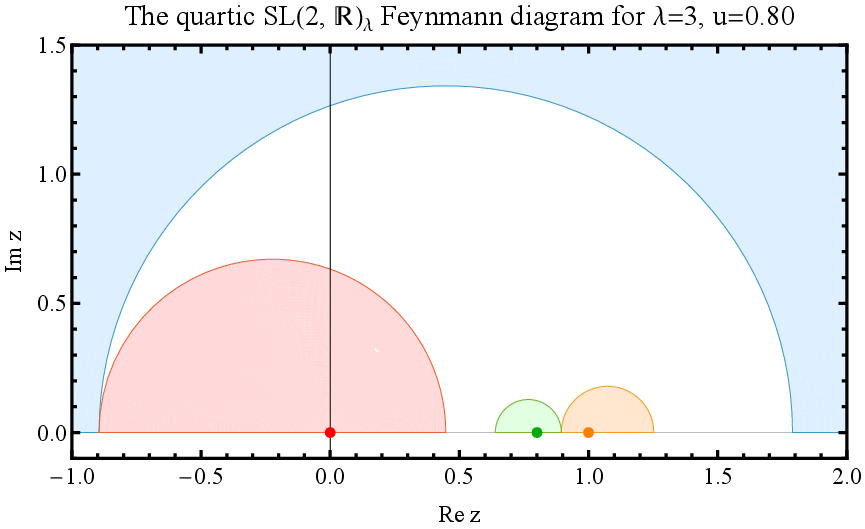}
	\caption{\label{fig:SL(2,R)F}The local coordinates for the quartic  $SL(2,\mathbb{R})_\lambda$ Feynman diagrams when $\lambda = 3$. We set $q=1$ for the figures on the top-right and bottom-left and $q = 1/2$ for the figures on the top-left and bottom-right.}
\end{figure} 

Observe that the cross ratios in~\eqref{eq:A.6} and~\eqref{eq:A.9} add up to $1$ when an equal sewing parameter $q$ is used. This is not a coincidence but a reflection of the cyclic symmetry of the Feynman diagrams. In order to see this we first note that it is~\emph{not} generically possible to cyclically-permute four punctures. The closest we can get is the global conformal map
\begin{align} \label{eq:A.10}
	\tau(z; u) = {z - u  \over z} \, ,
\end{align}
that takes $u \to 0 \to \infty  \to 1$. This forcefully maps the puncture at $z=1$ to $z = 1-u$, thus changing the cross ratio from $u$ to $1-u$. However, the sewing parameter $q$ remains invariant under this mapping, which provides an explanation for the observation above. This feature also necessitates the local coordinates $w_i^\pm(z)$ are permuted in the following manner: $\tau (z(w_i^-); u) = z(w_{i+1}^+)$. Here the subscript addition is modulo 4. Indeed, we explicitly checked that this is the case.

Replacing $u \to 1-u$ in~\eqref{eq:A.10} produces the map $\tau(z;1-u)$ that permutes $(1-u) \to 0 \to \infty  \to 1$ and takes $1$ to $u$. We can compose it with~\eqref{eq:A.10} to implement the pairwise exchange $\infty \leftrightarrow u$ and $1 \leftrightarrow 0$ of punctures:
\begin{align} \label{eq:A.11}
	\tau(\tau(z;u);1-u) = {u\, (z-1) \over z -u} \, .
\end{align}
Analogous to above, the local coordinates $w_i^\pm(z)$ get exchanged in an expected fashion under this map, i.e., $\tau(\tau(z(w_i^\pm);u);1-u) = z(w_{i+2}^\pm)$. We emphasize that checking the permutations under the maps~\eqref{eq:A.10} and~\eqref{eq:A.11} in this manner is sufficient to establish the cyclicity of the local coordinates since permuting again does not introduce any new information. These features regarding cyclicity are going to be important shortly.

Next, we observe that the relations~\eqref{eq:A.9} and~\eqref{eq:A.9a}, coupled with the range $0<q<1$, show that the surfaces obtained by sewing do not cover the entire moduli space of disks with four cyclically-ordered boundary punctures. Instead, they leave a nontrivial ``gap'', which is called~\emph{vertex region}:
\begin{align} \label{eq:A.13}
	0 <  u_-(\lambda) \equiv \bigg( {2 \lambda \over 1+\lambda^2} \bigg)^2
	\leq u \leq 
	\bigg({1-\lambda^2 \over 1+ \lambda^2}\bigg)^2 \equiv u_+(\lambda) < 1 \, .
\end{align}
We need to introduce suitable local coordinates $w_i^V(z;u)$ and ``fill'' this gap in open SFT. Specifying $w_i^V(z;u)$ in~\eqref{eq:A.13} is the question of defining classical quartic open string vertices. Fortunately, these local coordinates are not totally arbitrary. They should obey the following constraints:
\begin{enumerate}
	\item String vertices must satisfy geometric master equation so that the corresponding SFT is gauge invariant~\cite{Sen:2024nfd}. This constrains the local coordinates for the surfaces lying at the boundary of the gap~\eqref{eq:A.13} to be equal to those provided by the quartic Feynman diagrams with $q=1$:
	\begin{align} \label{eq:A.14}
		w_i^V(z;u = u_\pm(\lambda)) = w_i^\pm (z; q=1) \, .
	\end{align}
	The local coordinates for this particular situation is already shown in figure~\ref{fig:SL(2,R)F}.
	
	\item The local coordinates should be cyclically-invariant under the permutation of punctures in order to have a manifestly covariant action~\eqref{eq:2.1}. For us, this means that they must map to each other under~\eqref{eq:A.10} and~\eqref{eq:A.11} as
	\begin{subequations} \label{eq:A.15}
	\begin{align}
		&\tau\left(w_i^V(z; u); u\right) = w^V_{i+1}(z; 1-u)
		\, , \\
		&\tau\left(\tau\left(w_i^V(z; u),u\right),1-u\right) = w^V_{i+2}(z; u) \, ,
	\end{align}
		\end{subequations}
	where the subscript summation is modulo 4. We stress that the dependence on the moduli is non-trivial in these permutations.
	
	\item The local coordinates $w_i^V(z,u)$ should not overlap with each other in order to have a well-defined external states in the theory. This condition can be satisfied trivially by taking the stubs sufficiently large if necessary.
	
	\item The local coordinates $w_i^V(z;u) $ are given in terms of $SL(2,\mathbb{R})$ maps. This condition is purely for the computational convenience.
\end{enumerate}

It is possible to construct local coordinates based on these constraints. For now, we restrict our attention to $SL(2,\mathbb{R})_3$ vertices and comment on the generic stub parameters only at the end. The vertex region for $\lambda = 3$ takes the form
\begin{align} \label{eq:A.16}
	u_-(\lambda = 3) = {9 \over 25} \leq u \leq {16 \over 25} = u_+(\lambda = 3) \, ,
\end{align}
which we call~\emph{the interpolating region} to distinguish it from~\eqref{eq:A.13}. The question is how to specify the local coordinates $w_i^V(z;u)$ in~\eqref{eq:A.16} with the constraint listed above. To begin answering it, we first study the circular geometry provided by the local coordinates when $u = u_\pm(3)$ (or equivalently $q=1$) as shown in figure~\eqref{fig:SL(2,R)F}. These circles intersect each other at the points
\begin{align}
	\xi_i^- = - {3 \over 5} \, , {1 \over 5} \, , {3 \over 5} \, , {9 \over 5} \, ,
	\quad \quad
	\xi_i^+  = - {4 \over 5} \, , {2 \over 5} \, , {4 \over 5} \, ,  {8 \over 5} \, ,
\end{align}
when $u = u_\mp(3)$. Based on this geometry, it is natural to consider ``the interpolating circles'' corresponding to the punctures $z=\infty,1,u(t),0$ on the uniformizing UHP, with
\begin{align}
	u(t) = t u_+(3) + (1-t) u_-(3) = {9 \over 25} + {7 \over 25} t \, ,
	\quad \quad
	0 \leq t \leq 1 \, ,
\end{align}
such that the circles keep intersecting each other at the boundary points
\begin{align} \label{eq:A.18}
	\xi_i(t) = t \xi_i^+ + (1-t) \xi_i^-  \, ,
\end{align}
as we vary $u$ between $u_\pm(3)$ in order construct the quartic vertex. The resulting geometry when $u=1/2$ is shown in figure~\ref{fig:SL(2,R)V}. It is apparent that the interpolating circles are unique and reduce to the initial/final configuration of circles at $u = u_\mp(3)$ respectively. Furthermore, this geometry allows us to uniquely specify $SL(2,\mathbb{R})$ maps $z(w_i^V)$ that take the upper unit half-disks to the interior of the interpolating circles and the origins to the punctures. We declare that these are the ``seed'' local coordinates for $SL(2,\mathbb{R})_3$ quartic vertex. They satisfy the condition~\eqref{eq:A.14} and are non-overlapping by construction. 
\begin{figure}[t]
	\centering
	\includegraphics[scale=.59]{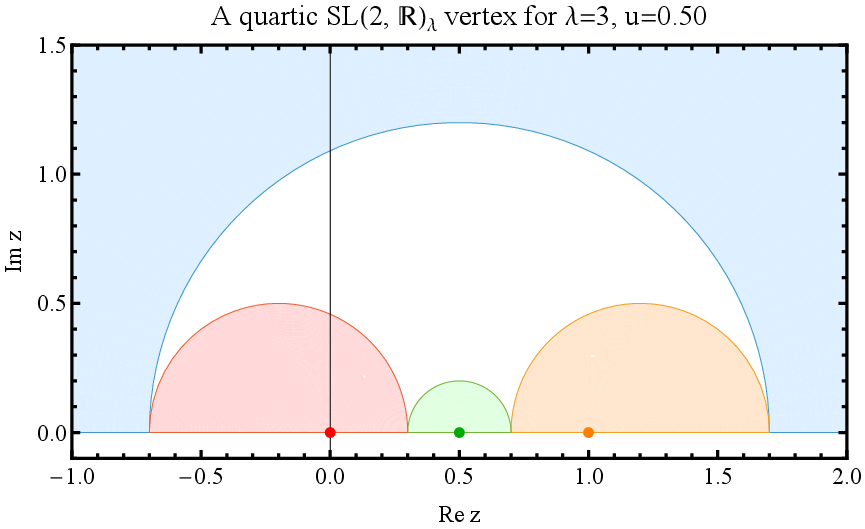}
	\includegraphics[scale=.59]{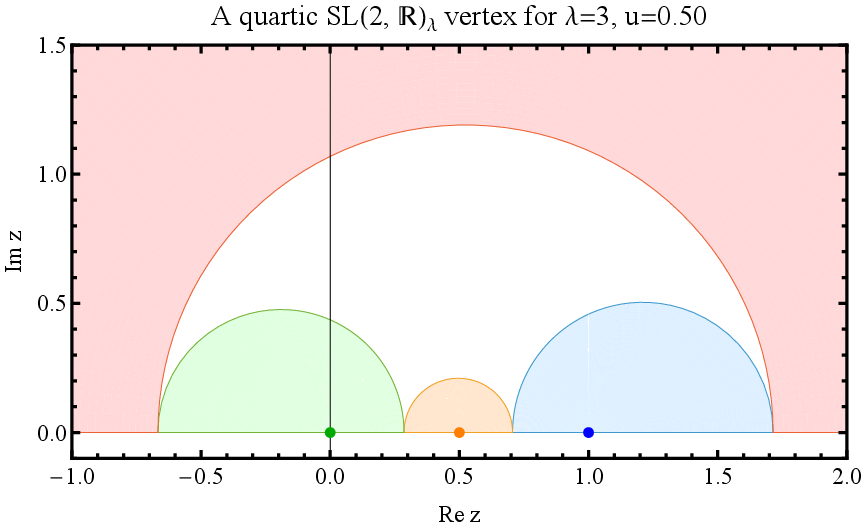}
	\includegraphics[scale=.59]{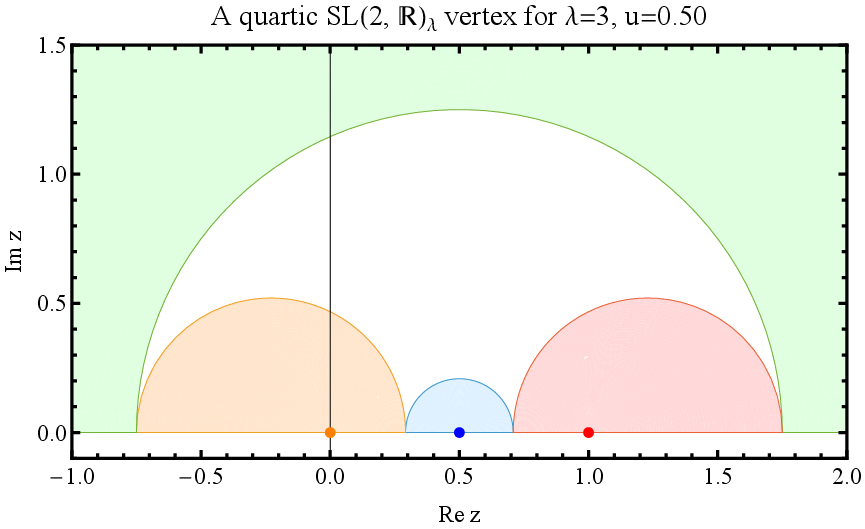}
	\includegraphics[scale=.59]{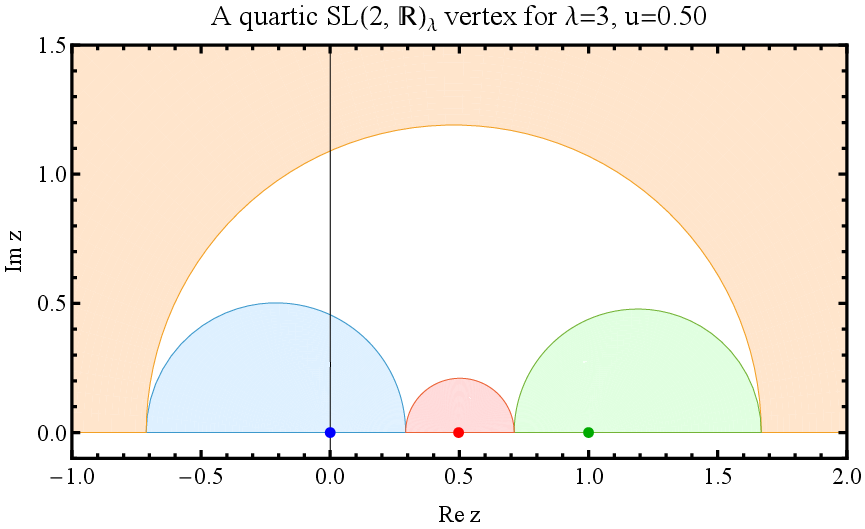}
	\caption{\label{fig:SL(2,R)V}The local coordinates for the cyclic copies of the quartic $SL(2,\mathbb{R})_3$ vertex when $u = 1/2$. The top-left figure shows the seed local coordinates. The coloring is consistent with the action of the cyclic permutation $\tau$~\eqref{eq:A.10} between different copies.}
\end{figure} 
Crucially, however, the seed local coordinates are~\emph{not} cyclically-invariant by themselves in the sense of~\eqref{eq:A.15}. Nevertheless, we can trivially satisfy the cyclicity constraint by averaging the moduli integrals over the copies of the seed local coordinates under the transformation~\eqref{eq:A.10}, which is the construction known as~\emph{generalized sections}~\cite{deLacroix:2017lif}. This procedure generates four inequivalent copies in total and they still satisfy the condition~\eqref{eq:A.14} thanks to the argument below~\eqref{eq:A.10}. Examples for the resulting local coordinates in different ``cyclic'' copies are shown in figure~\ref{fig:SL(2,R)V}.

We can extend the construction above to $\lambda > 3$ using the standard stubbing procedure. All we need to do is to include the part of the Feynman region of $SL(2,\mathbb{R})_3$ vertices with $ 9/\lambda^2 \leq q < 1$ (the so-called~\emph{stub region}) to the interpolating region~\eqref{eq:A.16} and scale each local coordinate $w_i$ by $\lambda / 3$ in the light of our remarks below~\eqref{eq:A.4}. The resulting vertex region is~\eqref{eq:A.13} and all the constraints remains satisfied. Introducing the quartic $SL(2,\mathbb{R})_\lambda$ vertices this way allows us to consider the simple field redefinition~\eqref{eq:4.46} and check the consistency of our evaluations. 

\begin{figure}[t]
	\centering
	\includegraphics[scale=.59]{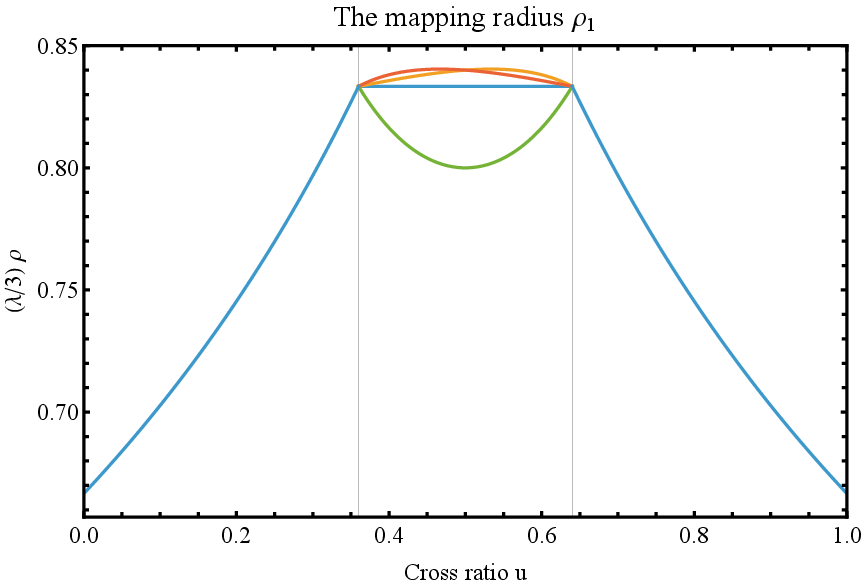}
	\includegraphics[scale=.59]{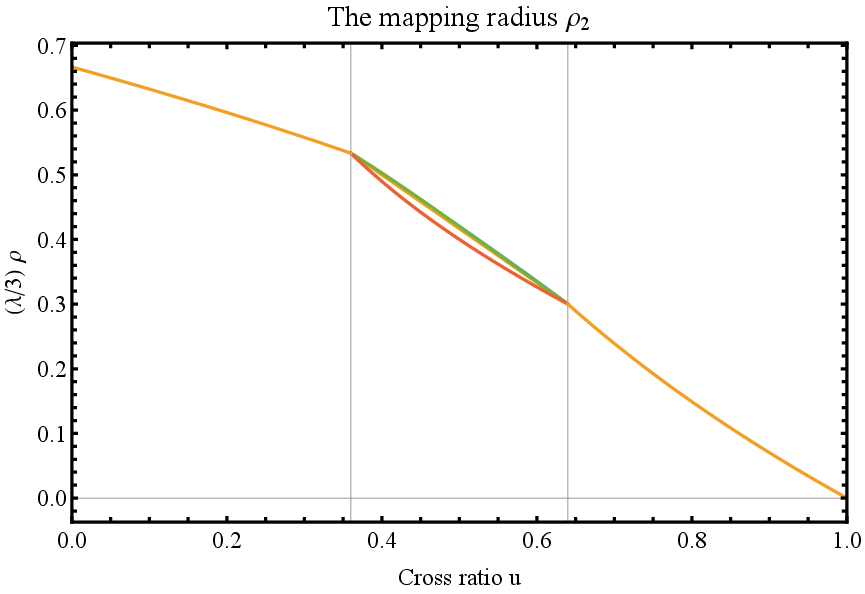}
	\includegraphics[scale=.59]{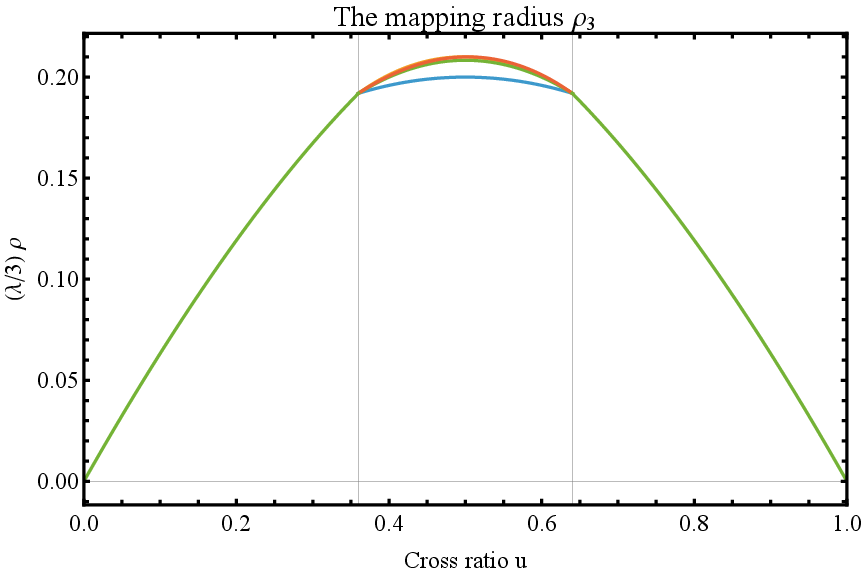}
	\includegraphics[scale=.59]{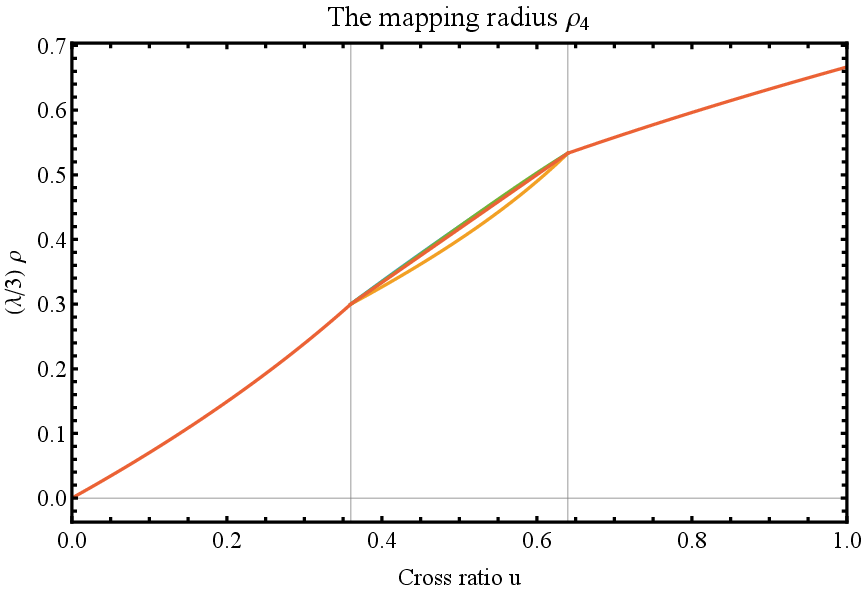}
	\caption{\label{fig:rho}The mapping radii $\rho_i = \rho_i(u)$ over the moduli space of disks with four cyclically-ordered boundary punctures. The coloring is consistent with previous figures. Note the imprint of using generalized sections for the interpolating region~\eqref{eq:A.16} in the figure.}
\end{figure}

Finally, all there is left to do is to read off the mapping radii $\rho_i(u)$~\eqref{eq:3.21} of the local coordinate maps resulting from the construction above given that they are the only relevant part of the local coordinate data in our application. It is somewhat impractical to list all of their expression here. Instead, we just show the behavior of $\rho_i = \rho_i(u)$ over the moduli space in figure~\ref{fig:rho}.

\section{$SL(2, \mathbb{R})$ Ellwood vertices} \label{app:B}

In this appendix we construct the local coordinate maps over the moduli spaces of disks with one bulk and $n$ cyclically-ordered boundary punctures relevant for the evaluation of the homotopy Ellwood invariant~\eqref{eq:5.18} for the open SFT using $SL(2,\mathbb{R})_3$ vertices. More details and explicit expressions can be found in the~\texttt{Mathematica} notebook in the Supplementary material.

\begin{figure}[t]
 	\centering
 	\includegraphics[scale=.59]{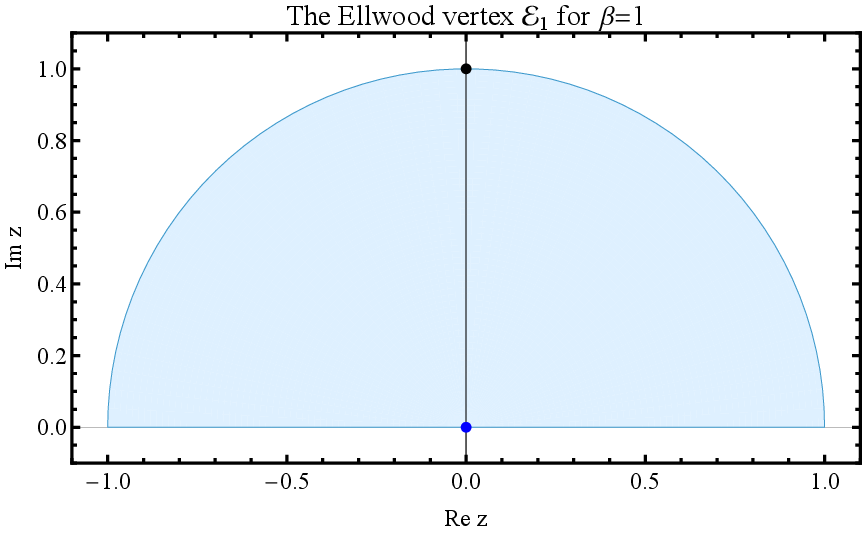}
 	\includegraphics[scale=.59]{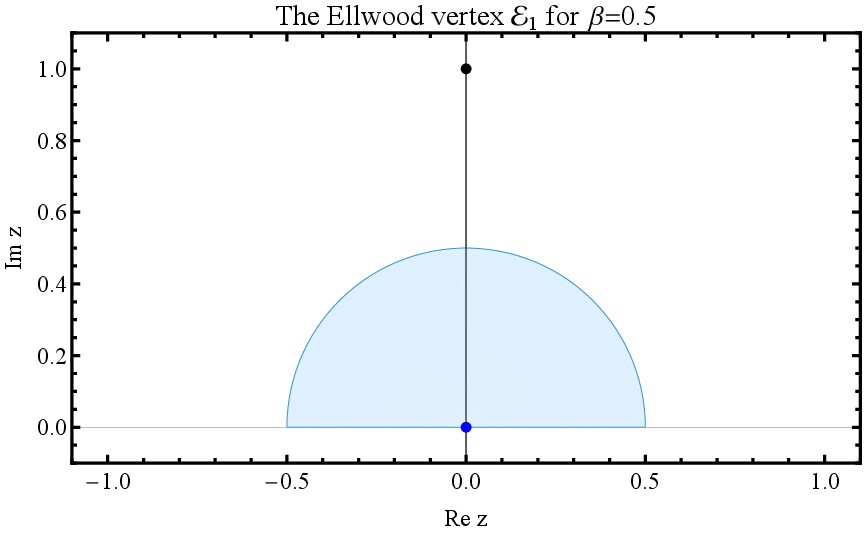}
 	\caption{The local coordinates for the first Ellwood vertex $\mathcal{E}_1$ when $\beta =1$ (\emph{left}) and $\beta = 1/2$ (\emph{right}). Note that the local coordinate around the bulk puncture at $z=i$ is singular.}
 	\label{fig:E1-coordinates}
 \end{figure}

We begin with $n=1$---the disk with one bulk and one boundary puncture. We respectively place these punctures at $z=i$ and $z=0$ on the uniformizing UHP and introduce the local coordinate around the boundary puncture with the following $SL(2,\mathbb{R})$ map:
\begin{align} \label{eq:B.1}
	w_1^{(1,1)} (z) = {1 \over \beta} z \, ,
\end{align}
where $0 < \beta \leq 1$ is a parameter that has to be constrained such that the local coordinates around the punctures do not overlap. Throughout this appendix we adopt the notation $w_k^{(m,n)} (z)$ for the local coordinate around the $k$th boundary puncture on the disk with $m$ bulk and $n$ boundary punctures. Two instances of~\eqref{eq:B.1} are shown in figure~\ref{fig:E1-coordinates}. Recall that we exclusively insert a weight-$(0,0)$ primary operator~\eqref{eq:5.8} at the bulk puncture in our application, which eliminates the need for keeping track of its associated local coordinate. This, in particular, permits us to take it to be singular and set $\beta = 1$. In this case~\eqref{eq:B.1} is a mere identity map and we have $\rho_1^{(1,1)}  = 1$ for its mapping radius. We adopt this extremely simple choice for the first Ellwood vertex~$\mathcal{E}_1$.
 
\begin{figure}[t]
	\centering
	\includegraphics[width=0.74\textwidth]{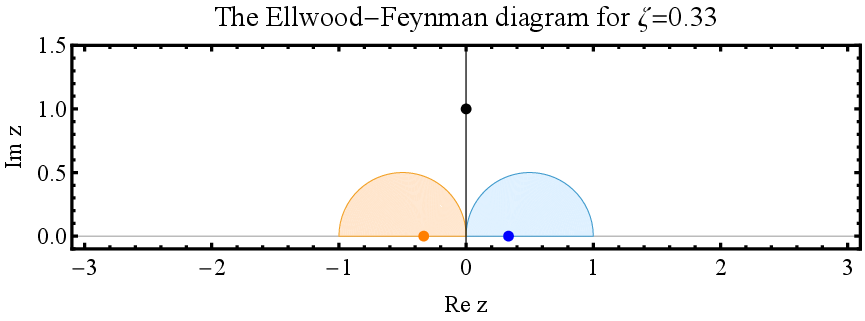}
	\includegraphics[width=0.74\textwidth]{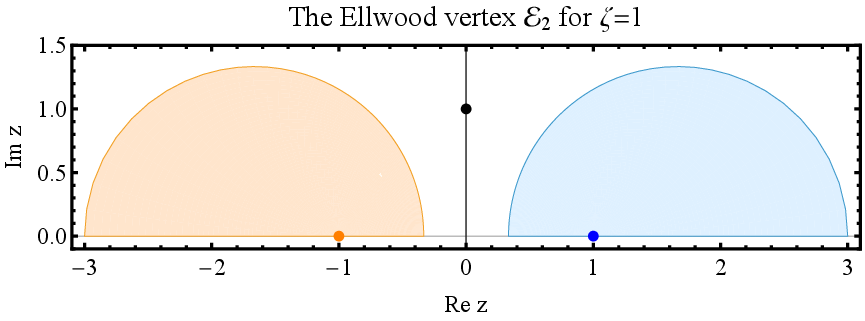}
	\includegraphics[width=0.74\textwidth]{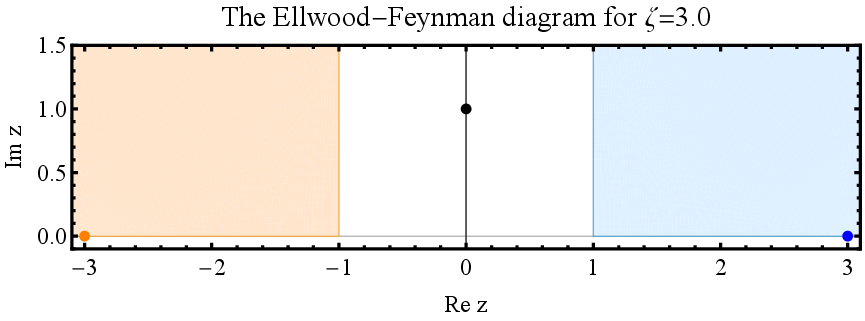}
	\caption{The local coordinates for the disk with one bulk and two boundary punctures. The top and bottom figures are the Ellwood-Feynman diagrams with $q=1$.}
	\label{fig:E2-coordinates}
\end{figure}

Next, we construct the ``Ellwood-Feynman'' diagrams containing two boundary punctures and a bulk puncture by sewing~\eqref{eq:B.1} to the $SL(2,\mathbb{R})_3$ cubic vertex~\eqref{eq:B.1}:
\begin{align} \label{eq:B.2}
 	w_1^{(1,1)} (z) w_3^{(0,3)} (z') = -q, 
 	\quad \quad 0<q<1 \, . \,
\end{align}
Solving for $z = z(z')$ and subsequently mapping the punctures at $z' = 1 , \infty$ to
\begin{align}
	z = \pm \zeta(q) \, , \quad \quad \zeta(q) = {q \over 3} \, ,
\end{align}
while fixing the bulk puncture at $z=i$, we obtain
\begin{align} \label{eq:B.3}
	z_1^{(1,2)} (w) = &  - q \, \frac{w+1}{w-3}, \quad \quad
	z_2^{(1,2)} (w) = q \, \frac{w-1}{w+3}\, ,
\end{align}
for the local coordinates around the boundary punctures respectively. We remind that there is another channel in the Feynman region whose associated local coordinates can be obtained by inverting~\eqref{eq:B.3} using $z \to - 1/z$ and swapping the order of the punctures. Examples for the resulting local coordinates for these Ellwood-Feynman diagrams are shown in figure~\ref{fig:E2-coordinates}.

The surfaces obtained like above cover the Feynman region of the moduli space of disks with one bulk and two boundary punctures. More specifically, the resulting surfaces cover the regions $0 < \zeta < 1/3$ and $3 < \zeta$ upon taking $0 < \zeta < \infty$ as our modulus. This leaves a gap
\begin{align} \label{eq:B.5}
	\zeta(q=1) = {1 \over 3} \leq \zeta \leq 3 = {1 \over \zeta(q=1)}\, ,
\end{align}
which is the integration region for the second Ellwood vertex $\mathcal{E}_2$. Similar to string vertices of previous appendix, the local coordinates in~\eqref{eq:B.5} should be defined so that the relevant geometric master equation is satisfied~\cite{Zwiebach:1997fe}. We ensure this by defining them using the following $SL(2,\mathbb{R})$ maps
\begin{align}
	z_1^{(1,2)} (w; \zeta) = \frac{a_1 (\zeta) + \zeta}{c_1 (\zeta) + 1}, 
	\quad \quad
	z_2^{(1,2)} (w; \zeta) = \frac{a_2 (\zeta) - \zeta}{c_2 (\zeta) + 1} ,
\end{align}
where the coefficients functions $a_i(\zeta), c_i(\zeta)$ ($i=1,2$) are determined by linearly interpolating the coefficients of~\eqref{eq:B.3} to their inverted cousins. An example for the resulting local coordinates is shown in figure~\ref{fig:E2-coordinates}. It is straightforward to establish that the local coordinates defined in this manner are cyclic and non-overlapping.

Finally, we consider disks with one bulk and three boundary punctures. We do not need to keep track of the local coordinates for these surfaces as explained below~\eqref{eq:5.32}. However, we still need to work out the Feynman decomposition of this moduli space---this is required for evaluating the moduli integral~\eqref{eq:5.30} using the SFT prescription~\eqref{eq:3.27a}. So, we begin by fixing the bulk puncture at $z=i$ and one of the boundary punctures at $z=0$ on the uniformizing UHP. The remaining punctures are placed at $u_1$ and $u_2$ and we assume the left handed cyclic ordering $(u_1,u_2,0)$ for them. In order to have a compact visualization for the moduli space we map this UHP to the disk frame with the Cayley transform
\begin{align} \label{eq:B.7}
	z_\text{Cayley} = f_\text{Cayley} (z) = & \frac{i-z}{i+z}.
\end{align}
The bulk puncture at $z=i$ is mapped to the center of the disk $z_\text{Cayley} = 0$ while the boundary puncture at $z=0$ is mapped to $z_\text{Cayley} = 1$. The remaining punctures are mapped to the unit circle,
\begin{align}
	e^{i \theta_1} = & f_\text{Cayley} (u_1), \quad \quad
	e^{i \theta_2 } = f_\text{Cayley} (u_2) \, ,
\end{align}
with the angles $\theta_1, \theta_2$ constrained as
\begin{align}
	0 \leq \theta_1 \leq 2\pi, \quad \quad
	0 \leq \theta_2 \leq 2\pi, \quad \quad
	\theta_2 \leq \theta_1 \, ,
\end{align}
according to our left handed conventions. This implies that the relevant moduli space takes a triangular shape. Figure \ref{fig:decomposition} illustrates this triangle and the associated angular conventions on the disk frame. Using different types of sewing fixtures, it can be shown that this triangle-shaped moduli space decomposes into four distinct regions shown in figure \ref{fig:decomposition}. In order:
\begin{enumerate}
	\item The~\textcolor{red}{red} region is the Feynman region obtained by sewing the second Ellwood vertex $\mathcal{E}_2$ to the cubic $SL(2,\mathbb{R})_3$ vertex. The positions of the punctures $(u_1,u_2)$ are given as a function of $(q,\zeta)$ in this region, where $q$ is the sewing parameter and $\zeta$ is in the range~\eqref{eq:B.5}.
	
	\item The~\textcolor{blue}{blue} region is the Feynman region obtained by sewing the first Ellwood vertex $\mathcal{E}_1$ to the quartic $SL(2,\mathbb{R})_3$ vertex. The positions of the punctures $(u_1,u_2)$ are given as a function of $(q,u)$ in this region, where $q$ is the sewing parameter and $u$ is in the range~\eqref{eq:A.16}. Since there are four cyclic copies of the quartic $SL(2,\mathbb{R})_3$ vertex, there are also four cyclic copies for each component of the~\textcolor{blue}{blue} region. An example for this is illustrated in the~\textcolor{purple}{purple} box in figure~\ref{fig:decomposition}.
	
	\item The~\textcolor{green}{green} region is the Feynman region whose diagrams contain two propagators. The positions of the puncture $(u_1,u_2)$ are given as a function of $(q_1,q_2)$, where $0 < q_i < 1$ are the sewing parameters associated with these two propagators. 
	
	\item The~\textcolor{yellow}{yellow} hexagonal patch is the vertex region for the third Ellwood vertex $\mathcal{E}_3$ that is left uncovered by the Feynman diagrams above. It has four distinct cyclic copies resulting from bordering different cyclic copies of the~\textcolor{blue}{blue} region.
\end{enumerate}
Treating the divergences in~\eqref{eq:5.30} associated with the first and second regions and evaluating the moduli integrals over them proceed exactly like~\eqref{eq:3.27a}. Doing these for the third region is also similar, but one should be mindful that there are divergences due to having two distinct propagators, which need to be handled independently from each other. Lastly, there remains an integration over the third Ellwood vertex $\mathcal{E}_3$, but this is manifestly finite and can be straightforwardly computed.

\section{Position zero mode correlator}  \label{sec:D}

In this appendix we derive a formula for the position zero mode correlator used during the evaluations of~\eqref{eq:4.41a}-\eqref{eq:4.40}. Instead of only working with these somewhat specialized situations, we keep the discussion general and focus on the correlators of the form
\begin{align} \label{eq:C.1}
	H(k_0;  k_1, \cdots, k_n) \equiv
	\left\langle
	e^{i k_0 \cdot \mathcal{X} } 
	\prod_{j=1}^n e^{i k_j \cdot X(u_j, u_j)} 
	\right\rangle^{X^\mu}_\UHP \, ,
\end{align}
containing $n$ boundary plane wave vertex operators ($u_1 > \cdots > u_n$) together with an insertion of $e^{i k_0 \cdot \mathcal{X}}$ in $X^\mu$ BCFT subject to Neumann boundary conditions. Here $\mathcal{X}^\mu$ is the operator
\begin{align} \label{eq:C.2}
	\mathcal{X}^\mu = \int_0^\pi d \theta \, g(\theta) 
	X^\mu \big(u(\theta), \overline{u (\theta)}\big)
	\quad \quad \text{with} \quad \quad
	\int_0^\pi g(\theta)  = 1 \, ,
\end{align}
where $u(\theta)$ for $\theta \in \left[0,\pi \right]$ parametrizes a curve in UHP whose ends are anchored on the real axis and $g(\theta)$ is a normalized weight function. Our objective is to express~\eqref{eq:C.1} in terms of a correlator of boundary plane wave vertex operators so that it can be evaluated using the standard CFT methods. We highlight that the form of $\mathcal{X}^\mu$ in~\eqref{eq:C.2} is inspired by trying to express the position zero mode $x_0^\mu$ in an arbitrary conformal frame. Later we specialize to the case where $\mathcal{X}^\mu$ is given by a conformal transformation of the position zero mode.

We begin our discussion with the following recursive formula for the~\emph{chiral} plane wave vertex operators on $\mathbb{C}$:
\begin{align} \label{eq:3.1}
	\left\langle 
	\prod_{j=0}^n e^{i k_j \cdot X(u_j)} 
	\right\rangle_{\mathbb{C}}^{X^\mu} 
	= (u_{01})^{{1 \over 2} k_0 \cdot k_1}
	\prod_{i=2}^n \left(
	{u_{0 i} \over u_{1 i}}
	\right)^{{1\over 2} k_0 \cdot k_i}
	\left\langle e^{i (k_0 + k_1 ) \cdot X(u_1)} 
	\prod_{j=2}^n e^{i k_j \cdot X(u_j)}  \right\rangle_{\mathbb{C}}^{X^\mu}  \, ,
\end{align}
where $u_{ij} = u_i - u_j $. After relabeling the momenta as $\{ \ell_0, \cdots \ell_N; k_1 \cdots, k_n \}$
and positions as $\{ s_0, \cdots s_N; u_1 \cdots, u_n \}$ and combining the vertex operators with momenta $\ell_0$ and $k_1$ according to~\eqref{eq:3.1}, we find
\begin{align}
	&\left\langle X^{\mu_0}(s_0) 
	\prod_{m=1}^N e^{i \ell_m \cdot X(s_m)} 
	 \prod_{j=1}^n e^{i k_j \cdot X(u_j)} 
	\right\rangle_{\mathbb{C}}^{X^\mu} 
	\\ \nonumber
	&\hspace{0.5in} = - {i \over 2} \left[
	k_1^{\mu_0} \log (s_0 - u_1)
	+ \sum_{m=1}^N \ell_m^{\mu_0}\log \left( {s_0 - s_m \over u_1 - s_m} \right)
	+ \sum_{j=2}^n k_j^{\mu_0} \log \left( {s_0 -u_j \over u_1 - u_j} \right)
	\right] 
	\\ \nonumber
	& \hspace{1.5in}
	\times \left\langle \prod_{m=1}^N e^{i \ell_m \cdot X(s_m)} 
	\prod_{j=1}^n e^{i k_j \cdot X(u_j)}   \right\rangle_{\mathbb{C}}^{X^\mu} 
	\\ \nonumber
	&\hspace{1in}
	- i {\partial \over \partial{k_{1,\mu_0}}}
	\left\langle   \prod_{m=1}^N e^{i \ell_m \cdot X(s_m)}
	 \prod_{j=1}^n e^{i k_j \cdot X(u_j)}  \right\rangle_{\mathbb{C}}^{X^\mu} \, ,
\end{align}
by differentiating with respect to $i \ell_{0, \mu_0}$ and setting it to zero. Analogously, differentiating with respect to the rest of $i \ell_{i, \mu_r}$ and setting them to zero yields
\begin{align}  \label{eq:C.5}
	&\left\langle  
	\prod_{m=0}^N X^{\mu_m} (s_m) 
	\prod_{j=1}^n e^{i k_j \cdot X(u_j)} 
	\right\rangle_{\mathbb{C}}^{X^\mu} 
	\\ \nonumber
	&\hspace{0.5in}= - {1 \over 2} \sum_{i=1}^N \eta^{\mu_0 \mu_i} 
	 \log  \left( {s_0 - s_i \over u_1 - s_i} \right)
	\left\langle 
	\prod_{\substack{m=1 \\ m \neq i}}^N X^{\mu_m} (s_m)
	\prod_{j=1}^n e^{i k_j \cdot X(u_j)} 
	\right\rangle_{\mathbb{C}}^{X^\mu} 
	\nonumber \\
	&\hspace{0.5in}- {i \over 2} \left[
	k_1^{\mu_0} \log(s_0-u_1)
	+ \sum_{j=2}^n k_j^{\mu_0} \log \left({s_0 - u_j \over u_1 -u_j}\right)
	\right]
	\left\langle
	\prod_{m=1}^N X^{\mu_m}(s_m) 
	\prod_{j=1}^n e^{i k_j \cdot X(u_j)}
	\right\rangle_{\mathbb{C}}^{X^\mu} 
	\nonumber \\
	&\hspace{0.5in}
	- i {\partial \over \partial{k_{1,\mu_0}}}
	 \left\langle
	\prod_{m=1}^N X^{\mu_m}(s_m) 
	\prod_{j=1}^n e^{i k_j \cdot X(u_j)}  
	\right\rangle_{\mathbb{C}}^{X^\mu} 
	\nonumber \, .
\end{align}
This is a recursive formula for the $X^{\mu_m}(s_m)$ insertions. For our purposes, however, we actually need a recursive formula for the bulk $X^{\mu_m}(s_m, \overline{s_m})$ insertions on UHP with the boundary plane wave vertex operators $e^{i k \cdot X(u, u)} $---inspect~\eqref{eq:C.1}. Using the doubling relations
\begin{align}
	e^{i k\cdot X(u,u)}  = e^{2 i k \cdot X(u)} \, ,
	\quad \quad 
	X^\mu(s, \overline{s}) = X^\mu(s) + X^\mu(\overline{s}) \, ,
\end{align}
we can obtain the desired recursive formula by adapting it from~\eqref{eq:C.5}:
\begin{align} \label{eq:C.7}
	&\left\langle  
	\prod_{m=0}^N X^{\mu_m} (s_m, \overline{s_m}) 
	\prod_{j=1}^n e^{i k_j \cdot X(u_j, u_j)} 
	\right\rangle_{\UHP}^{X^\mu} 
	\\ \nonumber
	&\hspace{0.5in}= - {1 \over 2} \sum_{i=1}^N \eta^{\mu_0 \mu_i} 
	\left[ \log  \left| {s_0 - s_i \over u_1 - s_i} \right|^2
	+\log  \left| {s_0 - \overline{s_i} \over u_1 - \overline{s_i}} \right|^2
	\right]
	\left\langle 
	\prod_{\substack{m=1 \\ m \neq i}}^N X^{\mu_m} (s_m, \overline{s_m})
	\prod_{j=1}^n e^{i k_j \cdot X(u_j,u_j)} 
	\right\rangle_{\UHP}^{X^\mu} 
	\\ \nonumber
	&\hspace{0.5in}- {i} \left[
	k_1^{\mu_0} \log|s_0-u_1|^2
	+ \sum_{j=2}^n k_j^{\mu_0} \log \left| {s_0 - u_j \over u_1 -u_j}\right|^2
	\right]
	\left\langle
	\prod_{m=1}^N X^{\mu_m}(s_m, \overline{s_m}) 
	\prod_{j=1}^n e^{i k_j \cdot X(u_j, u_j)}
	\right\rangle_{\UHP}^{X^\mu} 
	\\ \nonumber
	&\hspace{0.5in}
	- i {\partial \over \partial{k_{1,\mu_0}}}
	\left\langle
	\prod_{m=1}^N X^{\mu_m}(s_m, \overline{s_m}) 
	\prod_{j=1}^n e^{i k_j \cdot X(u_j, u_j)}  
	\right\rangle_{\UHP}^{X^\mu}  \, .
\end{align}

Now, we return back to~\eqref{eq:C.1} and take its derivative with respect to $i k_{0, \mu_0}$,
\begin{align}
	- i {\partial H \over \partial k_{0,{\mu_0}}}(k_0;  k_1, \cdots, k_n)
	= \left\langle  \mathcal{X}^{\mu_0}
	e^{i k_0 \cdot \mathcal{X} } 
	\prod_{j=1}^n e^{i k_j \cdot X(u_j, u_j)} 
	\right\rangle^{X^\mu}_\UHP  \, ,
\end{align}
in order to find a differential equation satisfied by the function $H$. Expanding $e^{i k_0 \cdot \mathcal{X} } $ in powers of $k_0$, we observe that its $N$-th term has the following form:
\begin{align}
	&
	\left\langle
	\mathcal{X}^{\mu_0} { (i k_0 \cdot \mathcal{X})^N  \over N!}
	\prod_{j=1}^n e^{i k_j \cdot X(u_j,u_j)} 
	\right\rangle_\UHP^{X^\mu}
	\\ \nonumber
	&\hspace{0.25in}=  {1 \over N!} \left[ \prod_{i=0}^N \int_0^\pi d \theta_i \, g(\theta_i) \right]
	 \left\langle
	X^{\mu_0} \left(u(\theta_0), \overline{u(\theta_0)}\right)
	\prod_{m=1}^N i k_0 \cdot X \left(u(\theta_m), \overline{u(\theta_m)} \right) 
	\prod_{j=1}^n e^{i k_j \cdot X(u_j, u_j)} 
	\right\rangle_\UHP^{X^\mu} \, .
\end{align}
We can apply the recursive formula~\eqref{eq:C.7} above to eliminate the $X^{\mu_0}$ insertion. This produces
\begin{align} \label{eq:C.10}
	&\left\langle
	\mathcal{X}^{\mu_0} { (i k_0 \cdot \mathcal{X})^N  \over N!}
	\prod_{j=1}^n e^{i k_j \cdot X(u_j,u_j)} 
	\right\rangle_\UHP^{X^\mu} 
	\\ \nonumber
	&\hspace{0.1in}=
	- i k_0^{\mu_0} {1 \over 2}
	\int_0^\pi d \theta \, g(\theta) \int_0^\pi d \theta'  \, g(\theta') \,
	\left[
	\log \left| {u(\theta) - u(\theta') \over u_1 - u (\theta')} \right|^2
	+ \log \left| {u(\theta) - \overline{u(\theta')} \over u_1 - \overline{u (\theta')}} \right|^2 
	\right]
	\nonumber \\
	&\hspace{1.1in}
	\times\left\langle { (i k_0 \cdot \mathcal{X})^{N-1} \over (N-1)!}
	\prod_{j=1}^n e^{i k_j \cdot X(u_j,u_j)} 
	\right\rangle_\UHP^{X^0}
	\nonumber \\
	&\hspace{0.1in} - \int_0^\pi d \theta \, g(\theta) \,
	\left[
	i k_1^{\mu_0} \log |u(\theta) - u_1|^2 
	+ \sum_{j=2}^n i k_j^{\mu_0} \log \left| {u(\theta) - u_j \over u_1 - u_j} \right|^2
	\right]
	\left\langle
	{( i k_0 \cdot \mathcal{X} )^N \over N!}
	\prod_{j=1}^n e^{i k_j \cdot X(u_j,u_j)} 
	\right\rangle_\UHP^{X^\mu}
	\nonumber \\
	&\hspace{0.1in}
	- i {\partial \over \partial{k_{1,\mu_0}}}
	\left\langle
	{(i k_0 \cdot \mathcal{X})^N \over N!}
	\prod_{j=1}^n e^{i k_j \cdot X(u_j, u_j)}  
	\right\rangle_\UHP^{X^\mu} \, , \nonumber
\end{align}
after using the symmetry among $\mathcal{X}$s for the first term. Introducing
\begin{subequations} \label{eq:C.11}
	\begin{align}
		&U_{00} \equiv \exp \left[ {1 \over 4}
		\int_0^\pi 
		d \theta  \, g(\theta) \int_0^\pi d \theta' \, g(\theta')
		\left(
		\log| u(\theta) - u(\theta')|^2 
		+ \log| u(\theta) - \overline{u(\theta')}|^2
		\right)
		\right] \, , \label{eq:C.11a}
		\\
		& U_{0 j} \equiv \exp \left[ 
		{1 \over 2}\int_0^\pi d \theta \, g(\theta)
		\log\left| u(\theta) - u_j \right|^2
		\right] \, , \quad \quad j = 1, \cdots , n \, ,
	\end{align}
\end{subequations}
and summing~\eqref{eq:C.10} over $N$, we find the desired differential equation:
\begin{align} \label{eq:C.12}
	&\left( {\partial \over \partial k_{0, \mu}} -  {\partial \over \partial k_{1, \mu}} \right) 
	H(k_0; k_1, \cdots, k_n)
	\\ \nonumber
	&\hspace{1in}= 2 \left(k_0^\mu \log {U_{00} \over U_{01}} 
	+ k_1^\mu \log U_{01}
	+ \sum_{j=2}^n k_j^\mu \log { U_{0 j} \over |u_{1j}| }
	\right) H(k_0; k_1, \cdots, k_n)
	\, .
\end{align}

In order to solve~\eqref{eq:C.12} we make the following ansatz
\begin{align} \label{eq:C.13}
	H(k_0; k_1, \cdots, k_n) = N(k_0 + k_1, k_1, \cdots, k_n) H(0; k_0 + k_1, k_2, \cdots, k_n) \, .
\end{align}
inspired by our goal of eliminating $\mathcal{X}^\mu$ in the correlators and the structure of the derivatives in~\eqref{eq:C.12}. Plugging it to~\eqref{eq:C.13}, we obtain an ordinary differential equation for the function $N$
\begin{align} \label{eq:C.14}
	{\partial N \over \partial q_{1,\mu}} (q_0, q_1, \cdots q_n)
	= - 2 \left(
	q_0^\mu \log {U_{00} \over U_{01}} 
	+ q_1^\mu \log {U_{01}^2 \over U_{00}} 
	+ \sum_{j=2}^n q_j^\mu \log { U_{0 j} \over |u_{1j}| }
	\right) N(q_0, q_1, \cdots, q_n) \, .
\end{align}
Moreover, this function satisfies
\begin{align} \label{eq:C.15}
	N(q_1, q_1, \cdots, q_n)  = 1 \, ,
\end{align}
upon setting $k_0 = 0$ in~\eqref{eq:C.13}, which provides the initial conditions for the differential equation~\eqref{eq:C.14}. Solving~\eqref{eq:C.14} accordingly, we find
\begin{align}
	N(q_0, q_1, \cdots, q_n) = \left( U_{00}\right)^{q_0^2}
	\left( { U_{00} \over U_{01} } \right)^{- 2 q_0 \cdot q_1 }
	\left( {U_{01}^2 \over U_{00}} \right)^{-q_1^2}
	\prod_{j=2}^n \left({ U_{0 j} \over |u_{1j}| }  \right)^{2  (q_0 -q_1 ) \cdot q_j }\, ,
\end{align}
and in extension
\begin{align} \label{eq:C.17}
	H(k_0; k_1, \cdots, k_n) = (U_{00})^{k_0^2} (U_{01})^{2 k_0 \cdot k_1}
	\prod_{j=2}^n \left({ U_{0 j} \over |u_{1j}| }  \right)^{2 k_0 \cdot k_j } 
	H(0; k_0 + k_1, k_2, \cdots, k_n) \, .
\end{align}
This relation demonstrates that the correlator~\eqref{eq:C.1} can be expressed in terms of a correlator containing only boundary plane wave operators---there are no $e^{i k_0 \cdot \mathcal{X}}$ insertion on the right hand side. Instead, we compensate the effects of $e^{i k_0 \cdot \mathcal{X}}$ insertion by transferring its momentum $k_0$ to the first plane wave vertex operator and including the prefactors~\eqref{eq:C.11}. Note that setting $k_1 = 0$ is admissible in~\eqref{eq:C.17}, which corresponds having no plane wave vertex operator at $z=u_1$ originally.

A further simplification is possible when we specialize to the case where $\mathcal{X}^\mu$ is a conformal transformation of the position zero mode operator $x_0^\mu$ by a biholomorphic map $f$ from the upper unit half-disk $D_{1/2}$ to the UHP,
\begin{align} \label{eq:C.18}
	\mathcal{X}^\mu = f \circ x_0^\mu = {1 \over \pi} \int_0^\pi d \theta X^\mu \left( 
	f(e^{i \theta}), \overline{f(e^{i \theta})}\right) \, ,
\end{align}
which is precisely the situation arising in our application. The origin of $D_{1/2}$ is mapped to the point $z = f(0) \equiv u_0$ on the UHP. It is natural to transfer the momentum $k_0$ of $e^{i k_0 \cdot \mathcal{X}}$ to this point using~\eqref{eq:C.17} while assuming that it initially does not carry any momentum. We call this point~\emph{virtual puncture} as in the main text.

It should not be too surprising that the specialized form of~\eqref{eq:C.18} leads to a simplification in the prefactors~\eqref{eq:C.11}. Consider $U_{0j}$ first. We have
\begin{align}
	\log U_{0 j} 
	= {1 \over 2 \pi} \int_0^{2 \pi} d \theta \left| f\left(e^{i \theta}\right) - u_j \right| \, ,
	\quad \quad
	j = 1, \cdots, n
	\, ,
\end{align}
after extending the biholomorphic function $f$ from the upper unit half-disk to the entire unit disk in an obvious manner. This integral can be easily evaluated using the Jensen's formula given that $f$ is invertible. We find
\begin{align} \label{eq:C.20}
	U_{0 j} = \begin{cases}
		\displaystyle \left| {u_{0j} \over f^{-1}(u_j) }\right| &\text{for} \quad u_j \in f \circ D_{1/2} \\[3ex]
		\displaystyle | u_{0j} | &\text{for} \quad u_j \notin f \circ D_{1/2}
	\end{cases} \, .
\end{align}
Interestingly, the integral produces different results based on whether the contour for $\mathcal{X}^\mu$ surrounds the $i$-th puncture on the UHP or not. 

Next, we rewrite $U_{00}$~\eqref{eq:C.11a} as
\begin{align} \label{eq:C.22}
	\log U_{00} &=
	{1 \over 4\pi^2} \int_0^\pi d \theta \int_{-\pi}^\pi d \theta'
	\left[ \log\left(f(e^{i \theta}) - f(e^{i \theta'})\right) 
	+ \log\left(f(e^{-i \theta}) - f(e^{i \theta'}) \right)
	\right]
	\\ \nonumber
	&= {1 \over 2 \pi} \int_0^\pi d \theta \oint\limits_{|\xi|=1} 
	{d \xi \over 2 \pi i} \, {1 \over \xi} 
	\left[ \log\left( f(e^{i \theta}) - f(\xi)\right) 
	+ \log\left(f(e^{-i \theta}) - f(\xi) \right)
	\right] \, .
\end{align}
Observe that the integrand is holomorphic for $|\xi| < 1$ since the arguments of the logarithms do not vanish in the interior of the disk $|\xi|=1$. Picking up the residue at $\xi = 0$ by the Cauchy's theorem,
\begin{align}
	\log U_{00} &= {1 \over 2 \pi} \int_0^\pi d \theta
	\left[ \log\left(f(e^{i \theta})-u_0\right) 
	+ \log\left(f(e^{-i \theta}) - u_0 \right) \right]
	\\ \nonumber
	&= {1 \over 2 \pi} \int_0^\pi d \theta
	\left[ \log\left({f(e^{i \theta})-u_0 \over e^{i \theta}} \right) 
	+ \log\left({ f(e^{-i \theta})-u_0 \over e^{-i \theta}} \right)
	\right] \, .
\end{align}
In the second line we added/subtracted $\log e^{i \theta}$ to eliminate the branch cuts in the resulting logarithms. Repeating the trick in~\eqref{eq:C.22} for the angle $\theta$ now,
\begin{align} \label{eq:C.24}
	\log U_{00}	&= \oint\limits_{|\xi| = 1}{d \xi \over 2 \pi i} \, {1 \over \xi}
	\log \left( { f(\xi) - u_0 \over \xi} \right) 
	= \log \partial f (0) 
	\quad \implies \quad
	U_{00} = \partial f(0)
	\, .
\end{align}
In other words, $U_{00}$ becomes the mapping radius of the biholomorphic map $f$. 

Finally, using the results~\eqref{eq:C.20} and~\eqref{eq:C.24} in~\eqref{eq:C.17} with suitably shuffled indices, we obtain a remarkably simple-looking formula:
\begin{align} \label{eq:C.24a}
	\left\langle
	f \circ e^{i k_0 \cdot x} 
	\prod_{j=1}^n e^{i k_j \cdot X(u_j, u_j)} 
	\right\rangle^{X^\mu}_\UHP 
	&= \partial f(0)^{k_0^2} 
	\prod_{u_j \in f \circ D_{1/2}} 
	\left| f^{-1}(u_j) \right|^{-2 k_0 \cdot k_j}
	\left\langle
	\prod_{j=0}^n e^{i k_j \cdot X(u_j, u_j)} 
	\right\rangle^{X^\mu}_\UHP  \, .
\end{align}
This is the relation employed in~\eqref{eq:4.41a}-\eqref{eq:4.40}. Observe that punctures outside the image of the upper unit half-disk $D_{1/2}$ do not contribute to the prefactors here---they just spectate.


\providecommand{\href}[2]{#2}
\end{document}